\documentclass[preprint]{elsarticle}

\usepackage{amssymb}
\usepackage{amsbsy,amsthm}
\usepackage{amsmath}
\usepackage{xcolor}
\usepackage{pdflscape}
\usepackage{afterpage}
\usepackage{capt-of}
\usepackage{fullpage}
\usepackage{mathtools}
\usepackage{mathtools}

\usepackage[normal,tight,center]{subfigure}
\usepackage{tikz}
\usepackage[ruled]{algorithm2e}
\usetikzlibrary{positioning,shapes}
\usepackage[normal,tight,center]{subfigure}
\usepackage{array}
\usepackage{calc}
\usepackage[thinlines]{easytable}
\usepackage{tabu}

\newcommand{\colr}[1]{\textcolor{red}{#1}}

\newcommand{\rtab}[1] {Table~\ref{tab:#1}}

\newcommand{\rsec}[1] {section~\ref{sec:#1}}

\newcommand{\rfig}[1] {Fig.~\ref{fig:#1}}
\newcommand{\eqn}[1]{Eq.~(\ref{eqn:#1})}

\newcommand{\eqnlabel}[1]{\label{eqn:#1}}
\newcommand{\figlabel}[1]{\label{fig:#1}}

\newcommand{\vo}[1]{\mathbf{#1}}

\def\A0 {\vo{A}_0}
\def\Id {\vo{I}}

\def\re {R_\lambda}

\def\la {\langle}
\def\ra {\rangle}
\def\eps {\epsilon}
\def\Om {\Omega}

\def\kp {\kappa}
\def\kt {\tilde{k}}
\def\mmm {{(1)}}
\def\nnn {{(2)}}

\def\dx {\Delta x}
\def\dt {\Delta t}
\def\ns {M}

\def\bi {\begin{itemize}}
\def\ei {\end{itemize}}
\def\bl {\begin{list}}
\def\el {\end{list}}
\def\ben {\begin{enumerate}}
\def\een {\end{enumerate}}
\def\be {\begin{equation}}
\def\ee {\end{equation}}
\def\bdm {\begin{displaymath}}
\def\edm {\end{displaymath}}
\def\bp {\begin{picture}(0,0)}
\def\ep {\end{picture}}
\def\la  {\langle}
\def\ra  {\rangle}
\def\ap  {{\approx}}

\def\Cr {\texttt{c}_r}
\def\Sr {\texttt{s}_r}
\def\Tl {\texttt{t}_\ell}

\def\rdml {r_{d,m}(L)}
\def\rdm {r_{d,m}}
\def\rcml {r_{c,m}(L)}
\def\rcm {r_{c,m}}

\def\rdma {r_{d,m}^a}

\def\rcma {r_{c,m}^a}

\newcommand{\U}[2] {{u_{#1}^{#2}}}
\newcommand{\ran}[1] {{\tilde{#1}}}
\renewcommand{\k}[1] {\ran{k}_{#1}}

\journal{Journal of Computational Physics}

\begin{document}

\begin{frontmatter}



\title{Direct Numerical Simulations of turbulent flows using high-order Asynchrony-Tolerant schemes: accuracy and performance}


%\author[label1]{Komal Kumari}
\address[label1]{Department of Aerospace Engineering, Texas A\&M University, College Station, TX 77843, United States}
\author[label1]{Komal Kumari}%\corref{cor1}}
%\ead{raktim@tamu.edu}
\author[label1]{Diego A.\ Donzis\corref{cor1}}
\ead{donzis@tamu.edu}
\cortext[cor1]{Corresponding author.}

\begin{abstract}
Direct numerical simulations (DNS)
are an indispensable tool for understanding the
fundamental physics of turbulent flows. Because of their steep
increase in computational cost with Reynolds number ($\re$),
well-resolved DNS are realizable only
on massively parallel supercomputers,
even at moderate $\re$.
However, at extreme scales,
the communications and synchronizations between processing elements (PEs)
involved in current approaches
become exceedingly expensive and are expected to be a major bottleneck to
scalability. In order to overcome this challenge,
we developed algorithms using the so-called Asynchrony-Tolerant (AT)
schemes that relax communication and synchronization constraints at a mathematical level, to
perform DNS of decaying and solenoidally forced compressible turbulence.
Asynchrony is introduced
using two approaches, one that avoids synchronizations and the other that avoids
communications. These result in periodic and random delays, respectively,
at PE boundaries.
We show that both asynchronous algorithms accurately resolve the
large-scale and small-scale motions of turbulence, including
instantaneous and intermittent fields.
We also show that in asynchronous simulations the communication time is a relatively
smaller fraction of the total computation time, especially at large processor
count, compared to standard synchronous simulations. As a consequence,
we observe improved parallel scalability up to $262144$ processors
 for both asynchronous algorithms.
\end{abstract}


\end{frontmatter}

%\begin{comment}
% \linenumbers

\section{Introduction}
Turbulence is the most common fluid state of motion
and is inherent in a large number of natural and physical
phenomena. The physics of turbulence can be modelled
mathematically by the Navier-Stokes (NS) equations under the
continuum limit. These equations are highly non-local and
non-linear and have resisted analytical analyses except for overly simplified
canonical problems. Numerical simulations, thus, serve as an
indispensable tool for understanding turbulence.
One of the most important characteristics of turbulence
is the inherent wide range of spatial and temporal scales.
This range of scales increases with the Reynolds Number ($Re_{\lambda}$), the ratio
between the inertial and viscous forces, which is typically very high in applications.
In accurate numerical simulations the computational domain
has to be large enough to accommodate the largest scales of motion in the flow
and the grid spacing small enough to resolve the so-called
Kolmogorov scale \cite{K41}, the smallest dynamically relevant
scale in a turbulent flow.
Furthermore, the simulation time should be sufficiently long to
capture the slow evolution of the largest scale while the time-step size
should be small enough to capture the fast Kolmogorov time scale characteristic
of the smallest scales.
Simulations that follow these stringent constraints and
consequently, accurately resolve the physics
of all relevant scales are known as Direct Numerical Simulations (DNS)
\cite{MM98,IGK2009}.
Using classical scaling relations based on Kolmogorov ideas \cite{K41}
for grid spacing and a CFL condition for
time-step size, the computational work
grows steeply as $Re_{\lambda}^{6}$, though more recent work
suggests $Re_{\lambda}^{8}$ if
all intermittent events are to be resolved \cite{Victor2005}.
Due to this steep power-law dependence, high fidelity DNS are
computationally prohibitively expensive and even with highly scalable codes
run on today's most powerful supercomputers, unachievable for
conditions of practical relevance. %For incompressible turbulence,
%DNS has been done for $Re_{\lambda}=1300$ at a grid resolution up to $N^3=16384^3$
%\cite{Iyer2019}. On the other hand, because of its
%additional computational complexity, the largest DNS of compressible
%turbulence in literature has reached
% $Re_{\lambda}=450$ at a grid resolution of $N^3=2048^3$ \cite{DS2013,SD2016}.
%There are computationally less expensive alternatives to DNS, for
%example, Large Eddy Simulations (LES)
%where only large scales are resolved and the small scales are
%modelled or Reynolds Averaged Navier-Stokes (RANS) which solves
%only for the averaged quantities. This allows the use of LES and RANS
%at realistic conditions but at the cost of
%physics of unresolved or modelled scales.
%Our goal is to push the limits of DNS to %by increase the realizable $Re_{\lambda}$
%and resolution and facilitate DNS at
%levels of complexity unachieveable by current state-of-the-art solvers and provide new
%insight into unprecedented levels of physical relevance.
%In order to do so, we identify the major bottlenecks to scalability, such as communications and
%synchronizations and introduce
%novel numerical schemes and paradigm shifting algorithms to mitigate these bottlenecks.
%This in turn achieve DNS at levels of complexity unachieveable
%by current state-of-the-art sovers. This in turn will provide new
%insight into unprecedented scales of relevance that will facilitate
% development of high fidelity turbulence models.% for RANS and LES.

Several numerical methods have been used for
DNS of the Navier-Stokes equations to study turbulence, depending
upon the complexity of the domain and the nature of problem of interest.
Spectral methods \cite{Canuto1988}, known for accurate
computation of derivatives, have been used extensively in incompressible
simulations. However, these present
challenges when extended to non-periodic boundary conditions.
An alternative to these methods, that is more amenable to the choice of boundary conditions, is the
compact difference schemes that have spectral like
resolution \cite{Lele1992}. These are widely used for simulations of
multi-scale phenomena like turbulence \cite{Mahesh1995, Petersen2010,JD2012,CD2019,KD2019}. However,
computation of derivatives using compact schemes involves a system of linear
equations. This imposes constraints on the computational domain since
each processor must have entire range of data in the direction of computation of
derivative. Such codes require multiple collective
communication calls, which in turn can make communication time quite
significant  \cite{CCW+2005,DYP2008,JD2012} for both compact and spectral implementations.
%The optimized implementation has been discussed in
%\cite{JD2012} with scalability upto 262,144 cores.

%Apart from spectral methods and compact schemes,
Explicit finite difference schemes have also been extensively used for
approximation of derivatives in partial differential equations (PDEs)
including in massive simulations of turbulent reacting flows
\cite{Chen2000, Chen2009}.
For explicit schemes, the derivative
at a grid point in the domain is approximated as a linear combination
of the values at its neighboring points only. Because of this local dependence,
different processors can work concurrently on different parts of the
domain.
However, at the processor boundaries, processors need to
communicate to obtain data from the neighboring processors
in order to compute the derivatives. Although these
are local communications as opposed to the collective communications for
compact or spectral schemes,
processors still incur in overheads due to the
need to communicate and synchronize at every time step
to meet accuracy requirements. While simulations have been successfully done
using hundreds of thousands of processors \cite{JD2012,Lee2013,Chen2016},
%irrespective of the choice of numerical methods,
the synchronizations and communication overheads, irrespective of the
choice of numerical methods,
pose a serious challenge to scalability at extreme scales \cite{Dongarra2011}. %lead to processor idling and wastage
%of computation resources. The communications also
%limit the scalability of the solver %and restrict the number of processors that
%can be used
%and, consequently, the resolution
%or realizable $Re_{\lambda}$. %at which the problem can be solved.
In order to overcome this bottleneck,
some work has focused on %researchers have attempted to
relaxing the synchronization requirements among the processors
and perform so-called asynchronous numerical simulations.
%Because of severe degradation of accuracy of the standard finite difference
%schemes in the presence of asynchrony, alternative approaches had to
%be
%Early work on relaxation of synchronization at
%processor boundaries is limited to lower
%order numerical methods and simple physical problems \cite{Amitai1992},
%\cite{Amitai1994}.
%Of the two
%techniques used in literature for the same,
%the governing PDE is modified \cite{Amitai1992,Amitai1994,Ankita2017} to
%account for errors due to asynchrony \textit{a priori} to simulations in the first one,
%while the numerical methods are modified to allow for asynchrony
%\cite{DA2014,Ansumali2014,AD2017} in the second one.
Early work in the literature dealt with asynchronous simulations
but severely limited to lower orders of accuracy and
restricted to certain class of PDEs \cite{Amitai1992,Amitai1994,Ansumali2014,Ankita2017}.
A new and more generalized approach, extensible to arbitrarily high orders of accuracy,
has been recently developed \cite{DA2014,AD2017}
to derive the so-called Asynchrony-Tolerant (AT) finite-difference schemes.

However, these studies investigated numerical accuracy and stability for
simplified model problems in low dimensions.
The ability of these schemes to accurately simulate realistic
three-dimensional turbulent flows have not been done before. Without
careful assessment of the numerical and parallel performance of these
schemes it is unclear whether they can indeed provide a path towards
exascale simulations in future massively parallel systems.
This is the main thrust of this paper.

In particular, we use AT schemes %the details of which can be
%found in \cite{AD2017}. %The need for new numerical arises
%because standard finite difference schemes degrade to first order of
%accuracy when used asynchronously {\cite{DA2014}}.
%The work in literature on asynchronous simulations is limited
%to simple 1D problems and limited to utmost second
%order of accuracy.
to perform, a first of a kind, asynchronous simulation
of three-dimensional compressible turbulence. Our focus is on the effect of
asynchrony on important turbulent characteristics such as evolution of the
turbulent kinetic energy, the spectra and PDFs of velocity gradients, enstrophy and
dissipation.
In order to conduct these simulations, in addition to the selection of
appropriate AT schemes, one needs to consider how asynchrony is introduced
which has implications in terms of both numerical and computational performance.
We propose two approaches for introducing asynchrony: one that avoids
synchronizations and the other that avoids communications. While the former leads to
reduction in processor idling time and results in random delays at processor boundaries,
the latter leads to periodic delays and reduction in the volume of communications.
Since power consumption, especially for data movement
is expected to be a major concern for the next generation exascale
machines, the reduced frequency of communications in communication avoiding
algorithm, make it a viable energy efficient alternative
to standard approaches.

%The results from both decaying and forced simulations show good
%agreement between asynchronous and standard synchronous well resolved simulations.
%We also show that the asynchronous
%simulations have better parallel scalability than their synchronous counterpart
%because of considerable reduction in overheads due to parallelization.

The rest of the paper is organized as follows. In section 2
we present the governing equations and the details of the
spatial and temporal discretization schemes. In section 3 we discuss
the implementation details and introduce the algorithms to allow
for asynchrony along with stability analysis. In section 4 we present the
numerical results for DNS of decaying and solenoidally forced
turbulence, showing the excellent agreement between standard synchronous
and asynchronous simulations for both
large and small-scale characteristics.
We conclude
the section with the discussion on the computational
performance of the asynchronous solver.
Conclusions and
scope of future work are discussed in section 4. The
appendix lists the AT schemes used in the paper and detailed
stability analysis of the schemes.

%%%%%%%%%%%%%%%%%%%%%%%%%%%%%%%%%%%%%%
\section{Governing equations and numerical schemes}
\label{sec:schemes}
The NS equations, which represent conservation of mass, momentum and
energy can be written as,
\begin{equation}
\frac{\partial{\rho}} {\partial{t}}
+\frac{\partial} {\partial{x_i}}(\rho u_i)=0,
\eqnlabel{cont}
\end{equation}
\begin{equation}
\frac{\partial} {\partial{t}}(\rho u_i)
+\frac{\partial} {\partial{x_j}}(\rho u_iu_j)=
-\frac{\partial p} {\partial{x_i}}
+\frac{\partial} {\partial{x_j}}(\sigma_{ij})
+\rho f_i,
\eqnlabel{mome}
\end{equation}
\begin{equation}
\frac{\partial} {\partial{t}}(\rho e)
+\frac{\partial} {\partial{x_i}}(\rho e u_i)=
-p\frac{\partial u_i} {\partial{x_i}}
+\frac{\partial} {\partial{x_i}}\left(k
\frac{\partial T}{\partial x_i}\right)
+\sigma_{ij}S_{ij},
%-\Lambda,
\eqnlabel{ie}
\end{equation}
with $\rho$ being the density, $u_i$ the $i^{th}$
component of velocity, $e$ the internal
energy per unit mass which depends upon temperature ($T$)
according to the perfect gas law, $k$ the coefficient
of thermal conductivity, $p$ the pressure, and $f_i$ the external forcing.
The viscous stress and the strain
rate tensors are given, respectively, by,
\begin{equation}
\sigma_{ij}=\mu \left(
\frac{\partial{u_i}}{\partial x_j}
+\frac{\partial{u_j}}{\partial x_i}
-\frac{2}{3}\delta_{ij}\frac{\partial{u_k}}{\partial x_k}
\right),
\end{equation}
\begin{equation}
S_{ij}=\frac{1}{2} \left(
\frac{\partial{u_i}}{\partial x_j}
+\frac{\partial{u_j}}{\partial x_i}
\right),
\end{equation}
where the dynamic viscosity, $\mu$, follows
Sutherland viscosity law.

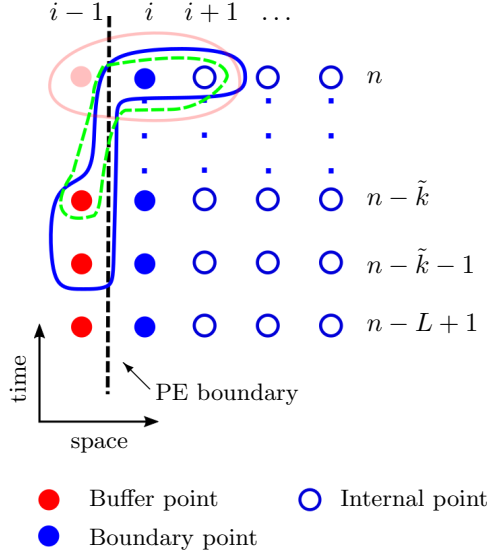
\begin{figure}[h]
\centering
\vspace{0.2cm}
\includegraphics[clip,width=0.25\textwidth]{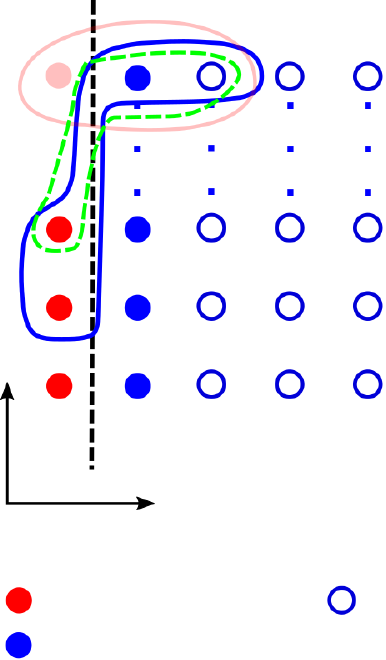}
	\begin{picture}(0,0)
        \put(5,176){$n$}
        \put(5,130){$n-\tilde{k}$}
        \put(5,104){$n-\tilde{k}-1$}
        \put(5,82){$n-L+1$}
%        \put(5,71){$n-L+1$}

        \put(-79,200){$i$}
        \put(-64,200){$i+1~~\dots$}
        \put(-115,200){$i-1$}

        \put(-130,55){\rotatebox{90}{\small{time}}}
        \put(-107,38){\small{space}}
        \put(-75,56){\small{PE boundary}}
	\put(-77,60){\vector(-1,1){10}}
        \put(-100,16){\small{Buffer~point}}
        \put(-100,1){\small{Boundary~point}}
        \put(-5,16){\small{Internal~point}}
%
%	\put(5,135){$n$}
%	\put(5,99){$n-\tilde{k}$}
%	\put(5,80){$n-\tilde{k}-1$}
%	\put(5,61){$n-\tilde{k}-2$}
%        \put(5,42){$n-L+1$}
%
%        \put(-67,155){$i$}
%        \put(-57,155){$i+1~~\dots$}
%        \put(-117,155){$\dots~~i-1$}
%        \put(-17,155){$i+3$}

	\end{picture}
\caption{Left boundary of a Processing Element (PE) with $L$ time
levels. The solid blue points are boundary points and
hollow blue points are internal points. The solid red
points are the buffer points communicated from the neighboring PE.}
\label{fig:domain}
\end{figure}

The above equations are solved numerically using
finite difference approximations for the spatial derivatives.
In order to do so, the physical domain is discretized into $N$ grid points in each direction
and this discretized domain is then decomposed into $P$ sub-domains, where $P$
the number of processing elements (PEs). \rfig{domain} shows the left boundary
(dashed black line) of one such PE in
1D, with internal points in hollow blue, boundary points
in solid blue and the points communicated from the neighboring PE,
known as the buffer points in solid red.
%Also, $N_B\cup N_I=N_T$, where $N_T$ is the total
%number of points in a PE and $N_B\cap N_I=\varnothing$.
For standard finite difference schemes, the derivative at the $i^{th}$ grid point
is a weighted average of the values at the neighboring points.
Mathematically, this approximation for second
derivative, at time level $n$, is given by
\begin{equation}
\left. {\partial^2 u\over \partial x^2}\right|_{i}^n \approx
{1\over\dx^2}\sum_{m=-\ns}^{\ns}{c_m u_{i+m}^n} + {\cal O}(\dx^{p}),
\eqnlabel{du}
\end{equation}
where $M$ is the stencil size in each direction and the $a_m$'s are
the weights that are computed using Taylor
expansion of $u_{i+m}$ in space such that the order $p$ in the
truncation error term is the highest.
%Since the derivative is computed at the $n^{th}$ time-level,
%all the points used for the computation of the derivative are at the same
%time level $n$.
For example, a standard synchronous implementation of a second order scheme ($M=1$) is
outlined in faded red in \rfig{domain} and given by
\begin{equation}
\left. {\partial^2 u\over \partial x^2}\right|_{i}^n =\frac{u_{i-1}^n-2u_i^n+u_{i+1}^{n}}{\dx^2},
\eqnlabel{d2s}
\end{equation}
where all points are at time level $n$.
%For a synchronous implementation of such scheme all $i+m$ points should
%be at time level $n$.
%This constraint is automatically satisfied
%for the internal points at all times. However,
Computation of derivatives at the boundary points requires updated
values
(i.e.\ at time level $n$)
at the buffer points which are communicated from the neighboring PE.
This forces the communications
across the PEs to synchronize and leads to additional overheads
due to processor idling.

To avoid this, one can instead allow
computations to proceed asynchronously, that is, without waiting for
the most updated
value at the PE boundaries.
This results in delayed values at ghost points.
Explicitly, the derivative is then computed as
\begin{equation}
\left. {\partial^2 u\over \partial x^2}\right|_{i}^n =\frac{u_{i-1}^{n-\tilde{k}}-2u_i^n+u_{i+1}^{n}}{\dx^2},
\eqnlabel{d2as}
\end{equation}
where $n-\tilde{k}$ is the latest available time level written in
terms of the delay $\tilde{k}$.
This scheme is schematically shown as a dashed green curve
in \rfig{domain}. Because of the
delay $(\tilde{k})$,
the accuracy of the standard scheme degrades severely.
In fact it can be shown \cite{DA2014} that the resulting scheme is
zeroth-order.
This prevents the use of the standard finite difference schemes
asynchronously
and necessitates the need for numerical methods that are resilient to
asynchrony.
%When the domain is decomposed into
%several subdomains assigned to different processors,
%all points within a processor are at same time level. This allows
%computation of derivatives at the internal
%points of the domain, where data at neighbouring points is
%available, using the standard finite difference schemes.
%However, at the boundary points, the processors have to
%communicate to share the updated values at each time-step.
%This in turn imposes synchronizations among the processors and can
%result in substantial processor idling.
%If we allow the use of delayed values at the boundary points,
%in the scheme defined in \eqn{du} then the order of
%truncation error term degrades to one. This has been shown
%in \cite{DA2014}.
Such family of schemes has been put forth in \cite{AD2017}.
These so-called Asynchrony-Tolerant (AT) schemes preserve the order
of accuracy, despite asynchrony and are described next.

\subsection{Spatial AT schemes}
AT schemes can be seen as a generalization
of standard finite differences, where
the computation of spatial derivatives
use function values of neighboring points in
both space and time. For example, the second derivative of a spatially and
temporally varying function, $u(x,t)$, at grid point $i$ and time
level $n$ can then be written as,
\begin{equation}
\left. {\partial^2 u\over \partial x^2}\right|_{x_i}^n \approx
{1\over\dx^2} \sum_{l=0}^{L} \sum_{m=-\ns}^{\ns}{c_{m}^{l}
u_{i+m}^{n-l}} + {\cal O}(\dx^{p}) .
\eqnlabel{at}
\end{equation}
%where $l$ is the time level and $m$ is the spatial location.
Here the weights $c_{m}^{l}$'s are computed by solving a
system of linear equations constructed by
imposing order of accuracy constraints
on the Taylor series expansion of $u_{i+m}^{n-l}$ in space and time.
The choice of stencil and the general methodology for the derivation
of these AT schemes has been explained
in detail in \cite{AD2017}. As an example, a second order AT scheme at the
left boundary \eqn{at22} with
stencil $M=1$ in space,
and a delay of $\tilde{k}$ can be written as,
\begin{equation}
\left. {\partial^2 u\over \partial x^2}\right|_{i}^n =
\frac{-\tilde{k}u_{i-1}^{n-\tilde{k}-1}+
(\tilde{k}+1)u_{i-1}^{n-\tilde{k}}-2u_i^n+u_{i+1}^{n}}{\dx^2},
\eqnlabel{at22}
\end{equation}
where we have used a diffusive CFL relation of the form $\dt \sim \dx^2$
to relate spatial and temporal resolutions.
This scheme is shown schematically
in \rfig{domain} with a solid blue curve. These schemes use
multiple consecutive time levels on the delayed side,
depending upon the order of accuracy.
%using Taylor series
%expansion in both space and time. The coefficients in the
%finite difference approximation are  no longer
%constant and depend upon the time level that is used.
%These schemes reduce to the standard finite-difference schemes, in the
%absence of delays $i.e$ when $l=0$, with
%the stencil for fourth order accurate scheme represented by the faded red line in
%\rfig{domain}.
%If $l>0$, then data at older time steps is used, implying there
%is a delay $l$ in the computation of the derivative. Due
%to the nature of derivation of these schemes, they preserve
%the order of accuracy when used asynchronously and have thus being
%rightly called Asynchrony-Tolerant (AT) schemes.
%The AT schemes, in
%order to conserve the order of accuracy, require data from
%multiple consecutive time levels, represented by the blue line in \rfig{domain}
%for the $4th$ order scheme.
%The choice of stencil and the general methodology for the derivation
%of these AT schemes has been explained
%in detail in \cite{AD2017}.
An interesting feature of this kind of AT schemes is that the coefficients
are a function of delay $\tilde{k}$ and they
reduce to standard coefficients in the absence of delays ($\tilde{k}=0$).
Note that these delays depend upon the machine characteristics such as
clock rate, network
latency, bandwidth and topology. Because of this,
the delays and, consequently, the coefficients are random
and computed dynamically at runtime.

In this work we use fourth-order AT schemes at processor boundaries
for spatial derivatives in each direction which require communication
across six faces of each PE in a 3D domain. At the internal points we use
standard fourth-order finite differences for spatial derivatives. Computation
of mixed derivatives is challenging as they require communication
across more neigboring PEs or communication of additional quantities such as gradients.
Both of these are detrimental to parallel performance. As an alternative, we limit our
communications per PE to six by
computing mixed derivatives %such as ($\partial\left(\partial
%u/\partial y \right)/\partial x$),
at the boundary points in three steps.
For example, for ($\partial\left(\partial
u/\partial y \right)/\partial x$), we first compute $\partial u/\partial y $ and
 $\partial u/\partial x $ using AT schemes at the boundaries.
Next we compute ($\partial\left(\partial
u/\partial y \right)/\partial x$) and
($\partial\left(\partial u/\partial x \right)/\partial y$)
using standard one sided finite difference schemes in $x$ and $y$
directiion, respectively. Since
%$u(x,t)$ is a continuous function,
($\partial\left(\partial
u/\partial y \right)/\partial x$) = ($\partial\left(\partial
u/\partial x \right)/\partial y$), %$i.e.$ order of computation of derivative does
we take the average of $\partial\left(\partial
u/\partial y \right)/\partial x$ and
($\partial\left(\partial u/\partial x \right)/\partial y$ to minimize
errors and use this value as the final approximation of the corresponding mixed
derivatives.
 %This is done to avoid communication at very step of AB schemes or
%communication across more than six neighbors.

%In the current paper we use fourth
%order AT schemes at the processor boundaries and standard fourth order
%finite differences at the internal points for all spatial
%derivatives.

%\colb{\\!!!!!!!!!!!!!!!!!!!!!!!!!!!!!!!!!!!!!!!!!!!!!!!!!!!!!!!1}
%and can affect the telescoping
%property of the original standard scheme. The AT schemes
%use a larger stencil size than the standard
%schemes and thus the number of $flops$ for the computation
%of derivatives at the boundary points increases. This will
%increase the total computation time, depending upon the percentage
%of grid points using AT schemes are the frequency of
%occurence of delays, which can be obtained from the $pdf$
%of the delays. Furthermore, since the computation at boundary points
%or boundary slabs in a 3D domain depends on multiple time levels, it increases
%the memory usage. This can lead to increased cache misses which will also affect
%the computation time. This, however, can be reduced by optimized
%algorithms for computation and storage of variables. We have used fourth order AT
%scheme at the boundaries for the simulations presented in this paper and
%the schemes have been tabulated in the appendix.

%An important aspect of standard finite difference is that these
%chemes satisfy the global conservation and are
Because the NS equations represent conservation laws, it is important
that the numerical discretization of these laws also satisfy the global
conservation. For example,
in a 1D form of conservation law,
\be
\frac{\partial u}{\partial t} +\frac{\partial f}{\partial x} =0,
\eqnlabel{convf}
\ee
 where $f(x,t)$ is the flux, the total variation of $u(x,t)$ over a domain $[0,1]$ depends only upon
the flux through  the boundaries. This can be expressed more
precisely by integrating
\eqn{convf} over the domain,
\be
\frac{d }{dt} \int_{0}^{1}u(x,t)dx=\int_{0}^{1}\left(\frac{\partial f}{\partial x}\right)dx
=f(1,t)-f(0,t), %=0
\ee
showing explicit dependence of variation in $u(x,t)$ only on the flux at the boundaries.
For periodic boundary condition \textit{i.e} $f(0,t)=f(1,t)$,
this flux is equal to zero. When the derivatives are approximated
numerically, it is desirable that the discrete form of the above
conservation law is
also satisfied to a given accuracy.
Consider a
%in 1D the change in $u(x,t)$ over the domain $[0,1]$
%depends only upon the flux of $u(x,t)$ at the boundaries. Mathematically,
%for periodic boundary condition \textit{i.e}
%when
%$u(x,t)|_{x=b,t=t}=u(x,t)|_{x=a,t=t}$,
%$u(a,t)=u(b,t)$,
%this is equivalent to,
%\be
%\int_0^1\left(\frac{\partial u}{\partial x}\right)dx=
%\left(u(x,t)|_{x=1}-u(x,t)|_{x=0}\right)=0.
%u(1,t)-u(0,t)=0.
%\ee
generalized spatial discretization given by \eqn{at},
for $N$ grid points and time level $n$, to yield
\be
\int_0^1\left.\frac{\partial f}{\partial x}\right|^ndx=
\sum_{i=1}^{N}
\left(
{1\over\dx} \sum_{l=0}^{L} \sum_{m=-\ns}^{\ns}{c_{ml}
f_{i+m}^{n-l}} \right).
\eqnlabel{tsf}
%\frac{u_2^n-(\kt_l^1+1)u_N^{n-\kt_l^1}+\kt_l^1u_N^{n-\kt_l^1-1} }{2\dx}
%+\frac{u_3-u_1}{2\dx}+\dots+
%\frac{u_{i+1}-u_{i-1}}{2\dx}+\dots
%\frac{u_N-u_{N-2}}{2\dx}+\frac{u_1-u_{N-1}}{2\dx}
\ee
For $M=1$, corresponding to an AT scheme with
leading truncation error term of order
$\mathcal{O}(\dx^{a})$ where $a=2$ when $\dt\sim\dx^2$ \cite{AD2017},
and a domain decomposed into 2 PEs such that
PE$^{(1)}$ holds gridpoints $i\in[1,N/2]$
and PE$^{(2)}$ holds gridpoints $i\in[N/2+1,N]$
and satisfies periodic boundary conditions, we can write \eqn{tsf} as
\begin{equation}
\begin{aligned}
\int_0^1 \left. \frac{\partial f}{\partial x}\right|^n dx=
&\frac{f_2^n-(\kt_{l}^{(1)}+1)f_N^{n-\kt_{l}^{(1)}}+\kt_l^{(1)}f_N^{n-\kt_l^{(1)}-1} }{2\dx}
+\sum_{i=2}^{N/2-1}
\left(
\frac{f_{i+1}^{n} -f_{i-1}^n}{2\dx} \right) \\&+
%+\frac{u_3-u_1}{2\dx}+\dots+
\frac{(\kt_r^{(1)}+1)f_{N/2+1}^{n-\kt_r^{(1)}}-\kt_r^{(1)}f_{N/2+1}^{n-\kt_r^{(1)}-1} -f^{n}_{N/2-1}}{2\dx}+
\frac{f_{N/2+2}^n-(\kt_l^\nnn+1)f_{N/2}^{n-\kt_l^\nnn}+\kt_l^\nnn f_{N/2}^{n-\kt_l^\nnn-1} }{2\dx}
\\
&+\sum_{i=N/2+2}^{N-1}
\left(
\frac{f_{i+1}^{n} -f_{i-1}^n}{2\dx} \right)
+\frac{(\kt_r^\nnn+1)f_{1}^{n-\kt_r^\nnn}-\kt_r^\nnn f_{1}^{n-\kt_r^\nnn-1} -f^{n}_{N-1}}{2\dx},
\end{aligned}
\end{equation}
where $\kt_l^\mmm$ and $\kt_r^\mmm$ are the delays on left and right boundary for PE$^{(1)}$ and
$\kt_l^\nnn$ and $\kt_r^\nnn$ are the delays on left and right
boundary for PE$^{(2)}$ and periodic
boundary conditions are used.
Because of the telescoping effect, the above expression can be
simplified to
%\colr{[DD: aren't we assuming $n$ is the same in all processors
%at a given time?]}
\be
\begin{aligned}
\int_0^1\left.\frac{\partial f}{\partial x}\right|^ndx=
&\frac{-(\kt_{l}^\mmm+1)f_N^{n-\kt_{l}^{\mmm}}+\kt_l^\mmm f_N^{n-\kt_l^\mmm-1} }{2\dx}
+
\left(
\frac{f_{N/2}^{n} -f_{1}^n}{2\dx} \right) +
%+\frac{u_3-u_1}{2\dx}+\dots+
\frac{(\kt_r^\mmm+1)f_{N/2+1}^{n-\kt_r^\mmm}-\kt_r^\mmm f_{N/2+1}^{n-\kt_r^\mmm-1} }{2\dx}+\\&
\frac{-(\kt_l^\nnn+1)f_{N/2}^{n-\kt_l^\nnn}+\kt_l^\nnn f_{N/2}^{n-\kt_l^\nnn-1} }{2\dx}
+\left(
\frac{f_{N}^{n} -f_{N/2+1}^n}{2\dx} \right)
+\frac{(\kt_r^\nnn+1)f_{1}^{n-\kt_r^\nnn}-\kt_r^\nnn f_{1}^{n-\kt_r^\nnn-1} }{2\dx}.
\end{aligned}
\eqnlabel{ts3}
\ee
For the standard sychronous case,
$\kt_l^\mmm=\kt_r^\mmm=\kt_l^\nnn=\kt_r^\nnn=0$,
that is, when delays are absent, all terms on the right-hand side of
\eqn{ts3} cancel each other
and the conservative property is trivially
satisfied.
In the presence of delays, on the other hand, this
is not immediately obvious from \eqn{ts3}.
Further simplification of this equation
can be done using a Taylor series expansion in time which
leads to similar cancellation of all low-order terms yielding a
residual of the order of $\dx^{3}$. More generally,
for larger $M$, that is,
for AT schemes of order $a=2M$ and $\dt\sim\dx^2$,
the residual is found to be
\be
\int_0^1\left.\frac{\partial f}{\partial x}\right|^ndx=\mathcal{O}(\dx^{a+1}).
\eqnlabel{ts4}
\ee
Thus, we conclude that
the AT schemes retain the conservative property
up to an order higher than the order of the scheme.
For the fourth-order schemes used below, conservation is satisfied
to $\mathcal{O}(\dx^5)$.

\subsection{Temporal schemes}
For the evolution of a system of PDEs in time,
spatial schemes need to be coupled with a temporal
scheme of appropriate order.
High order explicit temporal methods including
multistage Runge-Kutta (RK)
schemes and multistep Adams-Bashforth schemes, are very common
choices of temporal discretizations.
While RK schemes are known for their good stability characteristics,
the computation of a stage of RK requires communication of
previous stages across all neighbors at all times.
%This can be overcome by increasing the
%size of the message such that all the stages can be computed within
%the processor.
In a 3D domain this is equivalent to $26\times s$ communications and $s$ computations of
the right hand side of the PDE, per PE, per time-step, for an $s$-$stage$ RK scheme.
Consequently, RK schemes are computation, communication and synchronization intensive.
On the other hand, multi-step Adams-Bashforth (AB) schemes
offer more flexibility in terms of implementation
and require less communications. A general AB scheme with $T$
steps for an equation of the form $\partial u /\partial t = f$, can be written as,
\begin{equation}
u^{n+1}_i=u^{n}_i+ \dt \sum_{m=0}^{T-1} \beta_m f_i^{n-m},
\eqnlabel{time_ab}
\end{equation}
where the coeffcients $\beta_m$ depend upon the desired order of accuracy \cite{Stoer2013}.
% Here $f^{n-m}$ can be computed
%using AT schemes without affecting the order of accuracy of
%the AB schemes \cite{AD2017}.
%Unlike RK, which involves computation of $s$ at every time-step,
%for AB only $f^{n}$ is computed and $f^{n-m},m>0$ is used from
%previous step.
Not only can AB be efficiently implemented with only six
communications per PE per time-step, it only requires computation of
$f^{n}$ every time-step since $f^{n-m},m>0$ is used from
previous steps. Furthermore, the computation of $f^{n-m}$ using
AT schemes does not alter the order of accuracy of AB schemes \cite{AD2017}.
Thus, here we use second-order AB schemes for the temporal
evolution
in both synchronous and asynchronous simulations.

The CFL, relating time-step size to the
grid spacing, %$(\Delta t \sim\Delta x^{r})$,
can be used to determine the leading
order error term of a fully discretized PDE
%with both space and time discretization errors.
in order to ensure that global order of accuracy is preserved.
%Numerical simulations can either be carried at a fixed $cfl$ or a fixed
%time-step $\dt$.
For example, for a convective CFL ($r_c$), the
time-step $\dt$ is computed as,
\begin{equation}
\dt=\frac{r_c \dx}{u_{max}}
\end{equation}
where $u_{max}$ is the global maximum velocity.
Since this maximum is computed across all PEs,
it requires a collective blocking communication call at every time step and
leads to more synchronization overheads.
To avoid this, instead
of a CFL condition,
one can use a fixed $\dt$ \cite{Chen2009,Gruber2018}.
This is the approach we adopt here.
%for asynchronous simulations.
For consistency, synchronous simulations are also done at the same
fixed $\dt$.
%Use of a fixed time step in DNS is not an uncommon practice
%for DNS studies \cite{Chen2009,Gruber2018}.
In summary, we use fourth-order AT schemes in space for boundary points,
fourth-order finite difference schemes at internal points and
 AB schemes in time with a fixed $\dt$.

\section{Implementation}
\label{sec:imp}
%\colr{(Add cross derivatives)}\\
%The terminology that is used in the rest of the paper
%has been explaned using a simple 1D domain decomposition
%where every Processing Element (PE) has only two neighbors.
The compressible flow solver is parallelized using a 3D domain decomposition
and each PE is responsible for computations in a piece of this 3D domain.
Communications between PEs are localized to the nearest neighbors
only. In \rfig{domain} we see a
simple 1D domain decomposition
where every PE has $N_{T}$ grid points and two neighbors.
The number of internal points ($N_I$) $i.e.$ the points that use
standard (synchronous) finite
differences, with $M$ points in each direction
%, for computation of spatial derivatives
is equal to $N_T-2M$. The total number of boundary points
($N_B$) and buffer points ($N_{Bf}$) in this case are equal to $2M$.
Clearly $N_B\cup N_I=N_T$ and $N_B\cap N_I=\varnothing$.

%The number of buffer points ($N_{Bf}$) is usually equal to $N_b$ and the number
%of internal points $(N_I)$
%The number of boundary points, $N_B$,
%on each side of the PE boundary is equal to the stencil
%size ($M$) used in the spatial scheme.
%We also have $N_B\cup N_I=N_T$, where $N_T$ is the total
%number of points in the PE and $N_B\cap N_I=\varnothing$. As described
%before, in \rfig{domain} the
%faded red line shows the points required for the computation of
%derivatives synchronously and the solid blue curve represents
%an AT schemes that uses multiple delayed levels for computation
%of derivative at the boundary points.\\
Extending this idea to a 3D computational topology,
such that each PE has a total of twenty six neighboring PEs, we can compute
the total number of internal and boundary points.
Consider a general 3D domain with $N_x$,$N_y$, and $N_z$ grid points
and $P_x$, $P_y$, and $P_z$ processors in the $x$, $y$, and
$z$ directions,
respectively. Then the total number of
grid points $(N_T)$ per PE is
\begin{equation}
N_T=\frac{N_x N_y N_z}{P_x P_y P_z}.
\eqnlabel{NT}
\end{equation}
Using spatial schemes which require $M$ points on each side for all three directions,
it is easy to show that the number internal
points $(N_I)$ is
\begin{equation}
N_I=\left(\frac{N_x}{P_x}-2M\right)
\left(\frac{N_y}{P_y}-2M\right)
\left(\frac{N_z}{P_z}-2M\right).
\eqnlabel{NI}
\end{equation}
Since communications are done across all six faces of a PE,
\eqn{NT} and \eqn{NI} gives us the exact number of
boundary points $(N_B)$ or the points that use AT schemes for the
computation of spatial derivatives,
\begin{equation}
N_B=\frac{N_x N_y N_z}{P_x P_y P_z}-
\left(\frac{N_x}{P_x}-2M\right)
\left(\frac{N_y}{P_y}-2M\right)
\left(\frac{N_z}{P_z}-2M\right).
\eqnlabel{NB}
\end{equation}
We can then compute the percentage of points that use AT schemes,
\begin{equation}
N_B(\%)=100\left(\frac{N_B}{N_T}\right),
\end{equation}
which can be used as a metric of the extent in space in which
asynchrony affects the computations of derivatives directly.
%We have simulations with as high as $65\%$ of points
%being directly affected by asynchrony.

\subsection{Algorithm}
We solve the NS equation for five variables $(\rho,\rho u_1,\rho u_2,\rho u_3, \rho e)$
at every time step.
Since data at older time levels is used for AT
schemes, each PE stores $5\times(N_I+N_B+N_{bf})\times (L+\Tl)$
data points, where $L$ is the maximum allowed delay that can also
be used as a control parameter for error and stability as we show below
and $\Tl$ is the number of consecutive time-levels required
for the computation of derivatives by AT schemes.
We use two-sided non-blocking MPI
calls (\textit{MPI\_Isend, MPI\_Irecv}) for asynchronous communications
between the PEs across the six faces of the 3D computational domain.
In each direction, these communications are limited to immediate neighbors only.
 The status of these non-blocking communications is checked
using \textit{MPI\_test} and it is utilized to compute delay at each PE boundary.
%
%Since AT schemes
%require $\Tl$ number of
%consecutive time-levels for the computation of derivatives,
%the absence of $\Tl$ levels in some direction, forces
%the two neigboring PEs in that direction to synchronize.
%Thus, synchronizations are enforced only between
%immediate neighbors in required direction and are therefore local in nature.
%a local
%synchronization in that direction is enforced.
%This forced synchronization,
%done using \textit{MPI\_Waitall}, requires multiple message to
%complete and imposing synchronization penaltys larger than
%standard synchronous case where at every time only one message is
%completed.
%We can also wait
%only for the required number of oldest messages to complete,
To control the manner in which asynchrony appears we utilize
two control parameters $\Cr$ and $L$.
% that determine the
%frequency of communications and synchronizations.
The \textit{communication rate} $\Cr$
specifies the frequency of communication in each direction, that
is to say, PEs initiate communication calls
every $\Cr$ consecutive time steps.
% When $cr=T$,
%then the processors communicate with a period $T$. For $T>1$ these
%so-called communication avoiding algorithms, results in periodic
%delays at processor boundaries.
The second parameter is the
maximum allowed delay, $L$. %, which can be same or different across each boundary.
% Based on the value of $L$, the
%\textit{synchronization rate ($\Sr$)} which
PEs impose explicit synchronization
by invoking \textit{MPI\_Wait}
whenever instantaneous delays at PE boundaries cross this threshold $L$.
This synchronization is imposed only in the direction in which the
delay is larger than $L$ and is thus local in nature.
These two parameters determine the nature of delays.
For example, if $\Cr>1,L>1$ delays are periodic
 and if $\Cr=1,L>1$ then delays at PE boundaries are
random. In both the cases the delays are however bounded by $L$.
A synchronous simulations is realized when
 $\Cr=1$ and $L=1$.
Irrespective of $\Cr$ and $L$,
global communications and synchronizations
involving all PEs are done only for I/O.

\iffalse
\colr{[DD: I'm not sure it is clear what $\Sr$ does...]}
%depending upon the delay seen by a processor
%at the boundaries. For $sr>1$, we have the so-called synchronization avoiding
%algorithms that result in random delays between the processors.
These two parameters define the nature of delays.
For example, if $\Sr>1$ delays are random
\colr{[DD: for any $\Cr$?]}
while if $\Cr>1$
delays are periodic
\colr{[DD: for any $\Sr$?]}.
A synchronous simulations is realized when
$\Cr=1$ and $\Sr=1$. %. for the standard
%synchronous implementation.%We used profiling
%tool $TAU$ to compute the rate of
%forced synchronization. \colr{Add details(?)}.
\fi

%\subsection{Nature of delays}
%Depending upon the control parameters $(cr,sr)$ that specify the
%frequency of communications and frequency of synchronizations, we can
%have random or periodic delays. The effect of both of these
%is different in terms of performance and also in their
%accuracy in turbulence simulations. Here we study two general approaches to
%allow asynchrony in the computations, namely, synchronization avoiding and
%communication avoiding algorithms that lead to random and periodic delays,
%respectively, and are decribed next.
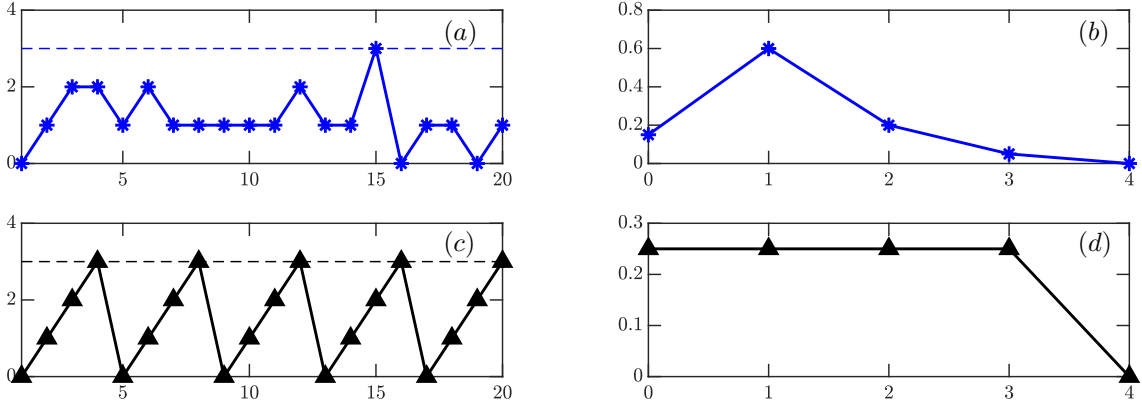
\begin{figure}[h]
\includegraphics[clip,width=0.5\textwidth]{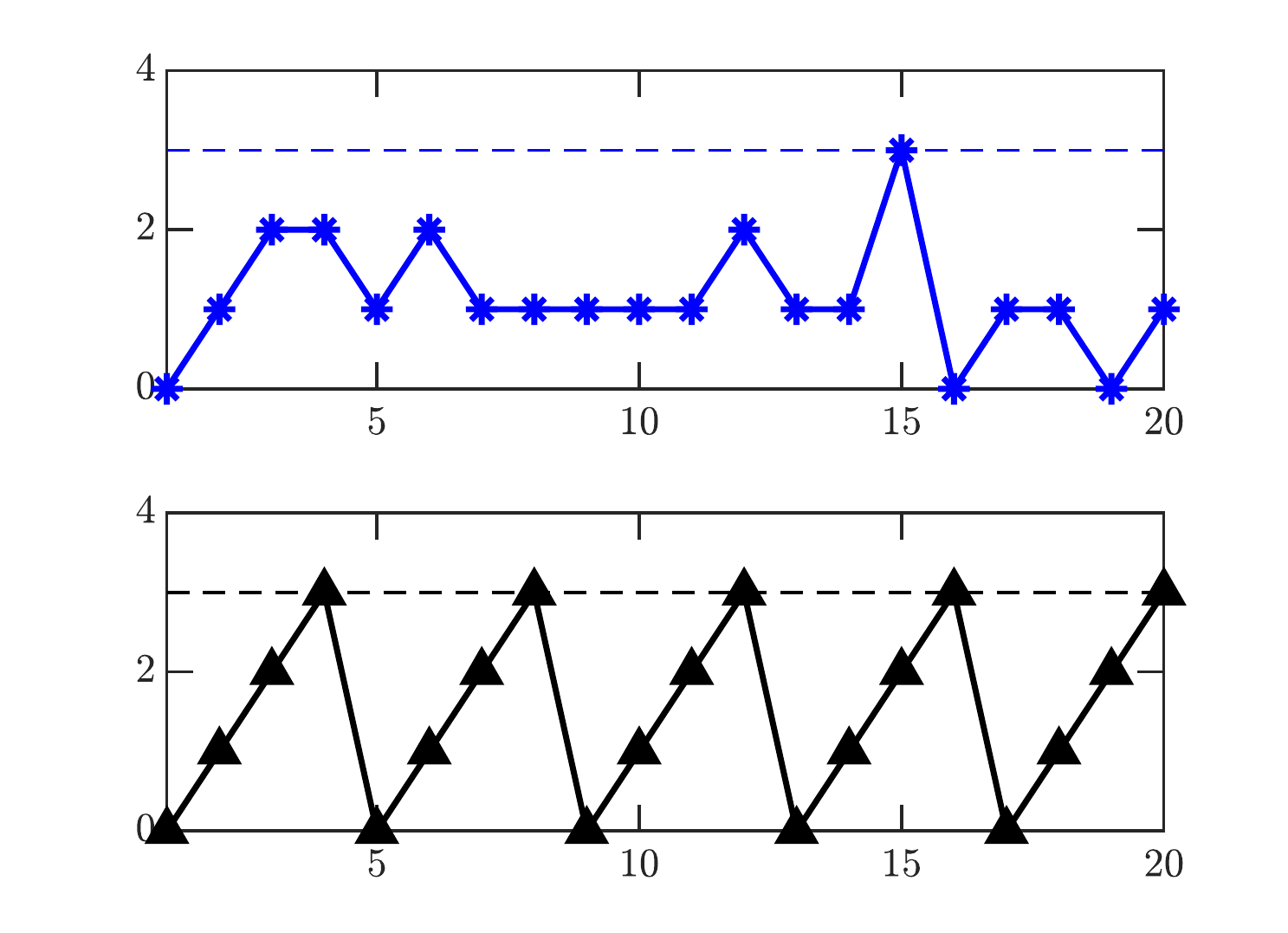}
\includegraphics[clip,width=0.5\textwidth]{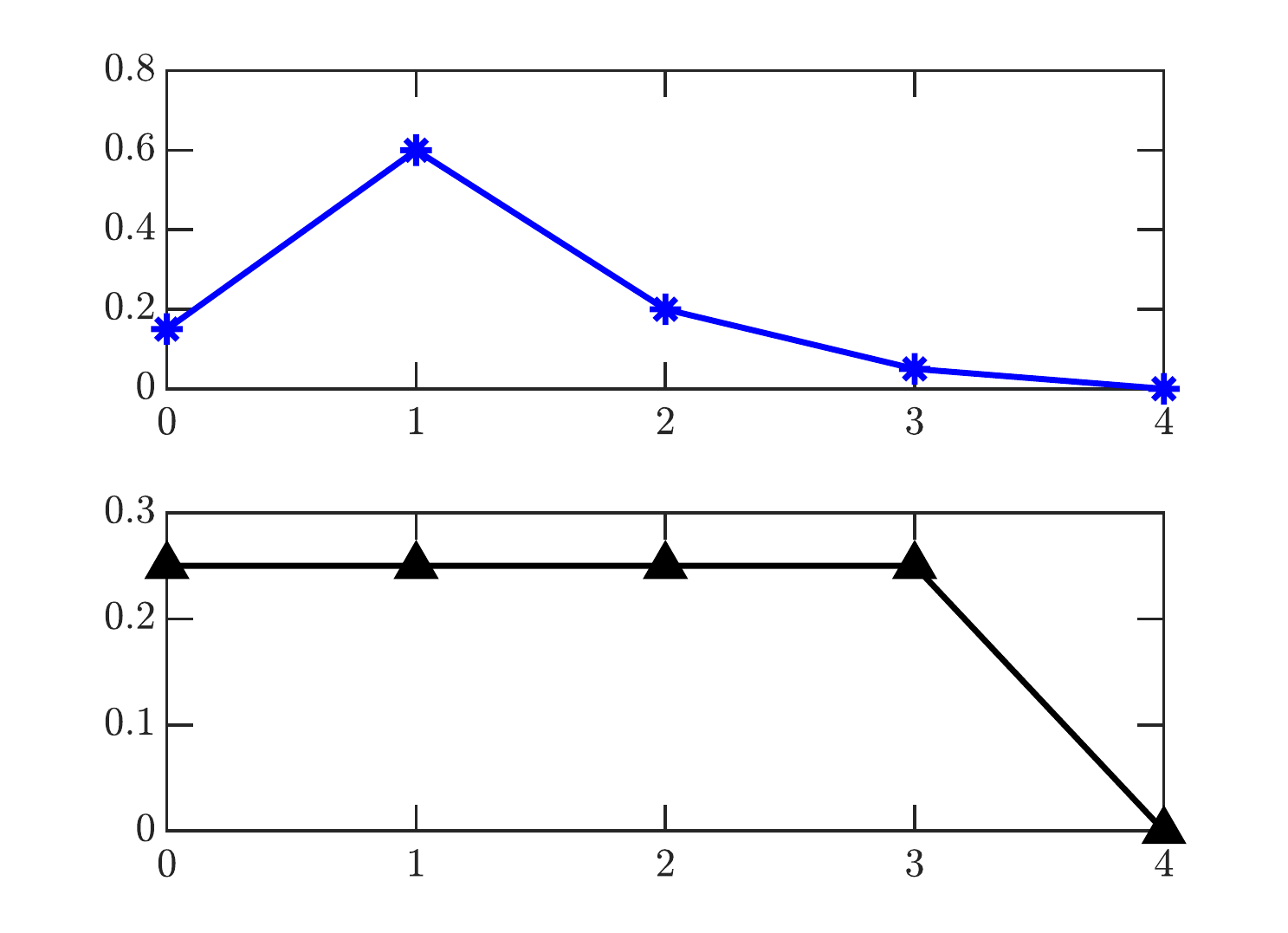}
\begin{picture}(0,0)
\put(190,158){$(a)$}
\put(430,158){$(b)$}
\put(190,78){$(c)$}
\put(430,78){$(d)$}
\end{picture}
\caption{Simulated time series of delays for (a) SAA with $\Cr=1,L=3$ and
(c) CAA with $\Cr=4,L=3$.
PDF of these simulated delays for (b) SAA and (d) CAA.
}
\label{fig:del_pdf}
\end{figure}

\subsubsection{Synchronization Avoiding Algorithm (SAA): random delays}
For $\Cr=1$ and $L>1$,  we have what we call a synchronization
avoiding algorithm: a local synchronization is applied
$if~and~only~if$ the delay ($\tilde{k}$) at a PE boundary is greater
than the maximum allowed delay $L$. %which in this case satisfies $L=\Sr$.
We use circular send ($U_{send}$) and receive $(U_{recv})$ buffers in %of size $M\times (L+\Tl)$ in
each direction for communicating to and from the neighboring PEs, respectively.
At each time step, PEs sent data at only one time level across the boundaries.
A generalized SAA is listed in Algorithm \ref{algorithm:SAA},
where computations can proceed without waiting for updated values whenever $\kt\le L$.
%The send buffer is used to pack the updated data at the boundary points $(U_B)$
%and the receive buffer is used to store the values at the buffer points $(U_{Bf})$.
When synchronization is not imposed,
communications can complete in the background,
facilitating overlap between communications and
computations. Despite the reduction
of synchronization overheads and PE idling, this method
does require communication at every step. The delay
observed at PE boundaries is
a function of machine characteristics, such as, network performance,
processor and memory speeds etc.,
and is therefore a random
variable, with a different value at each of the six PE boundaries.
%This distribution will vary for
%machines with different architectures and network topologies.
%The PDF of delays gives important information such as frequency
%of forced synchronization in an otherwise synchronous simulation
%and also the maximum delay observed in a machine.
Since the delays and, consequently,
coefficients of AT schemes may be different for each PE boundary, %Since AT schemes are used at the boundaries, the conservative property has a residual ($\mathcal{O}(\dx^5)$) as described in section 2.1.
%Numerical simulations show that this effect is negligible for values of $L$
%that must satisfy stability (\rsec{stab}).
some additional numerical errors can be introduced due to random nature of these delays.
Numerical simulations show that this effect is
negligible for values of $L$ that satisfy stability (\rsec{stab}).
%For example, for all simulations in this work,
%the instantaneous average density is within $\pm0.003\%$ of its exact
%analytical value. %Larger variations often lead
%to instabilities and culmination of the numerical simulation but
%this does not happen if the $CFL$ is within the stability region.
%The stability is discussed later.

A typical time series of random delays for SAA is shown in \rfig{del_pdf}(a)
along with its PDF in \rfig{del_pdf}(b). %The figure illustrates delay
%values at a PE boundary as a function of time step.
In this example, the delay is bounded by $L=3$ as shown by the
dashed blue line in \rfig{del_pdf}(a).
The statistical moments of the distribution of delays have an effect on the accuracy of the
solution \cite{AD2017}. Since statistical characteristics of the delays can be controlled by
forced synchronization, $L$ becomes a parameter for error control.
%The error of AT schemes depend upon statistical moments of these delays \cite{AD2017}
%which can be computed from the PDF shown in \rfig{del_pdf}(b).\\

\begin{algorithm}[H]
\SetAlgoLined
Synchronous Loop: Initialize $L+\Tl$ levels of $U,U_{send}\leftarrow U_{boundary},U_{buffer}\leftarrow U_{recv}$\\
Asynchronous Loop:\\
 \For{$n=L+\Tl+1,\cdots,st$}{
 $U^{n+1}=f(U^n,U^{n-1},...,U^{n-T+1})$\\
$U_{send}\leftarrow U^{n+1}_{boundary}$\\
Send data across 6 faces: $MPI\_Isend$ \\
 \For{$face=1:6$}{
Check communication status:$MPI\_Test$\\
Compute delay $(\tilde{k})$\\
  \eIf{\upshape{delay }$(\tilde{k})\le L$}{
    $U_{buffer} \leftarrow U_{recv}$
   }{
   Force synchronization: $MPI\_Wait$\\
  Update delay $(\tilde{k})$\\
   $U_{buffer} \leftarrow U_{recv}$\\
  }
   Compute coefficients of the AT schemes (Appendix B)%$\leftarrow g(\tilde{k})$
}
  }
 \caption{Synchronization Avoiding Algorithm (SAA). Here $U^{n}$ is the variable array
at time level $n$, $U_{send}$ is the send buffer, $U_{recv}$ is the receive buffer,
$U^n_{buffer}$ is the data at buffer points, $U^n_{boundary}$ is the data at boundary points
and $f$ evaluates the discretized NS equation using AT schemes in space and AB schemes in time
for number of time steps equal to $st$.}
 \label{algorithm:SAA}
\end{algorithm}

%\colr{|||||||||++++++++++++++++++++++++++++++++++++++||||||||}
\subsubsection{Communication Avoiding Algorithm (CAA): periodic delays}
As alternative to communicating at every time step,
we propose the so-called communication avoiding algorithm, in which
the PEs communicate periodically every $\Cr>1$ steps. As a result, the delay
changes periodically from $0$ (no delay) to a maximum allowed delay
$L$ which satisfies $L=\Cr-1$. Because of this periodicity,
the delay across all the PE boundaries is the same in every direction.
%and therefore the conservative
%property of the spatial schemes is preserved.
Since PEs communicate every $\Cr$ time steps,
the send and receive buffers now have data at $min(\Tl,\Cr)$ time levels.
This multiple time level data is required for computation of derivatives using AT
schemes at the communication avoiding time steps.
%The delay across all the PE boundaries is same and therefore the conservative
%property of the spatial schemes is preserved.
%We can also generalize periodic delays and allow for some degree of randomness in
%the delays.
 We have listed a generalized implementation of
CAA in Algorithm \ref{alg:CAA}, where the delay is incremented by one when
PEs do not communicate.
A typical time series of delays bounded by $L=3$ (dashed black line)
is shown in \rfig{del_pdf}(c) for CAA with $\Cr=4$.
The delay in this case is deterministic and the PDF shown in \rfig{del_pdf}(d)
has a uniform distribution. Both delay and its PDF are independent of the machine
characteristics and depend only upon the control parameters, contrary to
SAA where the delay is random and its PDF
is machine specific. CAA reduce the total latency time
by a factor of $\Cr$ in comparison to synchronous avoiding or
standard synchronous algorithms and are therefore particularly effective in latency-dominated
machines. Furthermore, because of the reduction in frequency of communications,
the energy consumption for these algorithms is also expected to be reduced.
One drawback of the communication avoiding algorithms is the larger
size of send and receive buffers that could adversely affect
performance for bandwidth-dominated machines.\\

\begin{algorithm}[H]
\SetAlgoLined
Synchronous Loop: Initialize $L+\Tl$ levels of $U,U_{send}\leftarrow U_{boundary},U_{buffer}\leftarrow U_{recv}$\\
Compute $\ell=min(\Tl,\Cr)$\\
Asynchronous Loop:\\
 \For{$n=L+1,\cdots,st$}{
 $U^{n+1}=f(U^n,U^{n-1},...,U^{n-T+1})$\\
 \For{$face=1:6$}{
%$U_{send}\leftarrow U^{n+1}_{B}$\\
%
%$MPI\_Isend$ and $MPI\_Irecv$ \\
%Check communication status ($MPI\_Test$) and compute delay $(\tilde{k})$\\
  \eIf{($mod(n,\Cr)==0$)}{
$U_{send}\leftarrow U^{n+1}_{boundary},...,U^{n-\ell}_{boundary}$\\
    $MPI\_Isend$ and $MPI\_Irecv$ \\
    $U_{buffer} \leftarrow U_{recv}$\\
 $\tilde{k} \leftarrow 0$
   }{
   Update delay: $\tilde{k}\leftarrow \tilde{k+1}$
  }
  Compute coefficients of the AT schemes (Appendix B)%$\leftarrow g(\tilde{k})$
}
 }
 \caption{Communication avoiding algorithm. Here $U^{n}$ is the variable array
at time level $n$, $U_{send}$ is the send buffer, $U_{recv}$ is the receive buffer,
$U^n_{buffer}$ is the data at buffer points, $U^n_{boundary}$ is the data at boundary
 points and $f$ evaluates the discretized NS equation using AT schemes in space and AB schemes in time
for number of time steps equal to $st$.}
 \label{alg:CAA}
\end{algorithm}

\subsection{Maximum delay $L$ and stability}
\label{sec:stab}
The maximum allowed delay $L$ is an important control parameter as it determines
the error and stability of the AT schemes as well as the computational
performance of the solver. As shown in \cite{AD2017},
the error due to asynchrony in AT schemes is a function of
statistical moments of delays which depend upon the architecture
of the machine, communication links and patterns, latency, bandwidth and
clock speed. Since the asynchronous error grows with $L$ \cite{AD2017}, very large values of $L$
can affect the accuracy of simulations. Furthermore, the
memory requirement of all stored variables, the size of send and receive buffers and the
rate of synchronizations and communications are also directly affected by the choice of parameter $L$.
It is therefore critical that $L$ be chosen judiciously in simulations and this choice
can be based on two main factors that are described next.

First, $L$ has implications in terms of the computational implementation
of the solver. Increasing $L$ increases the number of times levels that need
to be stored which increases memory requirement. %in data which
%leads to a non-sequential memory access with strides that depend upon $L$.
%This increased cache miss rate adversely affects the
%computational performance.
At the same time, if $L$ is too
small then synchronization will be forced more often than required and
asynchrony will not be leveraged efficiently.
In practice, one can run a short simulation with a very large $L$ and
obtain the PDF of the delays ($\tilde{k}$). From this data, one can
calculate an appropriate $L$ by requiring $P(\tilde{k}>L)\lesssim c$, that is to
say, one would expect forced synchronization $c\%$ of the time.
Thus, $c$ exposes tradeoff between performance and accuracy through the degree of asynchrony.
For example, at $c= 0$ the simulation is completely asynchronous, $i.e.$,
synchronization is never imposed, which is detrimental to accuracy if $L$ is large.
%In our simulations
%we use $c=5\%$. %Smaller $c$ would further reduce forced synchronizations but that
%will also increase $L$
%To overcome this challenge, we
%obtain the PDF of delay distribution for a
%machine for a large initial delay of $L_i$
%and use it to determine optimal $L$ such that $P(\tilde{k}>L)\lesssim0.05$.
%Here the choice $0.05$ is arbitrary and indicates the probability of delays larger
%than $L$.

\begin{figure}[h]
\centering
\includegraphics[clip,width=0.45\textwidth]{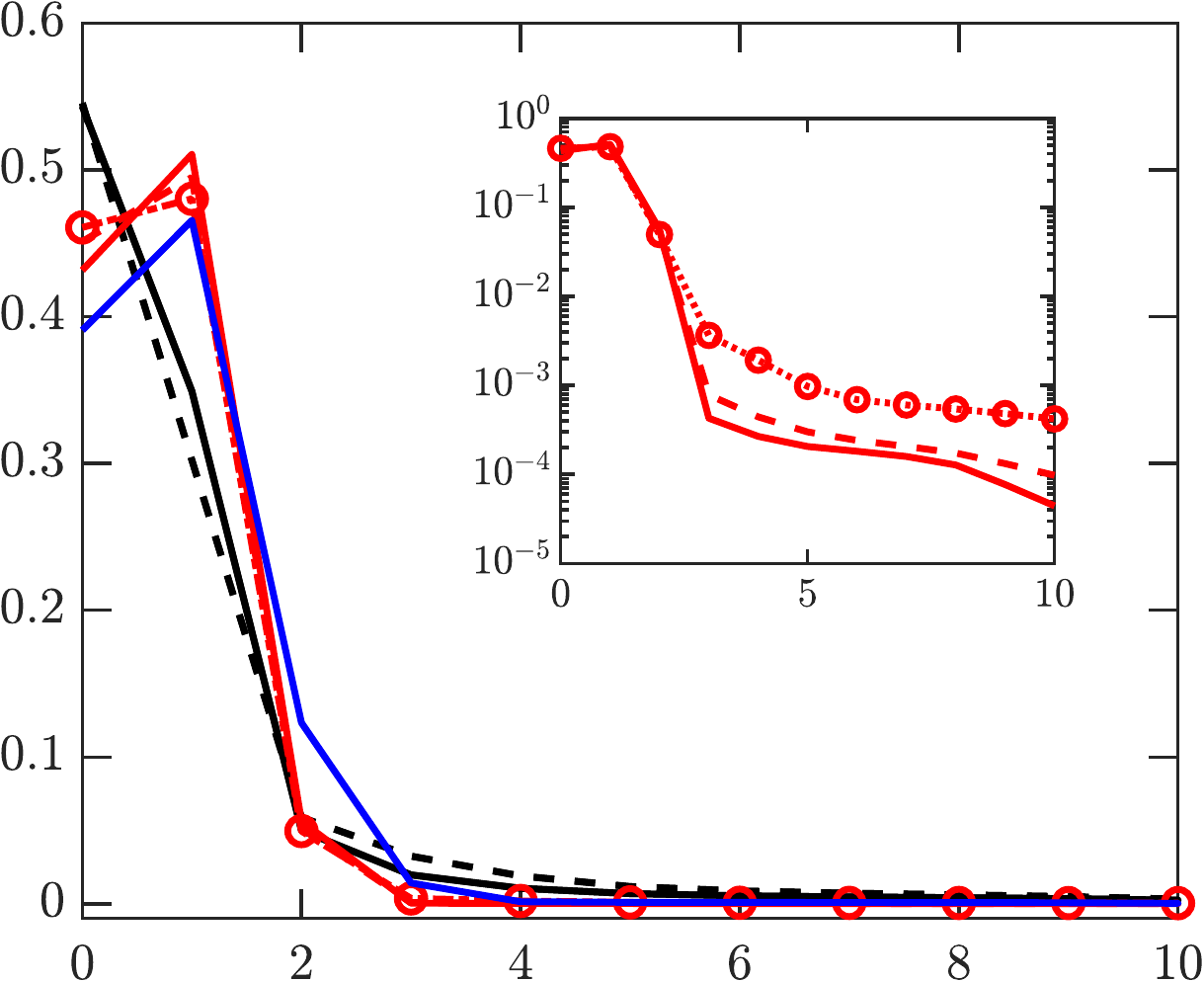}
\begin{picture}(0,0)
\put(-105,-10){L}
\put(-230,40){\rotatebox{90}{Probability of delay}}
\put(-55,80){\vector(1,2){15}}
\end{picture}
\caption{PDF of delays on Stampede2 (black), Frontera (red) and Lonestar5 (blue)
with maximum allowed delay of $L=10$. Different lines
are $P=4096$ (solid red, solid blue), $P=8192$ (solid black, dashed red) and $P=16384$ (dashed black)
and $P=262144$ (dotted-circles). Inset is PDF of delays on Frontera in linear-log scale
with arrow denoting increasing P.
}

\label{fig:del_machines}
\end{figure}
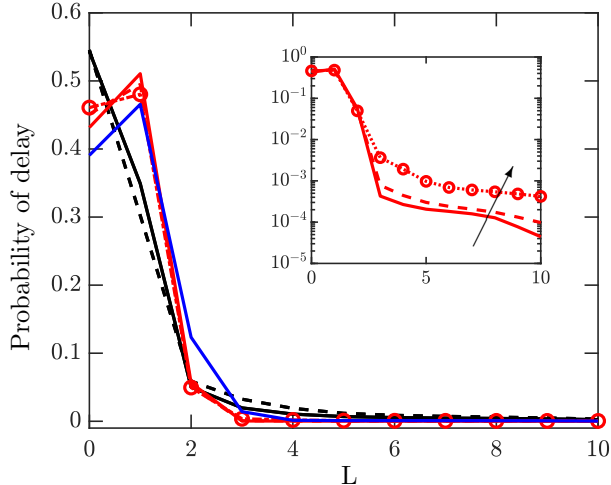

For illustration purposes, in \rfig{del_machines} we show
 PDF of delays $(\tilde{k})$ on three large systems at Texas Advanced Computing Center (TACC),
namely, Stampede2,  Frontera and Lonestar5 for $L=10$ and different processor counts.
From the black lines, we can clearly see that the probability of delays
decreases with increasing delay and
$P(\tilde{k}>3)\lesssim 0.05$ on Stampede2. This implies that
for a simulation with $L=3$, synchronizations will be forcefully imposed less than $5\%$ of
the time. % and thus $L=3$ is a good choice.
The trend is consistent even if we double the number of processors from
$P=8192$ (solid black) to $P=16384$ (dashed black). For Frontera (red lines in \rfig{del_machines}) we
see that the probability of $\tilde{k}=1$ is higher than the
probability of $\tilde{k}=0$ for all the three cases.
%$P=4096$ (solid red), $P=8192$ (dashed red) and $P=262144$ (red circles).
This points to a slow network that is expected to adversely affect the
scaling for standard synchronous simulations.
We see similar behavior for Lonestar5 (blue), with
probability of $\tilde{k}=1$ being the maximum. %, along with higher
%probability for larger delays. %However, we do see that Lonestar5 also has a higher
%probability of $\tilde{k}=2$ as compared to the other two machines.
For both Frontera and Lonestar5, $P(\tilde{k}>3)\lesssim 0.05$, for all the processor counts
shown in \rfig{del_machines}. Thus, $L=3$ is a reasonable choice for these three machines.
Note that for CAA this is equivalent to a reduction in the volume of communications by
a factor of four. This reduction will be particularly critical
when the PE count is high as envisioned in the exascale machines.

It is also worth noting the non-vanishing probability of delays as large as
$L=10$ in \rfig{del_machines} (inset) for Frontera,
indicating that at least some fraction of communications were synchronized.
While this is not of much consequence at low processor counts,
for an increasingly large number of PEs, even a small probability
of large delays can account for severe overheads. %lead to larger errors or even instabilities.
For example, for a seemingly low probability of
$P(\kt=10)\sim\mathcal{O}(10^{-4})$,
at a processor count of $\mathcal{O}(10^5)$, at least $\mathcal{O}(10^1)$
processors
%\colr{[DD: why $10^3$ processors? Shouldn't it be in this
%case $10^{-4}\times 10^{5}\sim 10^1$?]}
see a delay of $L=10$ at
the boundaries and are forced to synchronize at every time step.
Considering that the probability of large delays increases with increasing number of PEs
as seen in \rfig{del_machines} (inset),
a much larger fraction of processors would see large delays
in the next generation exascale machines
where the number of PEs is expected to be of
$\mathcal{O}(10^{6})$-$\mathcal{O}(10^{9})$
with increased architectural inhomogeneity. %, a much larger fraction of processors
%would see large delays.
Thus, even with a large value for the maximum allowed delay,
a significant number of PEs would be subject to forced synchronization at extreme scales.
However, these synchronizations would still be extremely small in comparison to the
standard synchronous algorithms that require all PEs to synchronize at all times.
%Thus, in the absence of asynchronous computing, the
%communication and synchronization overheads at such extreme scales would be
%insurmountable.

The second and equally important factor to be considered in the choice of
$L$ is numerical stability.
%Because of the methodology of derivation of AT schemes,
%these schemes mathematically preserve the order
%of accuracy. However,
Asynchrony and the associated random nature of delays and
coefficients introduce random numerical errors.
These error can trigger instabilities, especially if the delay ($\tilde{k}$),
bounded by $L$, is very large.
We will discuss this effect for a simple 1D diffusion equation,
\be
\frac{\partial u}{\partial t}=\alpha \frac{\partial^2 u }{\partial x^2}
\eqnlabel{diff}
\ee
where $\alpha$ is the diffusivity constant and $u(x,t)$ is the velocity
field. This equation is discretized using a second order AT scheme in space
 and forward Euler in time. Following \cite{AD2017}, we can discretize \eqn{diff}
at the $i^ {th}$ grid point
with delay $\tilde{k}_l$ at the left boundary and
$\tilde{k}_r$ at the right boundary as, %, respectively, of a PE $p$ as,
\be
u_i^{n+1}=u_i^n+\frac{\alpha\Delta t}{\Delta x^2}
\left((\tilde{k}_l+1)u_{i-1}^{n-\tilde{k}_l}
-\tilde{k}_l  u_{i-1}^{n-\tilde{k}_l -1}
-2u_i^{n}+(\tilde{k}_r +1)u_{i+1}^{n-\tilde{k}_r}
-\tilde{k}_r u_{i+1}^{n-\tilde{k}_r -1}\right).
\eqnlabel{dis_at}
\ee
For the above discretization we have considered an extreme case scenario
where $P=N$ and $N_T=1$ $i.e.$ every PE has only one grid point. It
can be shown that \eqn{dis_at} preserves the order of accuracy despite
delays on both boundaries. Next we define
$U^{n}:=[u_0^n,u_1^n,...,u_N^n]$ and $V^{n}:=[U^{n},~...~,~U^{n-\tilde{k}-1}]^T$, where
$\tilde{k}=max(\tilde{k_l},\tilde{k_r})$. Using these definitions,
we can write the matrix form of the evolution equation as,
\be
V^{n+1}=\mathbf{A}(\tilde{k}_l,\tilde{k}_r)V^n
\ee
where the coefficient matrix is
\be
\mathbf{A}(\tilde{k}_l,\tilde{k}_r)=
\begin{bmatrix}
\A0 &  \vo{A}_1 &\dots & \vo{A}_{\tilde{k}} & \vo{A}_{\tilde{k}+1} \\
\Id & \vo{0}&\dots & \vo{0} & \vo{0}\\
\vdots &\vdots& \vdots & \vdots & \vdots \\
\vo{0} & \vo{0}&\dots & \Id & \vo{0}
\end{bmatrix}.
\ee
While this equation is very general, we specialize this system
to same delay on both sides ($\tilde{k}=\tilde{k}_l=\tilde{k}_r$) for all processors,
which can be thought as a worst case scenario. Defining $r_d=\alpha \dt /\dx^2$ as the diffusive CFL,
we can then write,
\[
\textbf{A$_0$}(\tilde{k})
=
\begin{bmatrix}
1-2r_d & \mathcal{L}_1^{0}r_d & 0&\dots &0& \mathcal{L}_1^{0}r_d \\
\mathcal{L}_1^{0}r_d & 1-2r_d & \mathcal{L}_1^{0}r_d & \dots &0&0 \\
\vdots & \vdots &\vdots & \vdots &\vdots & \vdots\\
\mathcal{L}_1^{0}r_d & 0 &0&\dots & \mathcal{L}_1^{0}r_d& 1-2r_d \\
\end{bmatrix},
\hspace{5mm}
\textbf{A$_{\tilde{k}}$}(\tilde{k})
=
\begin{bmatrix}

0 & \mathcal{J} & 0 & \dots & 0 & \mathcal{J}\\
\mathcal{J} & 0 &  \mathcal{J} & \dots & 0 & 0\\
\vdots & \vdots & \vdots &\vdots & \vdots & \vdots \\
 \mathcal{J}&0 & 0 & \dots &  \mathcal{J}&0\\
\end{bmatrix}
\]
with $\mathcal{J}$ defined as,
\be
\mathcal{J} =\mathcal{L}_1^{m}r_d (\tilde{k} +1)
-\mathcal{L}_2^{m}r_d \tilde{k}
\eqnlabel{jdef}
\ee
which is used to set the coefficient as $r_d (\tilde{k} +1)$ for
$\vo{U}^{n-\kt}$ and $-r_d \tilde{k}$ for $\vo{U}^{n-\kt-1}$.
For this we use
%to pick the correct coefficient ($r_d (\tilde{k} +1)$ or $-r_d \tilde{k}$).
$\mathcal{L}^m$, which is the Lagrange polynomial of order $L$,\\
\be
\mathcal{L}_1^{m}(\tilde{k})=\prod_{l\ne m}^{L}\frac{\tilde{k}-l}{m-l}
~~~~~~~~~~~~~~~\mathcal{L}_2^{m}(\tilde{k})=\prod_{l\ne m}^{L}\frac{\tilde{k}+1-l}{m-l}.
\ee
By definition, $\mathcal{L}_1^{m}(\tilde{k})$
takes value 1 if $m=\tilde{k}$ and zero otherwise. Similarly
$\mathcal{L}_2^{m}(\tilde{k})$  is 1 if $m=\tilde{k}+1$ and 0 for
other values of $m$. The number of Lagrange polynomials
is equal to the number of time levels in the AT scheme, which for
the second-order scheme used here is equal to two.
In the absence of delays we have, $\vo{A}= \A0 $,
which is equivalent to the standard second-order finite difference system.

For stability, the
spectral radius of $\vo{A}(\tilde{k})$ should be bounded by unity to ensure that the
numerical perturbations do not grow unboundedly in time. Because of the
complexity of the system, the spectrum has
to be computed numerically. Again as a worst case
scenario \cite{AD2017}, we assume a Dirac delta distribution of delays, such that,
$\tilde{k}=L$ at all points.
For a given $L$, we compute
the maximum or critical $r_d$ for which all the eigenvalues of the
evolution matrix are less than unity. This is the largest value for
which the numerical scheme is stable, and is denoted by
$\rdml$.
%The maximum $r_d$
%that satisfies this constraint is referred to as $\rdms$ for
%$L=0$ and $\rdmal$ for $L>0$.
The results of this analysis are shown in
\rfig{stability_cfl}. In the synchronous limit ($L=0$), we obtain the
well known stability limit for a second order central
difference scheme in space with forward Euler in time, $\rdm(0)=0.5$ \cite{hirsch}.
As we increase $L$, this stability limit decreases as can be seen
from the solid red circles in \rfig{stability_cfl}(a). % but the relative effect
%decreases as $L$ increases.
Similar analysis was also done for the advection-diffusion
equation which has both first and second derivatives and thus
both convective ($r_c=c\dt/\dx$) and diffusive CFLs are used to determine
stability. Here again we fix the delay $\tilde{k}=L$ and compute
the stability limit in the $r_c$-$r_d$ plane.
The procedure was repeated for different values of $L$.
The result is plotted in \rfig{stability_cfl}(b).
For $L=0$, we get the well known stability bound, $\rcm(0)\le 2(\rdm(0))^2
\le 1$ \cite{hirsch}.
%where $\rcm(0)$  and $\rdm(0)$ are the maximum diffusive and
%convective CFLs, respectively, such that all the eigenvalues of the
%evolution matrix $\vo{A}_0$ are bounded by unity.
As $L$ is increased we see that both
$\rcml$  and $\rdml$ decrease resulting in a smaller
stability region.
Thus, for stability a time-step smaller than that for $L=0$ is required
whenever $L>0$. We do point out that these stability bounds are
based on worst case scenario assumptions and are thus strict. In more
realistic
scenarios (\ref{sec:Vnat}), the effect of asynchrony on stability is
relatively weaker.

%The direct implication of this reduced stability for large
%$l$ is the need for a smaller time-step in order for an asynchronous simulation to be stable.

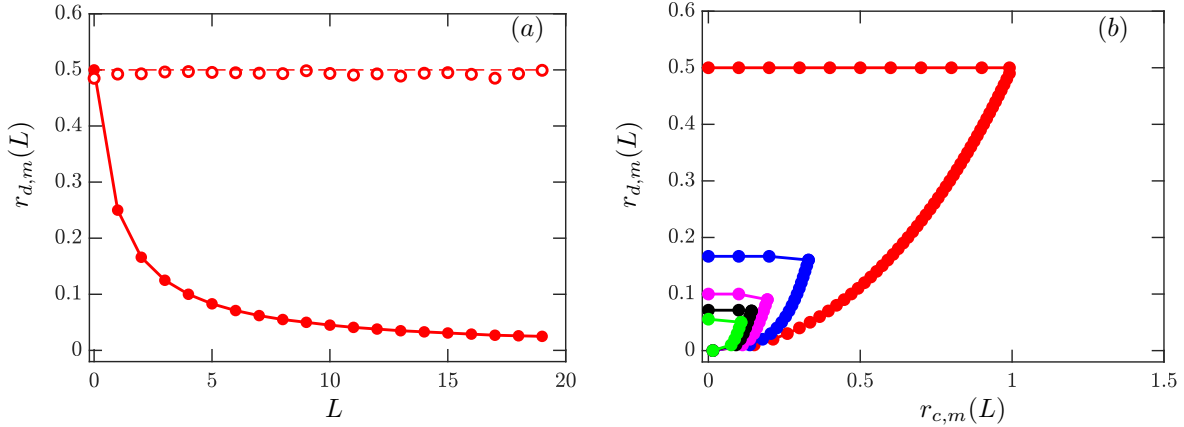
\begin{figure}[h]
\begin{center}
\includegraphics[width=0.42\textwidth]{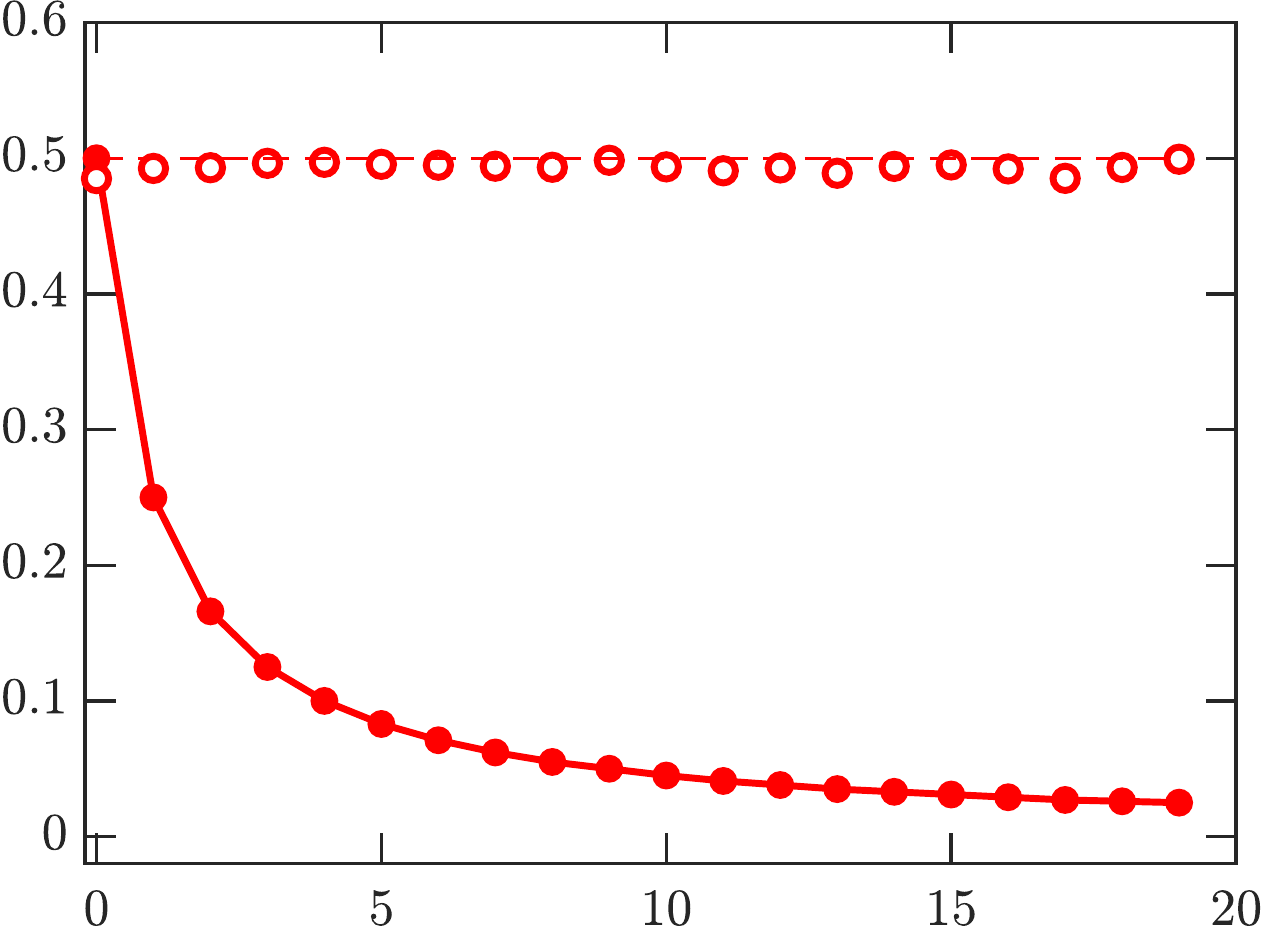}
\hspace{1cm}
\includegraphics[width=0.41\textwidth]{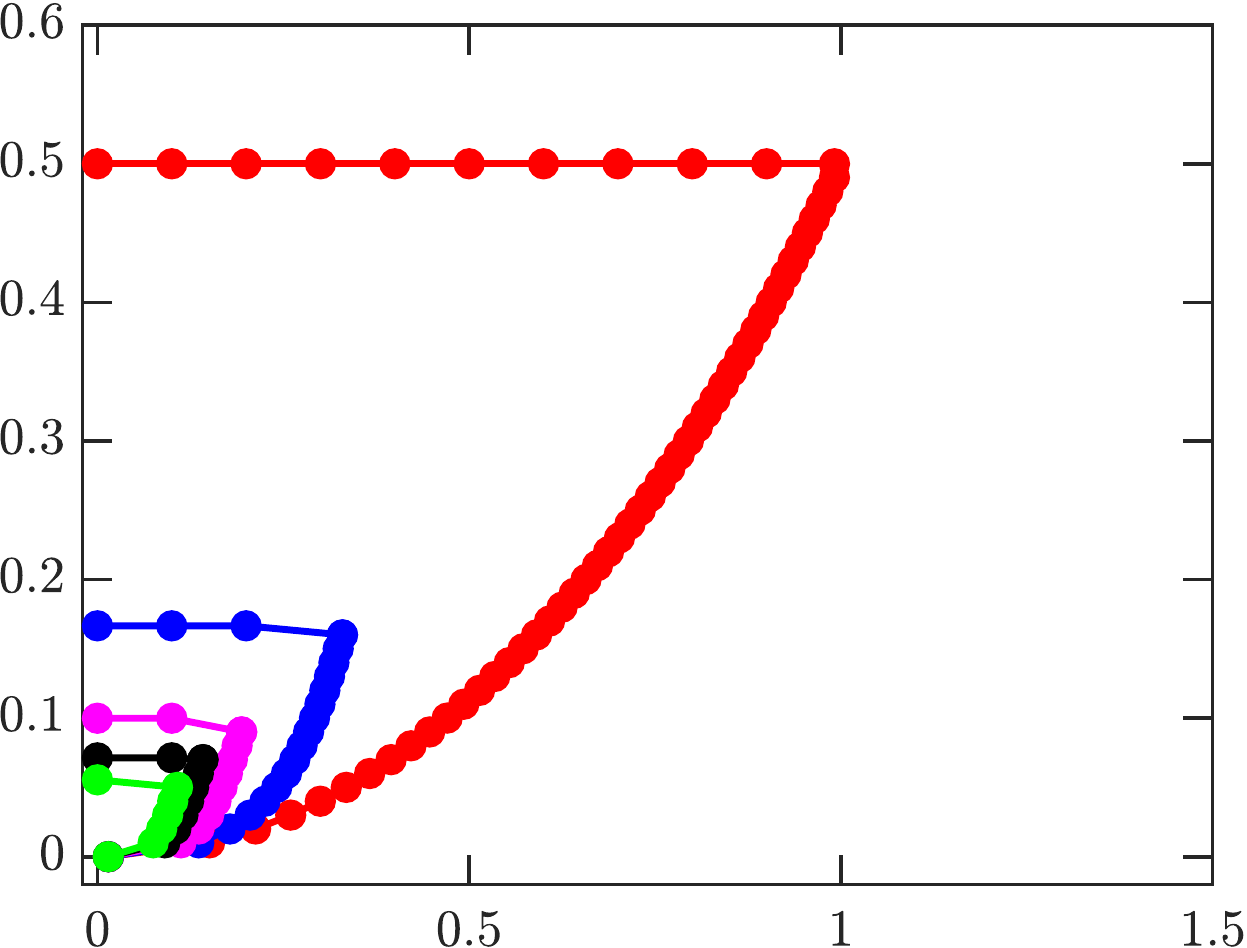}

\begin{picture}(0,0)
\put(-40,145){$(a)$}
\put(182,145){$(b)$}
%\put(-35,105){$(e)$}
%\put(-35,45){$(f)$}
\put(-230,80){\rotatebox{90}{$r_{d,m}(L)$}}
\put(0,80){\rotatebox{90}{$r_{d,m}(L)$}}
\put(-110,1){$L$}
\put(115,1){$r_{c,m}(L)$}

\end{picture}

\caption{(a) Variation of stability limit $r_d$ (solid) and $\widetilde{r_d}=(L+1)r_d$ (hollow)
with $L$ for diffusion equation.
(b) Stability limit in $r_c$-$r_d$ plane for advection-diffusion equation for
 $L=0$ (red), $L=2$ (blue), $L=4$ (magenta), $L=6$ (black) and $L=8$ (green).
}
\figlabel{stability_cfl}
\end{center}
\end{figure}

%While the above numerical analysis gives us the exact stability limit for
%simple 1D problems, we still need to quantify the decrease in the stability limit with $L$.
%Since this decrease, as we can see from \rfig{stability_cfl}(a), is
%non-linear in terms of $L$,
%without loss of generality,
In order to characterize the reduction in stability limits,
it is of interest to
obtain the stability limit in an asynchronous simulations
from the known stability limit of a synchronous implementation.
This can be written as
\begin{equation}
\rdml=\rdm(0)/f(L),
\eqnlabel{rdml}
\end{equation}
where the yet unknown
function $f(L)$ characterizes the
effect of delays.
Clearly, $f(0)=1$.
%In order to obtain the correct functional form of $f(L)$, we look
Some guidance on a plausible functional form for $f(L)$ can be
obtained by a careful examination
of \eqn{dis_at} where we observe that, in the presence of delays,
$r_d$ at the boundary points always appears in
conjunction with functions of delays that are also the coefficients of the
AT scheme.
%terms linear in $\kt$ and of the form
In the present case, from \eqn{jdef} we have
$r_d(\kt+1)$ and $-r_d\kt$
in the evolution matrix $\vo{A}(\kt)$.
%that directly affect the stability limit.
Since both terms are linear in the delay,
it is natural to expect that, for $\kt=L$,
stability, and thus $f(L)$, would be a linear function of $L$.
In fact, %a non-linear \colr{[DD: what do you mean with ``non-linear fit''?]}
%fit for
a best fit approximation for
$\rdml=\rdm(0)/f(L)$ does yield a linear relation $f(L)\approx L+1$.
Both $\rdm(0)/(L+1)$ (solid line) and $\rdml$ (solid circles)
are plotted in \rfig{stability_cfl}(a) and are in excellent
agreement with each other. Furthermore, we can
re-arrange $\rdml=\rdm(0)/f(L)$ to read as
$\rdma=\rdml\times f(L) =\rdm(0)$. This
implies that with a correct approximation for $f(L)$,
we can express stability in terms of an \textit{effective asynchronous}
CFL ($\rdma$), which is independent of delay $L$ and essentially equal to
the synchronous stability limit ($\rdm(0)$). The numerical
data do support this argument as can be seen from
\rfig{stability_cfl}(a) where $\rdma$ (hollow circles)
are constant for all $L$ and close to $\rdm(0)=0.5$ (dashed line).

We also computed the stability limit for the schemes used
for the turbulence simulations in this work, namely,
fourth-order AT schemes coupled
with AB2 in time. This is shown in \rfig{stability_cfl3}(a) for the
diffusion equation. In this case,
$r_d$ appears multiplied by
the coefficients in this fourth-order AT scheme (Appendix B)
in the discrete equation
%(e.g.\ $(L^2+3L+2)/2$),
which are seen to be quadratic in $L$.
Then, based on the argument above, we expect
$f(L)$ also to be quadratic in $L$.
%in the effective
%time-step $\widetilde{\Delta t}=f(L)\dt$ or $\widetilde{r_d}\approx f(L) r_d$,
%is quadratic in $L$ since coefficients in the
%fourth-order AT scheme (Appendix B),
%for example, $(L^2+3L+2)/2$ etc. that would multiply $r_d$
%in the discrete equation, are also quadratic.
From \rfig{stability_cfl3}(a) we can see that that
$\rdml$ (solid circles) decreases with $L$ and
is in good agreement with $\rdm(0)/f(0.74L^2+0.47L+1)$ (solid line).
Moreover, $\rdma$ (hollow circles) is close to $\rdm(0)\approx0.18$ (dashed
line) for all $L$.
This again supports the proposed rescaling in \eqn{rdml}.

One can understand this effect more intuitively as follows.
When there is a delay at the PE boundaries, data from multiple delayed
time levels is used at these points for computation of derivatives.
As a result the effective time-step, as seen by the numerical scheme,
increases. This effective time-step is essentially
equivalent to $\dt_L=\dt\times f(L)$ and is apparent when $\rdma$
is written as
\be
\rdma=\rdml\times f(L)=\frac{\alpha(\dt \times f(L)}{\dx^2}=\frac{\alpha (\Delta t_L)}{\dx^2},
%\text{   or, }
%\w_c\approx\frac{c(f(L)\dt)}{\dx}
%\approx\frac{c (\widetilde{\Delta t})}{\dx}.
\eqnlabel{rdtilde}
\ee
For fixed grid spacing $\dx$, this
increase in time-step is compensated by a decrease in $\rdml$
to ensure stability. On the other hand, $\rdma$
which is already expressed in terms of $\dt_L$, remains approximately
constant
with $L$ and is equal to $\rdm(0)$.

\begin{figure}[h]
\begin{center}
\includegraphics[width=0.41\textwidth]{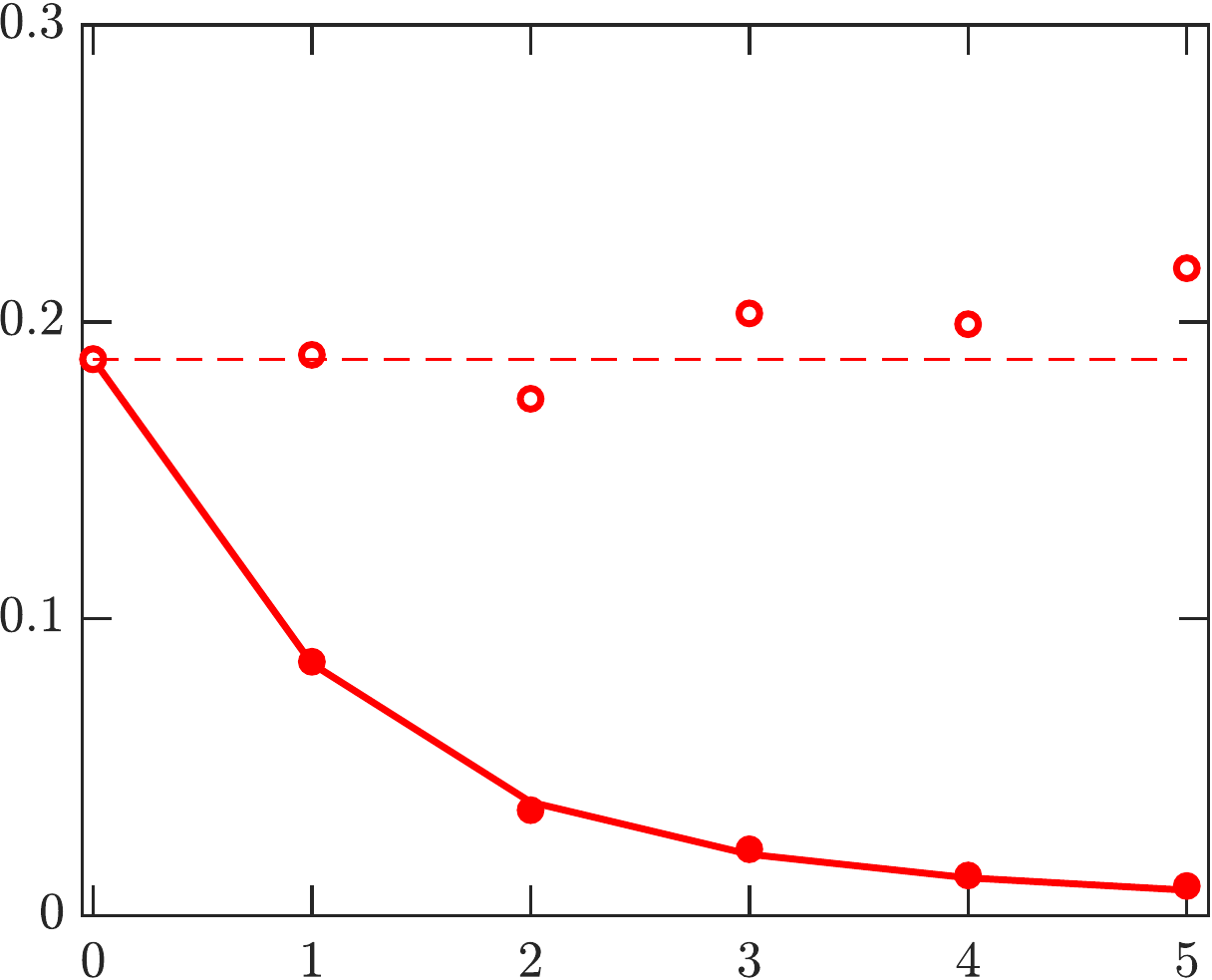}
\hspace{1cm}
\includegraphics[width=0.425\textwidth]{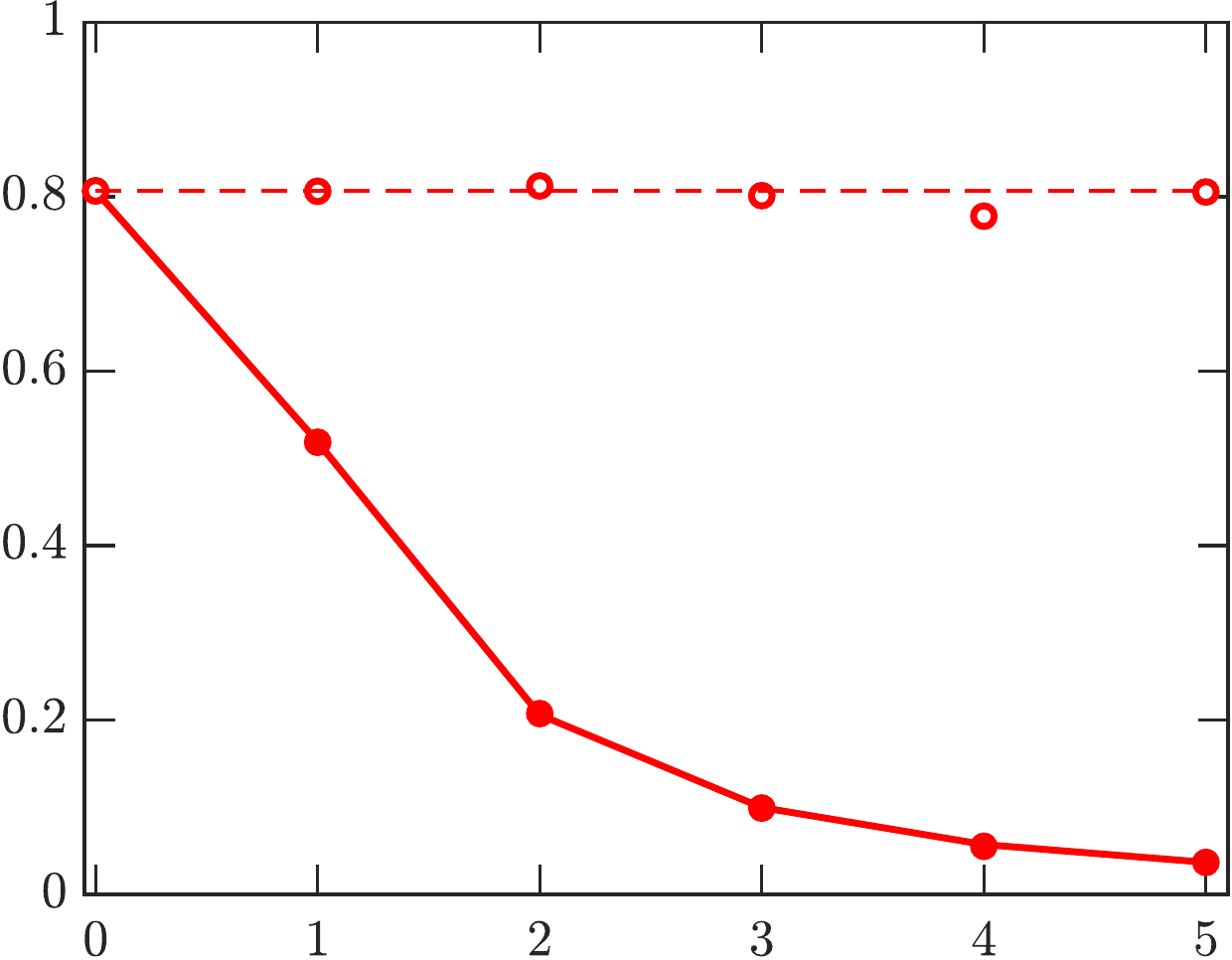}

\begin{picture}(0,0)
\put(-40,147){$(a)$}
\put(182,147){$(b)$}
%\put(-35,105){$(e)$}
%\put(-35,45){$(f)$}
\put(-230,80){\rotatebox{90}{$\rdml$}}
\put(0,80){\rotatebox{90}{$\rcml$}}
\put(-110,1){$L$}
\put(115,1){$L$}

\end{picture}

\caption{Variation of stability limit (a) $r_d$ (solid) and $\widetilde{r_d}=
(0.74L^2+0.47L+1)r_d$ (hollow)
with $L$ for diffusion equation and (b) $r_c$ (solid) and
$\widetilde{r_c}=(0.90L^2-0.35L+1)r_c$ (hollow)
 for NS equation, using fourth-order AT scheme in space and AB2 in time.
}
\figlabel{stability_cfl3}
\end{center}
\end{figure}
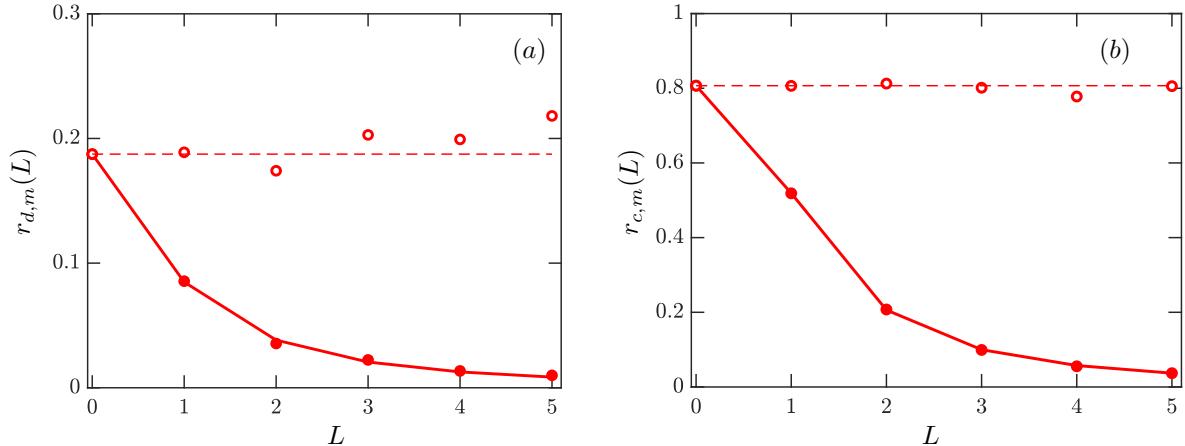

For a complex system of equations, such as the Navier-Stokes
equations,
an analytical stability analysis is difficult. However, stability limits
can be computed numerically either by gradually increasing the CFL until the
system becomes unstable or by using the bisection method.
We obtained the stability limit for decaying turbulence at $Re_{\lambda}\approx 35$,
by imposing a fixed delay $L$ at all the six faces at every time step.
Since both diffusive and convective terms are
present in the NS equations, the time-step is determined by the smallest
physical time scale, which for the simulations presented
is always the latter. Thus, the stability
limit is obtained in terms of a convective CFL ($r_c$) and is
shown in \rfig{stability_cfl3}(b) with $\rcml$ (solid circles)
decreasing with $L$. As before,
this effect is
accurately captured by $\rcml=\rcm(0)/f(L)$ (solid line),
where $f(L)\approx0.90L^2-0.35L+1$. Here again,
$\rcma=\rcml\times f(L)$ (hollow circles) is
seen to be a constant
consistent with the synchronous limit
$\rcm(0)\approx0.8$ for all $L$.
%Here again
%$f(L)$ is a quadratic function of $L$.
%write $\widetilde{\dt}=f(L)\dt$, where $f(L)$ is quadratic for fourth-order
%AT scheme.

In general, this analysis shows that
the stability limit for the AT schemes
decreases with $L$ as $\rdm(0)/f(L)$. This dependency
can also be expressed using the \textit{effective asynchronous} CFL $(\rcma$ or $\rdma)$
which satisfies the same limit as the synchronous case $(\rcm(0)$ or $\rdm(0))$ and
uses an effective time-step
($\dt_L=f(L)\times\dt$). Here $f(L)$,
which is of the same order in $L$ as the coefficients in the
corresponding AT scheme, gives a quantitative measure of the
effect of delays on the stability limit. For example, for a large
value of $f(L)$, in order to keep $\rcma=f(L)\times \rcml$
%close to a
constant,
$\rcml$ needs to be small. This implies that a small $\dt$ is
required for the asynchronous simulation to be stable, which
in turn can
increase the computational cost.
However, we note that while simulations of turbulent flows at
$r_c=1$ are prevalent in literature,
recent studies have shown that for adequate temporal
resolution, a much
smaller $r_c$  should be used \cite{PK2018}.
Thus, the CFL (or $\dt$)
dictated by those resolution requirements, could be
much smaller than the reduced stability
limit discussed above.
%We can therefore use a small $r_c$ (or time-step)
%for both synchronous and asynchronous
%simulations without additional overheads due to lower stability
%bound for the AT schemes.

Summarizing the results from this section,
the maximum allowable delay ($L$) is chosen such that the
PEs incur in minimal
overheads because of forced synchronization and communications,
without additional computational cost to ensure stability.

\section{Numerical results}
\label{sec:results}
We have implemented the synchronous and asynchronous
numerical methods and algorithms described in the previous sections
to perform DNS of decaying and forced isotropic turbulence
%were performed for
at different Reynolds numbers to assess the effect (or lack thereof) of asynchrony.
%and
%and characterize effects (or lack thereof)
%of asynchrony% was studied
%on turbulence
%characteristics.
The resolution used for both synchronous and asynchronous implementations
is $\eta/\dx\approx0.5$ or $\kappa_{max}\eta\approx1.5$,
where $\eta=(\nu^3/\langle \epsilon \rangle)^{1/4}$
is the Kolmogorov length scale, $\nu$ is the kinematic viscosity and $\kappa_{max}=\sqrt{2}N/3$
is the highest resolvable wave number for commonly used
pseudospectral simulations in a cubic domain of length $2\pi$ on each
side
and $N^3$ points \cite{Canuto1988,PK2018}.
This resolution has been shown to lead to well-resolved simulations for the conditions and
quantities of interest presented here \cite{SD2016,Wang2017}.
As discussed in \rsec{schemes},
the time-step size $\Delta t$ is fixed
at a value that yields an initial CFL of ${\cal O}(0.1)$
consistent with the recommendation in \cite{PK2018}.
To facilitate comparisons
both synchronous and asynchronous simulations use the same
time step.
We use periodic boundary condition in all directions.
The initial velocity field is a stationary state obtained by forcing
the large scales of motion as done in \cite{DS2013,SD2016} and is same for both synchronous and
asynchronous simulations. The important simulation parameters including resolution,
percentage of points directly affected by asynchrony $(N_B\%)$,
$Re_{\lambda}$, and simulation time in terms of eddy turnover time $T_e=\mathcal{L}/u_{rms}$,
where $\mathcal{L}$ is the integral length scale and $u_{rms}$ is the root mean square
of velocity fluctuations, are tabulated in \rtab{DNS}.
The level of compressibility is commonly defined in terms
of the turbulent Mach number
$M_t=\langle u_i u_i \rangle^{1/2}/c$, where
$c$ is the mean speed of sound, $u_i$ is the velocity fluctuation,
$\langle \cdot \rangle$ is the average computed
across the entire domain and summation convention is used.
For the simulations in this paper $M_t\ap 0.3$ which represents
a case where dilatational effects start becoming important
\cite{SD2016}.
%Here $\eta=(\nu^3/\langle \epsilon \rangle)^{1/4}$
%is the Kolmogorov length scale,
%$\kappa_{max}$ is the blaah and $T_e$ is the eddy turnover time.
%Both decaying and forced cases are evolved using synchronous
%and asynchrounous schemes and various quantities
%are compared to assess the effect (or lack thereof) of
%asynchrony.
%initial condition in each of these cases was obtained from the
%forced, stationary state simulations. These fully developed
%turbulent velocity fields of the initial state were then allowed to evolve
%using synchronous and asynchronous schemes (both random and communication
%avoiding) and the results were compared.

For the rest of this section, we will refer the
synchronous simulations using standard finite differences as SFD. The asynchronous
simulations using AT schemes with random delays will be referred to as SAA
and that with periodic delays will be referred to as CAA. We also
have tenth-order compact schemes (C10) with third order RK scheme in time for
one of the cases for comparison purposes to highlight that
our finite difference simulations are comparable
to the most well resolved simulations of compressible turbulence in literature \cite{Wang2010,JD2012,DS2013,Wang2017}.
%Also, unless otherwise mentioned, the red line with circles is SFD,
%the blue line with or without asteriks is ATR and the black like with
%triangles is ATP.

\begin{table}[h]
\begin{center}
\begin{tabular}{ c c c c c c}
\hline
\hline
\multicolumn{6}{c}{$Decaying$}\\
\hline
  $N^3$ & $N_B(\%)$ & $Re_{\lambda}(0)$& $\eta(0)/\dx$ & $\kp_{max}\eta(0)$ & $t/T_e(0)$  \\
    $256^3$ &   57.8 &    100  &  0.5 &     1.4 & 24 \\
    $512^3$ &   50.8 &   145  &  0.5&    1.5 & 24 \\
\hline
\hline\\
\multicolumn{6}{c}{$Forced$}\\
\hline
$N^3$ & $N_B(\%)$ & $Re_{\lambda}$ & $\eta/\dx$ & $\kp_{max}\eta$ & $t/T_e$  \\
    $64^3$ &   57.8  &     35 &  0.5 & 1.6 & 10   \\
    $256^3$ &  57.8  &    100 &  0.5 & 1.8 & 19\\
\hline
\hline
\end{tabular}
\end{center}
        \caption{DNS parameters: number of grid points $N^3$,
percentage of boundary points $N_B\%$, Taylor Reynolds number $Re_{\lambda}$,
resolution $\eta/\dx$ and $\kp_{max}\eta$ and normalized simulation time $t/T_e$.
Normalization is done using the initial values
($Re_{\lambda}(0),\eta(0),T_e(0)$) for the decay cases and
using average computed over stationary state for the forced case.}
\label{tab:DNS}
\end{table}

\subsection{Decaying turbulence}
\subsubsection{Low order statistics in physical space}
It is important for any numerical scheme to accurately
capture
the large scale behavior of the system.
%We first look at the decay of the
An important and widely studied \cite{Kida1992, Samtaney2001}
large scale quantity in fluid turbulence is the
mean turbulent kinetic energy per unit mass defined as,
\begin{equation}
K=\frac{1}{2}\langle \rho u_iu_i \rangle.
\end{equation}
%where $u_i$ is the velocity fluctuation, $\langle .\rangle$
%is the average across the entire domain and summation convention is used.
In the absence of energy input to the system, $K$
decays in time as shown in \rfig{tke}($a,c$), where $K$
is normalized by its initial value $K_0$ and time is
normalized by initial eddy turnover time,
$T_e(0)=\mathcal{L}/u_{rms}$.
After an initial transient, the decay obeys a power-law in time observed %wevident form
as a straight line on a log-log scale in \rfig{tke}($a,c$).
%for $Re_{\lambda}(0)\approx100$, where large $t/T_e(0)$ is large.
The decay exponent is seen to be consistent with that found in the
literature for similar conditions \cite{Samtaney2001,McComb2018}.
The excellent agreement between SFD, CAA and SAA in \rfig{tke}($a,c$) at all times shows that
asynchronous implementations %It is clear that asynchronous implementations
% capture %this physical phenomena
have accuracy comparable to SFD. %the standard numerical methods.
%The same conclusion can be drawn by looking at other
%quantities such as the %We have also shown the evolution of
%\rfig{tke}(b,c) show the evolution of the

The rate at which kinetic energy is dissipated is
given by $\langle \epsilon \rangle=2\left\langle \sigma_{ij}S_{ij} \right\rangle$.
Because most of the contribution to dissipation comes from small scales
(or high wavenumbers) it is %\begin{equation}
%\langle \epsilon \rangle =2
%\left\langle \sigma_{ij}S_{ij} \right\rangle.
%\eqnlabel{diss}
%\end{equation}
%a small scale quantity and
therefore sensitive to how accurately high wavenumbers are resolved by the numerical methods.
%normalized by its initial value.
The decay of $\langle \epsilon \rangle$ (normalized by its initial value)
is shown in %We have shown its decay in
\rfig{tke}($b,d$) for SFD, CAA and SAA  with no observable differences.
%Also shown in \rfig{tke}($a,b$) in magenta is the evolution obtained
%for tenth order compact scheme with RK3 in time
%We see that after
%an initial constant value, the dissipation rate
%decays. A close agreement is also seen between the asynchronous and
%the synchronous decay of dissipation.
%Thus, there is no observable difference in the computation of
Thus, we find that the %The decay of both $K$
%and $\langle \epsilon \rangle$ show us that
asynchronous implementations are able to
capture the evolution of low-order large and small  scale quantities with accuracy comparable
to the standard finite differences.
Also shown in \rfig{tke}($a,b$) is the evolution obtained
for C10 (magenta line), which is identical to
the evolution obtained for both asynchronous and synchronous finite difference.
%for tenth order compact scheme with RK3 in time

%large scale quantities averaged quantities for asynchronous simualtions
%when compared against the synchronous case.
%This is consistent for both ATP and ATR and also
%for different Reynolds numbers plotted in \rfig{tke}.

\begin{figure}[h]
\centering%center}
\subfigure{\includegraphics[width=0.42\linewidth]{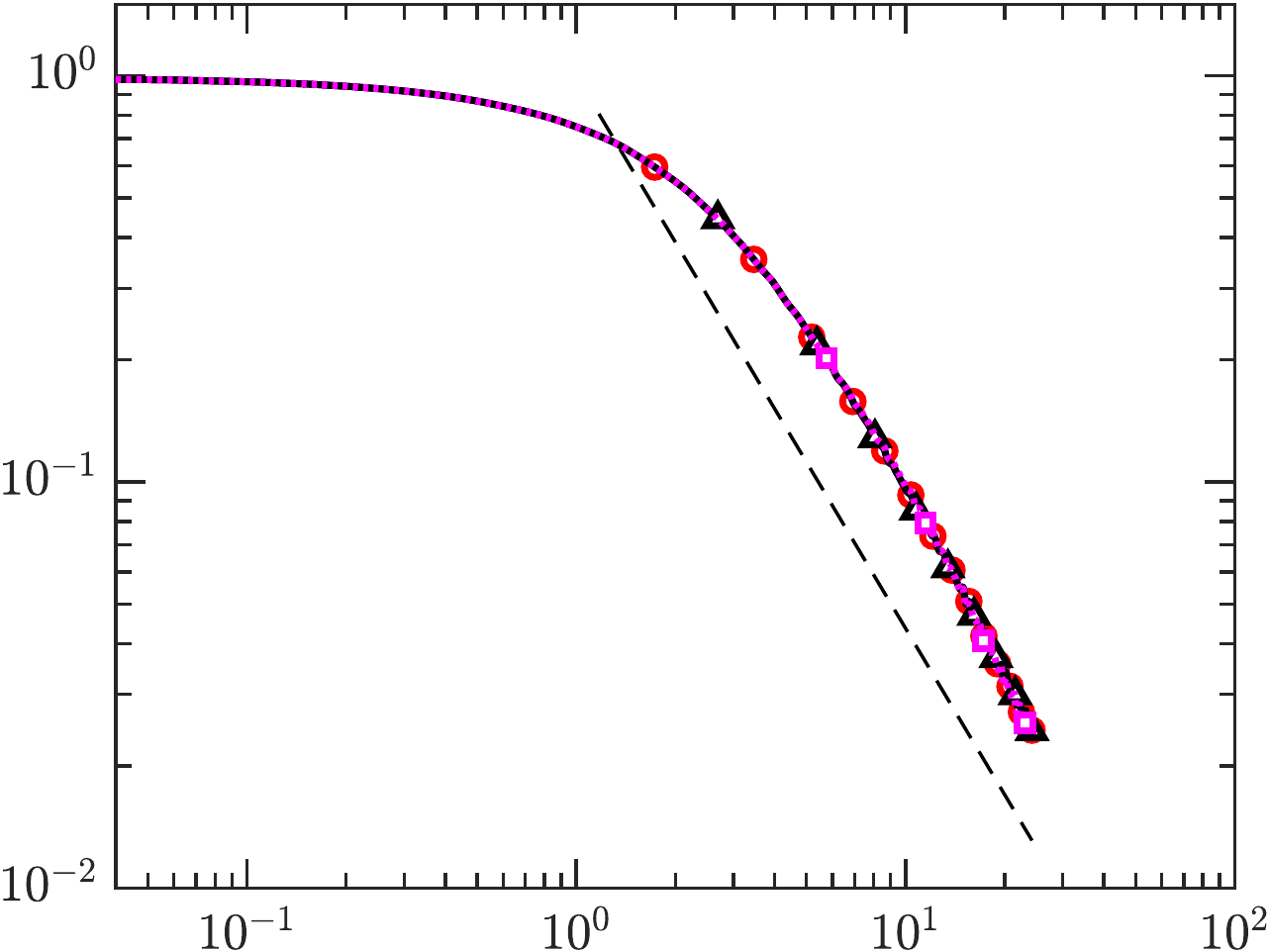}}
\hspace{1cm}
\subfigure{\includegraphics[width=0.42\linewidth]{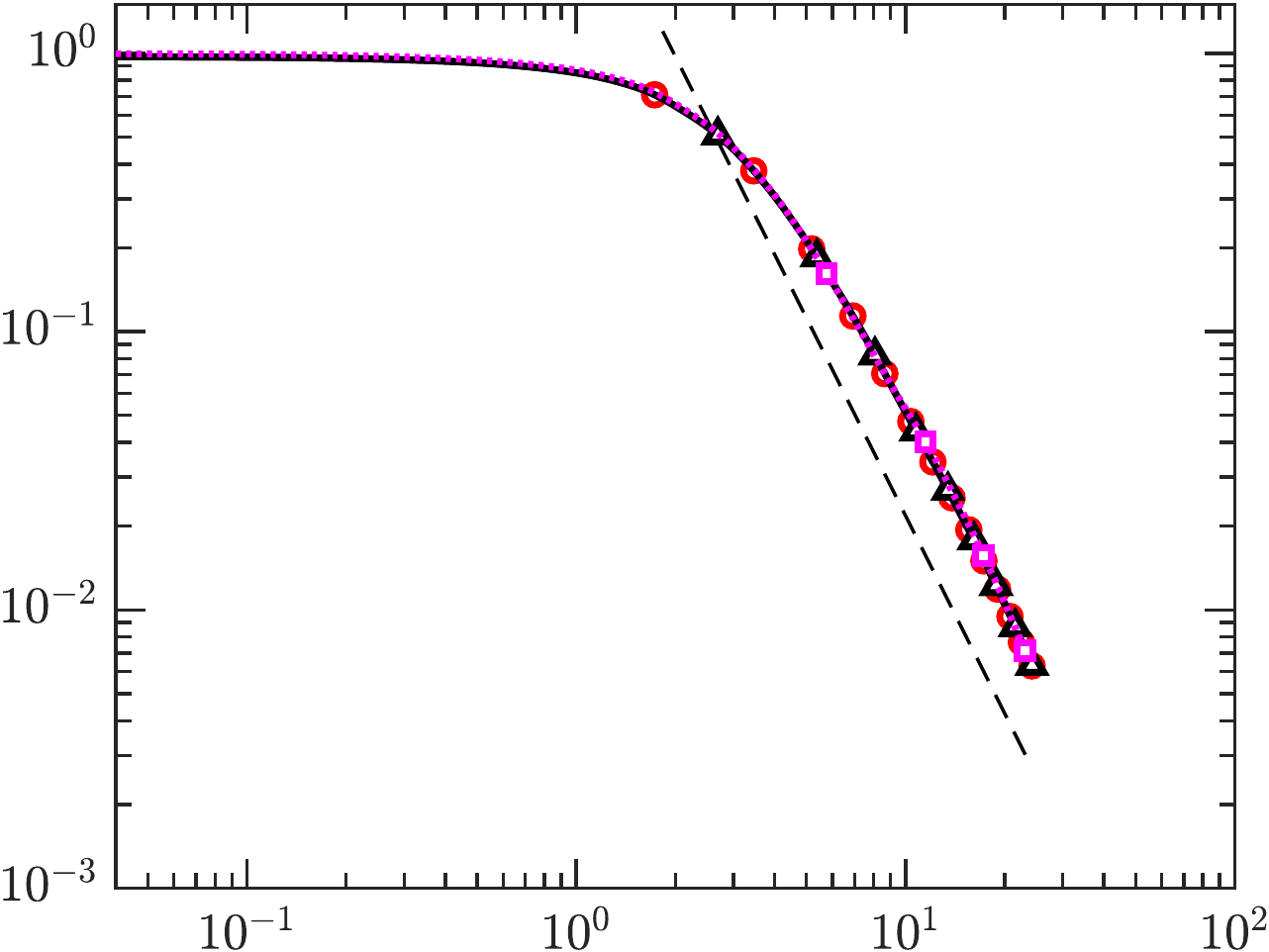}}

\vspace{0.2cm}
\subfigure{\includegraphics[width=0.42\linewidth]{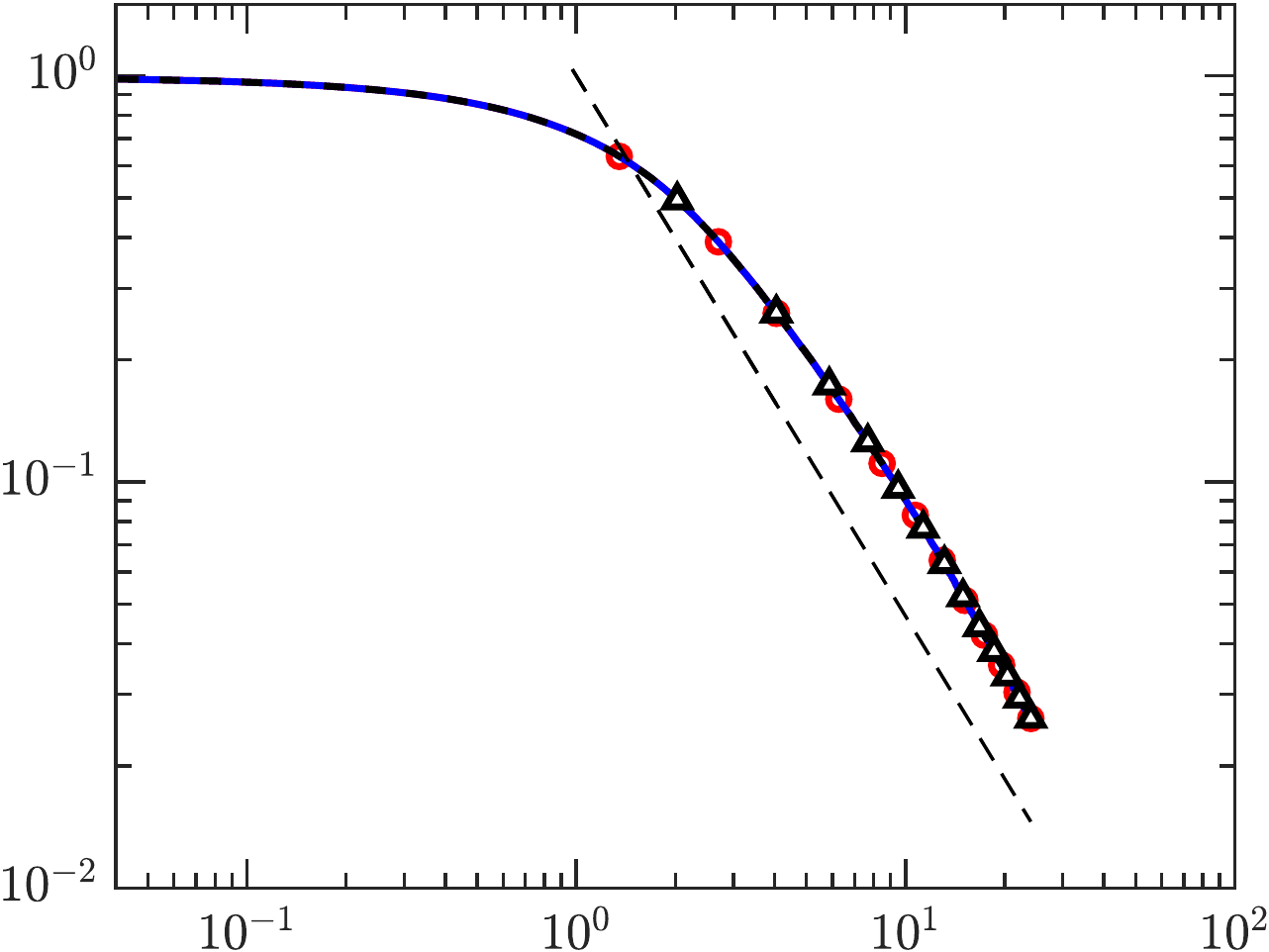}}
\hspace{1cm}
\subfigure{\includegraphics[width=0.40\linewidth]{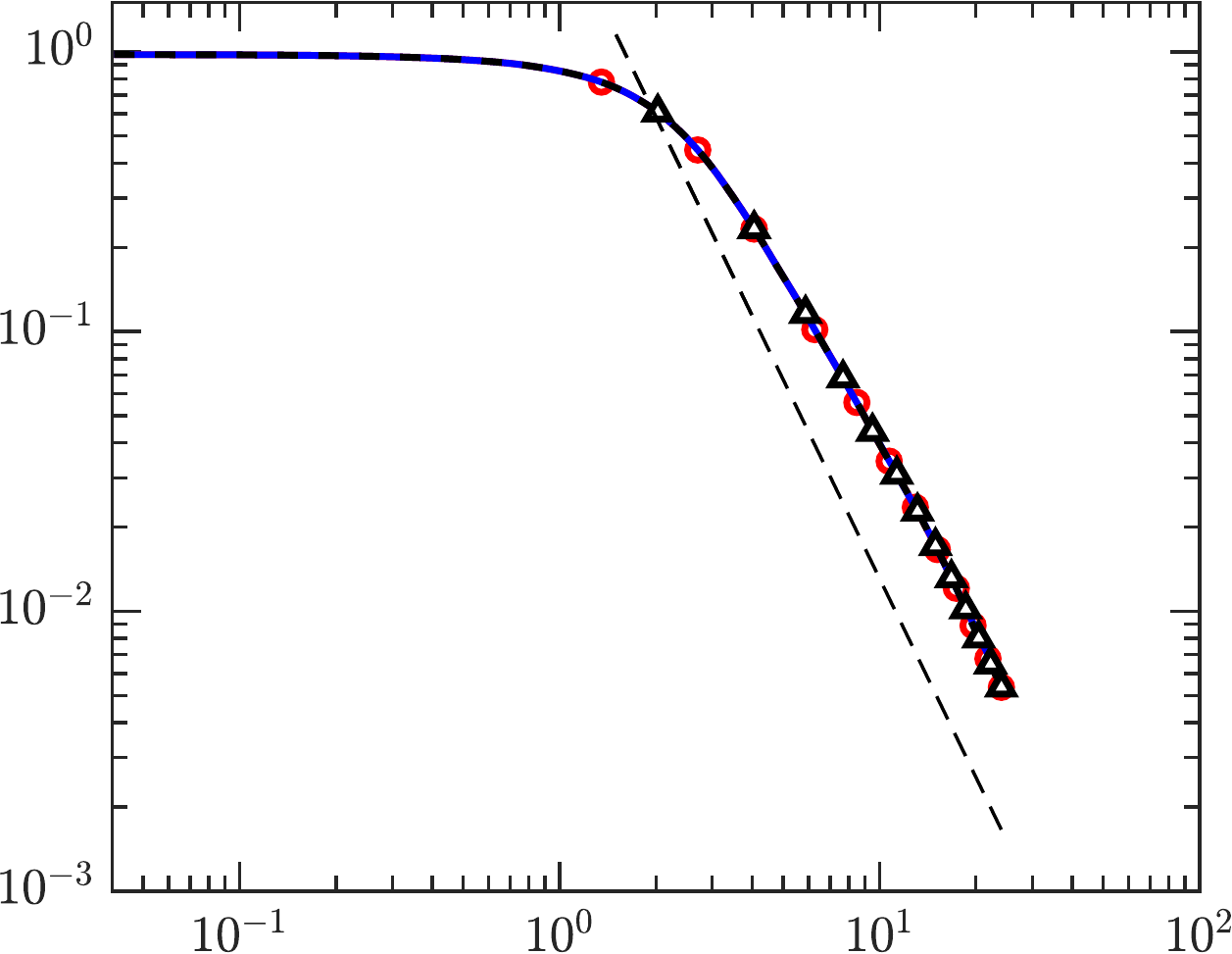}}

\begin{picture}(0,0)
\put(-40,305){$(a)$}
\put(180,305){$(b)$}
\put(-230,240){\rotatebox{90}{$K/K_0$}}
\put(0,230){\rotatebox{90}{$\langle \epsilon \rangle / \langle \epsilon_0\rangle$}}
%\put(-115,145){$t/T_e(0)$}
%\put(85,145){$t/T_e(0)$}

\put(-40,145){$(c)$}
\put(180,145){$(d)$}
%\put(-35,105){$(e)$}
%\put(-35,45){$(f)$}
\put(-230,80){\rotatebox{90}{$K/K_0$}}
\put(0,70){\rotatebox{90}{$\langle \epsilon \rangle / \langle \epsilon_0\rangle$}}
\put(-125,1){$t/T_e(0)$}
\put(95,1){$t/T_e(0)$}

\end{picture}

\caption{Evolution of space averaged
turbulent kinetic energy normalized by the
initial turbulent kinetic energy $K_0$ (left) and
evolution of space averaged
dissipation rate normalized by the initial dissipation rate $\epsilon_0$ (right) for
$Re_{\lambda}(0)\approx100$ ($a,b$) and $Re_{\lambda}(0)\approx145$ ($c,d$).
Different lines are: SFD (red-circle), CAA (black-triangle) and
SAA (blue) both with $L=3$. The black-dashed line corresponds
to $K/K_0\propto (t/T_e(0))^{-1.4}$ in $(a)$
and $\langle \epsilon \rangle/\langle \epsilon_0\rangle\propto (t/T_e(0))^{-2.4}$ in ($b$).
Magenta line in ($a,b$) is C10.
}
\figlabel{tke}
%\end{center}
\end{figure}
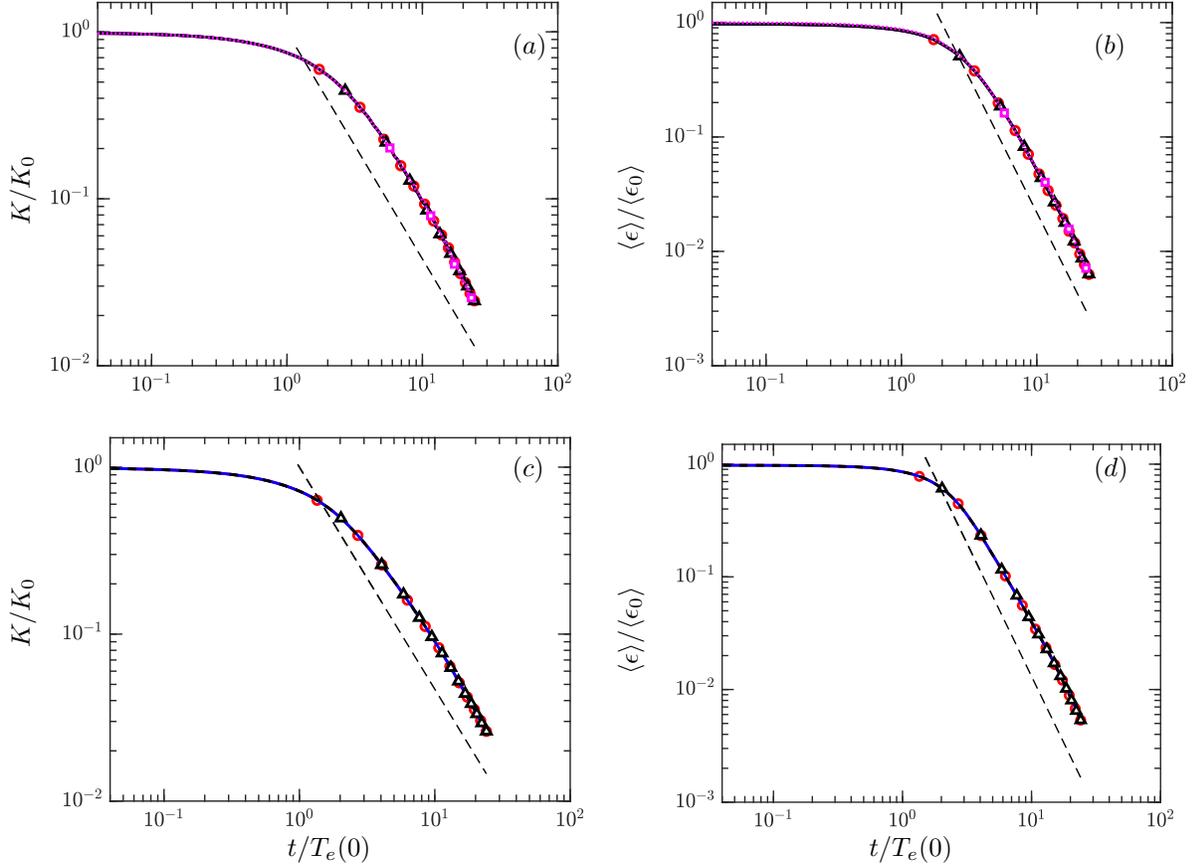

\subsubsection{Low order statistics in spectral space}
Fluid turbulence comprises a wide range of interacting scales \cite{pope2000}.
%Therefore it is important for numerical methods
%to accurately capture this wide range of scales.
%Turbulent flows are characterized by non-linear effects or
%triadic interactions which result in the transfer of energy from large
%scales to small scales or an energy cascade. According to the
The energy distribution across these scales is characterized by the
%An important consequence of this is the energy distribution across
%different scales that can be quantified using the %scale by scale
energy spectrum, which according to Kolmogorov self-similarity hypothesis (K41) \cite{K41}
%the smallest scales are universal at large enough
%Reynolds Number. An important implication of this theory is the scaling
%obtained for the energy spectrum in the inertial range
%According to \cite{K41},
is given by,
\begin{equation}
E(\kp)=C\langle \epsilon \rangle^{2/3}\kp^{-5/3}f(\kp\eta),
\eqnlabel{ener}
\end{equation}
where $C$ is the Kolmogorov constant, $\kappa$ is the wavenumber and
$\eta=(\nu^3/\langle \epsilon \rangle )^{1/4}$ is the Kolmogorov length
scale \cite{K41} and $f(\kp \eta)$ is a universal function. This has been compared
against simulations and experiments extensively and shown to be a good
representation of the spectrum across different flows and Reynolds numbers
for incompressible \cite{Sreeni1997} and compressible flows \cite{DS2013,Kida1990,Kida1992}
at low $M_t$. In the so-called inertial range ($1/\mathcal{L}\ll k\ll 1/\eta$), $f(k\eta)=1$
and the classical $5/3$ scaling for the energy spectrum \cite{Lele1994, Sreeni1997, Ish2009}
can be seen as a flat region in the compensated energy spectrum,
\begin{equation}
\frac{E(\kp)}{\langle \epsilon \rangle^{2/3}\kp^{-5/3}}=C,
\eqnlabel{ener1}
\end{equation}
which becomes wider with an increase in
Reynolds number. The height of this flat region gives the Kolmogorov
constant which has been estimated to be $C=1.6$
from simulations and experiments in incompressible turbulence \cite{Sreeni1995,PK1997,DSN2010}.
This value has been shown to be consistent for compressible simulations \cite{DS2013}.
At high wavenumbers, $f(k\eta)$ is a decaying
exponential %which is independent of Reynolds number
\cite{Kraichnan1959,Kraichnan1967,
SirovichEtAl1994,FoiasEtAl1990}
which may retain a weak Reynolds number effect at very high
wavenumbers \cite{Khurshid2018a}.
%and the energy spectrum is universal \cite{K41, Sreeni1995,PK1997}.
%For simulations in this paper, the turbulent Mach number $M_t=\langle u_i u_i \rangle^{1/2}/c$
%($c$ being the mean speed of sound) which is the measure of compressibility is
%$M_t\approx0.3$. At this $M_t$, the spectrum for compressible simulations
%are shown to be consistent with that obtained for incompressible simulations \cite{DS2013}.

%\cite{Sreeni}.
% is large enough. This

In \rfig{spec}($a,c$) we show the compensated energy spectrum
%is plotted
at $t/T_e(0)\approx1\text{ and }4$ for $Re_{\lambda}(0)\approx 100 \text{ and }145$ for SFD, CAA and SAA
implementations. % along with the tenth order compact scheme with RK3 in time (magenta) in
%\rfig{spec}$(a)$. %For both Reynolds numbers,
%$M_t(0) \approx0.3$ where $M_t=\langle u_i u_i \rangle^{1/2}/c$
%($c$ being the mean speed of sound) is the turbulent Mach number
%which is the measure of compressibility. Since $M_t$ is low,
%the spectrum scales smiliar to incompressible
A plateau in this normalization
corresponding to the inertial range can be seen at short times over a narrow range of scales.
%, spanning across  a few wavenumbers. %This range is clearly visible
%in \rfig{spec} for a small range of wavenumbers due to
%the low Reynolds number.
%However, because of low Reynolds number, this plateau extends to
%a very small range of $k\eta$
%in \rfig{spec}.
Because of the decrease in $Re_{\lambda}$ with time due to the decay,
the inertial range becomes less prominent at later times.
%less prominent.
We also see that the high
wavenumbers are universal as expected from \eqn{ener}.
%the data collapses at both instances for both the
%Reynolds conditions.
%Due to the cascade of energy and the dissipation
%at the smallest scales, $\eta$ increases with time, shifting the
%spectrum to the left.
Both SAA and CAA retain the universality
at small scales and accurately capture the evolution of inertial and large scales. %range and the
%time evolution of the spectrum.
We see a virtually perfect agreement even at the smallest scales (inset in\rfig{spec}($a,c$))
for CAA as well as SAA with SFD. %we do however see some errors for SAA. %These errors are
%considerably small and concentrated in high wavenumber, which in the
%given cases is beyond $k\eta$ of $1.5$.
%Typically, for low
%order statistics, simulations
%with small scale resolution of $k\eta\approx1.5$ are
%found to be well resolved \cite{SD2016}. The energy spectrum obtained here for the asynchronous algorithms
%are in an excellent agreement with the standard synchronous spectrum
%for all wavenumbers with very minor deviations beyond $k\eta \approx 2$.
Moreover, the energy spectrum is also identical to the one obtained with C10
 (magenta line in \rfig{spec}($a$)) from some of the most
well-resolved simulation of compressible turbulence \cite{DS2013,SD2016}.
%we can conclude that the asynchronous algorithms accurately capture the physics of
%the scales of interest. %Furthermore, the small deviations in
%the high wavenumbers can be reduced further by using higher order numerical
%methods, for example, spectra obtained using sixth order schemes is
%shown in \rfig{ord6}(a). Since all numerical methods are
%prone to such aliasing errors at high frequencies, it is not surprising that
%AT schemes have similar tendency.

%Next we look at the fluctuations in the thermodynamic variables such as pressure. % that become more
%prominent as the turbulent Mach number inceases.
Similar to the
energy spectrum, K41 also predicts a scaling in the
inertial range for pressure fluctuations \cite{Monin1975, Gotoh2001} which reads,
\begin{equation}
E_p(\kp)=C_p\langle \epsilon \rangle^{4/3}\kp^{-7/3}.
\eqnlabel{pres}
\end{equation}
The inertial range can be identified as the plateau in the
compensated pressure spectrum plot, if $Re_{\lambda}$ is high enough.
Since $M_t\approx0.3$ for our simulation is
fairly low, the pressure
spectrum should be similar to the incompressible spectrum
\cite{DS2013, Gotoh2001}. This is indeed observed
in \rfig{spec}($b,d$) for $Re_{\lambda}(0)\approx100 \text{ and } 145$ at
 $t/T_e(0)\approx 1 \text{ and }4$ for the universal part of the spectrum.
A horizontal dashed line at
$C_p=8$ is also included for reference obtained from incompressible flows \cite{Gotoh2001}.
These spectra are consistent with those in the literature at similar
conditions \cite{DS2013} with a collapse at the high wave-numbers
similar to the energy spectrum.
The data for CAA and SAA agree closely with that for SFD at both times
for both $Re_{\lambda}$. However, for $k\eta \ge1.5$, SAA spectrum has a
small pileup at the high wavenumbers.
This difference in the spectrum for CAA and SAA can be attributed to the
difference in the nature of delays which for the former is deterministic and random for
the latter.
%For the CAA the delays are deterministic %and identical on both
%PE boundaries in each direction, the conservative property of standard schemes is
%retained. % and the instantaneous average dilatation
%$(\langle u_{i,i}\rangle)$ is very small and comparable to SFD.
%but for the SAA the delays are random and
%different AT scheme is used at each PE boundary. %, the conservative property
%is not exactly satisifed resulting in an instantaneous density change of $<\pm0.003\%$.
The randomness associated with SAA can lead to
numerical errors that are absent in CAA
and can cause a small pileup of energy at the high wavenumbers
as seen in the pressure spectrum in \rfig{spec}$(b)$ for SAA. %, this effect is negligible
%in the pressure spectrum in \rfig{spec}$(d)$
%affects the
%instantaneous average dilatation which, albeit small, is relatively larger than
%SFD and results
The differences in \rfig{spec}$(b)$ are magnified because of
the prefactor $k^{7/3}$ but they are concentrated only in a few wavenumbers
and represent an extremely small contribution to \textit{e.g.}, pressure variance.
We have also performed simulations at higher $M_t\ap 0.6$ and found
that this small pileup disappears. Thus, this seems to be a low-$M_t$ effect which
can be explained by noting that as $M_t$ increases, there is stronger interaction
between the so-called solenoidal and dilatational velocity components
\cite{1907.07871}
which can help mix these already small perturbations at PE boundaries for SAA.
For CAA, no pileup is observed at any $M_t$.

%\colb{+++++++++++++++++++++++++++++++++++++++}

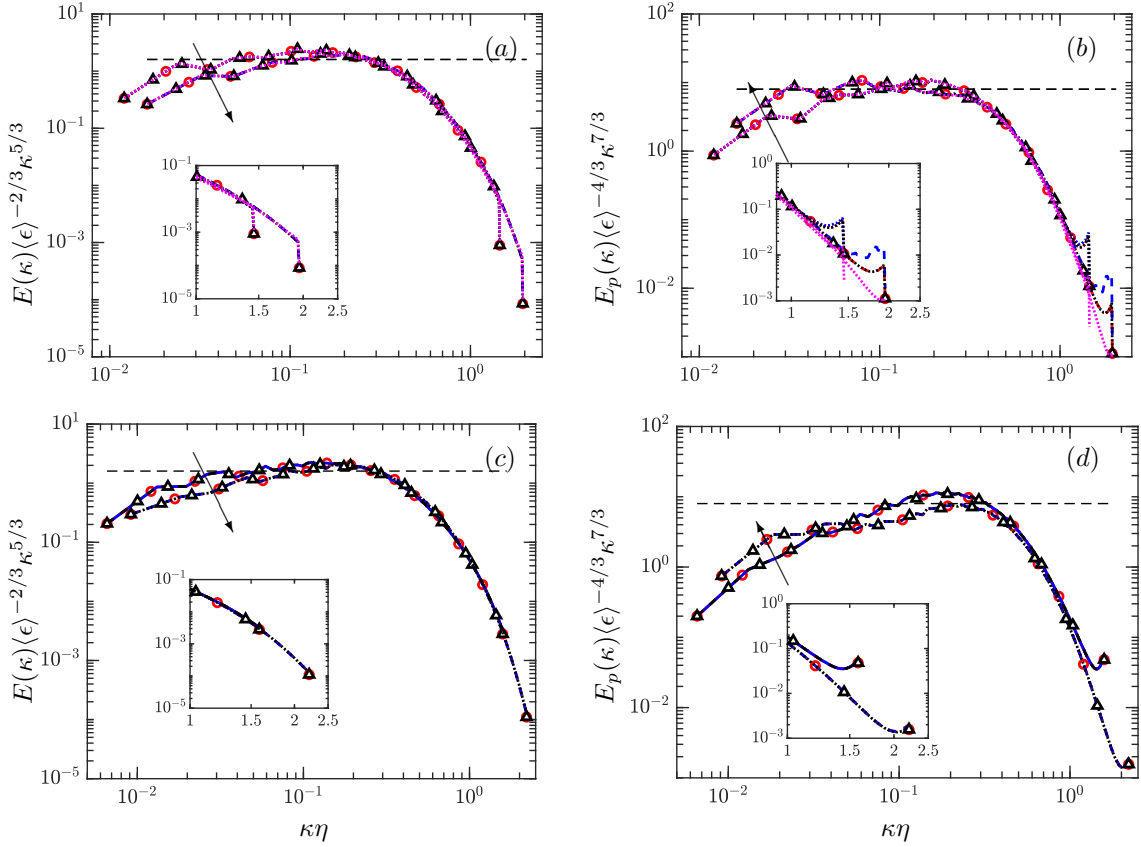
\begin{figure}[h]
\centering%center}
\subfigure{\includegraphics[width=0.4\textwidth]{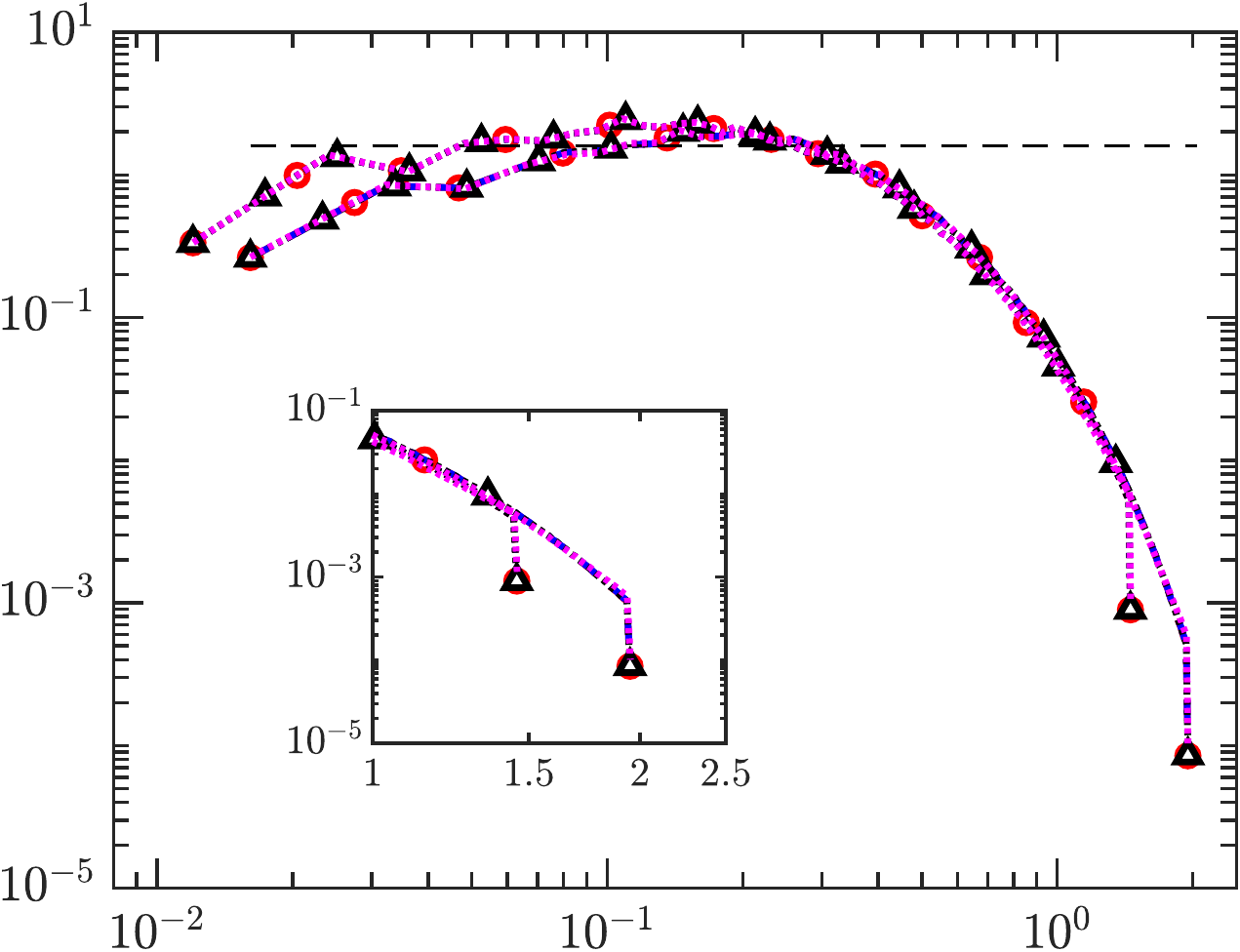}}
\hspace{1cm}
\subfigure{\includegraphics[width=0.4\textwidth]{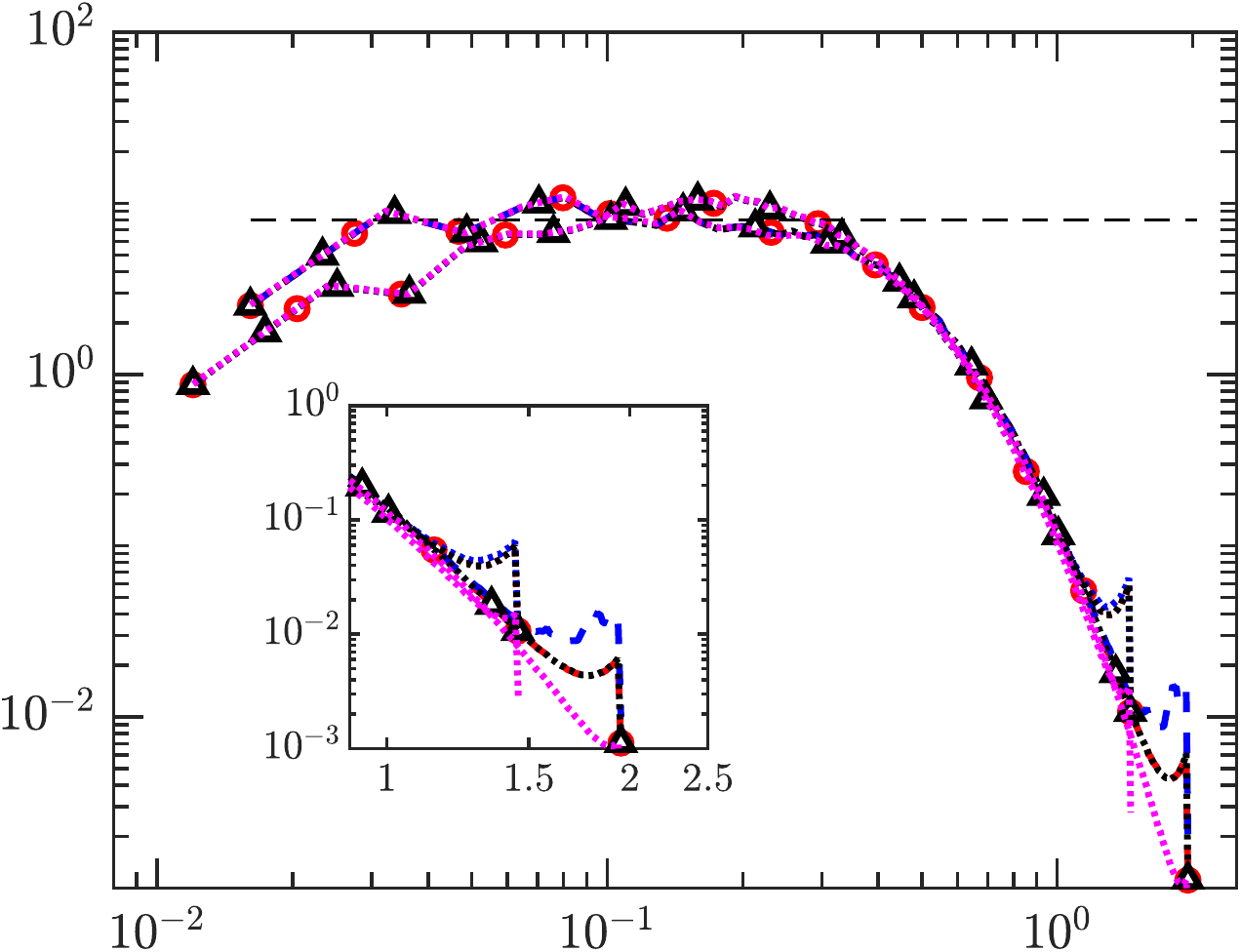}}

\subfigure{\includegraphics[width=0.4\textwidth]{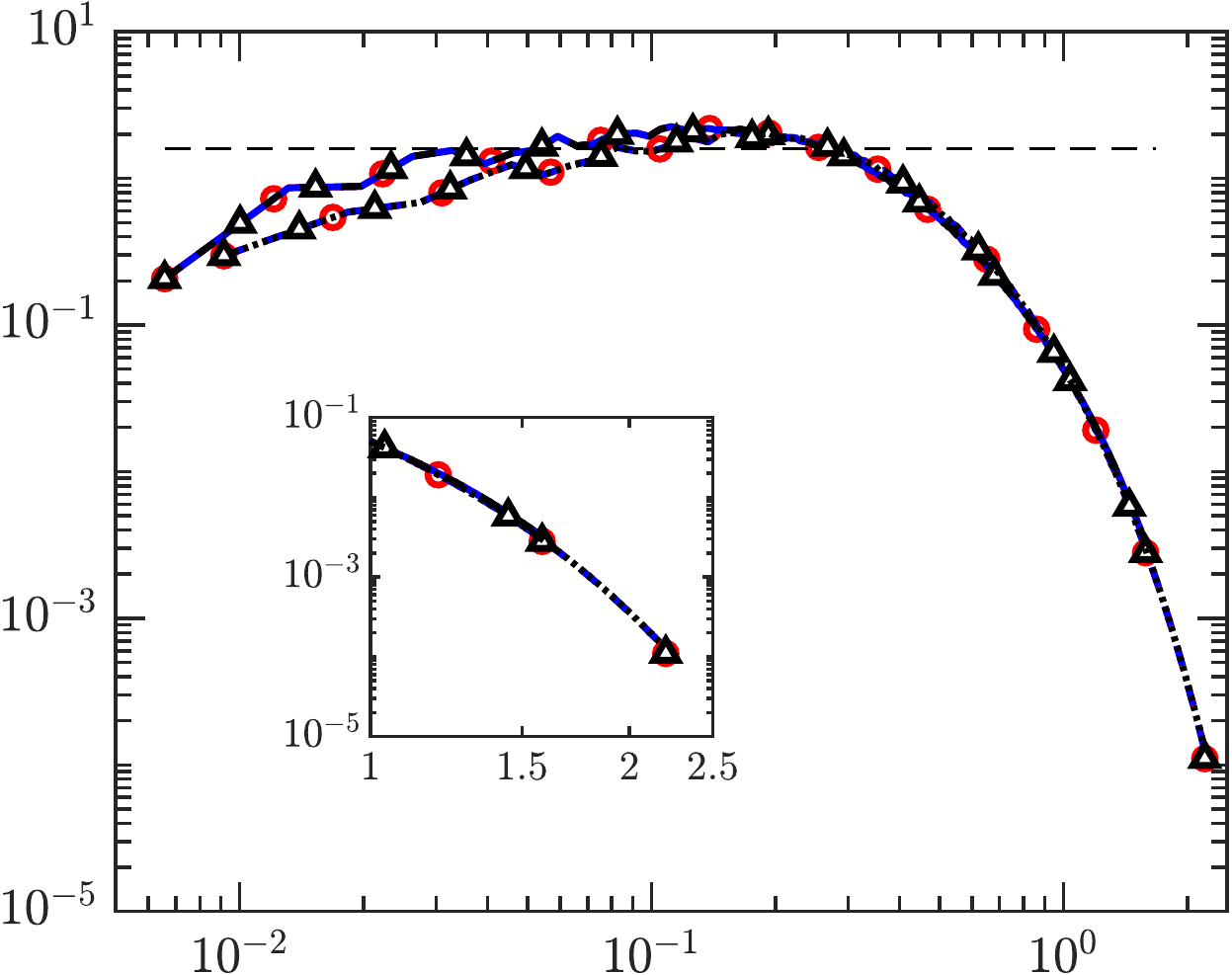}}
\hspace{1cm}
\subfigure{\includegraphics[width=0.41\textwidth]{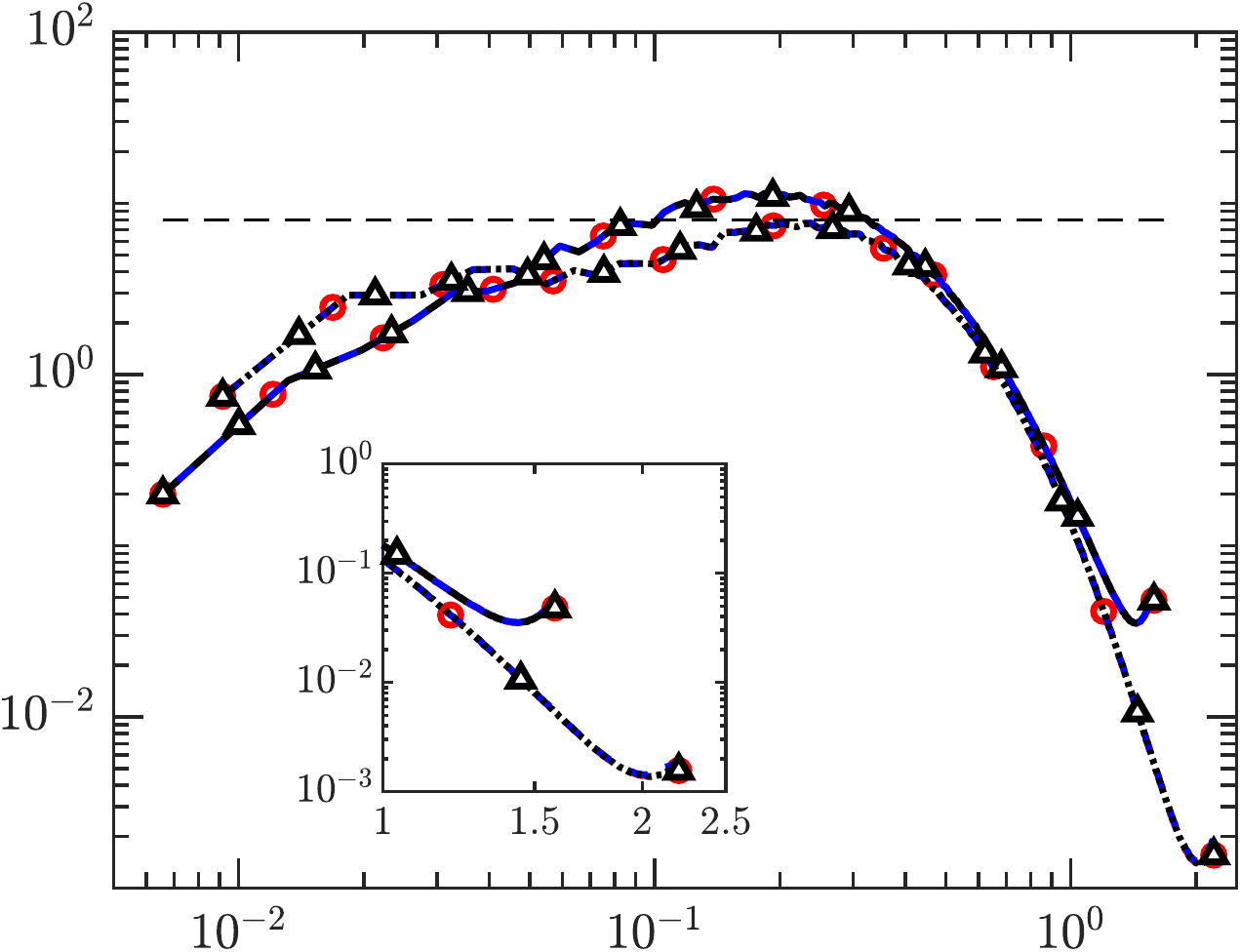}}

\begin{picture}(0,0)
\put(-40,295){$(a)$}
\put(180,295){$(b)$}
\put(-40,140){$(c)$}
\put(180,140){$(d)$}

\put(-150,300){\vector(1,-2){15}}
\put(75,255){\vector(-1,2){15}}
\put(-220,200){\rotatebox{90}{$E(\kp)\langle \epsilon\rangle^{-2/3}\kp^{5/3}$}}
\put(0,200){\rotatebox{90}{$E_p(\kp)\langle \epsilon\rangle^{-4/3}\kp^{7/3}$}}
\put(-110,0){$\kp\eta$}
\put(110,0){$\kp\eta$}

\put(-150,145){\vector(1,-2){15}}
\put(75,95){\vector(-1,2){12}}
\put(-220,50){\rotatebox{90}{$E(\kp)\langle \epsilon\rangle^{-2/3}\kp^{5/3}$}}
\put(0,50){\rotatebox{90}{$E_p(\kp)\langle \epsilon\rangle^{-4/3}\kp^{7/3}$}}

\end{picture}

\caption{Compensated energy spectrum (left) and
compensated pressure spectrum (right) for $Re_{\lambda}(0)\approx100$ ($a,b$) and
$Re_{\lambda}(0)\approx145$ ($c,d$) at $t/T_e(0)\approx 1$ and $4$.
Different lines are: SFD (red-circle), CAA (black-triangle) and
SAA (blue) with $L=3$. The arrow denotes increasing time. Magenta
line in ($a,b$) is C10.
}
\figlabel{spec}
%\end{center}
\end{figure}

A general conclusion one can draw from both
energy and pressure spectrum plots, is that the dynamics
of the flow at the scales of interest is accurately captured despite asynchrony even though
there are some very small deviations at the high wavenumbers in the pressure spectrum for SAA.
%While jumps in the gradients introduced at the PE boundaries
%due to asynchrony introduce pressure fluctuations, visualization shows that
%these are local and short lived and consequently the overally dynamics, as
%seen from the the energy spectrum, is not affected.
Furthermore, we see from \rfig{spec}$(b)$ that the pressure
spectrum for SFD itself is not identical to the spectrum obtained for C10
at higher wavenumbers. Thus, it is not
unexpected that asynchronous schemes present a different behavior at high
wavenumbers. The errors in SAA, though already very small,
can be mitigated if
higher order schemes or higher resolution is used.
As an example, in \rfig{ord6}(b), the compensated
pressure spectrum is shown for $Re_{\lambda}(0)\approx100$ at
$t/T_e(0)\approx4$ using fourth and sixth
order AT scheme
(included in the appendix)
for SAA. While the SAA with fourth-order AT scheme (solid blue) peels off at $\kp\eta \approx1.5$,
SAA with sixth-order AT scheme (faded-blue square), follows the SFD
spectrum till the highest $\kp\eta$.
\begin{figure}[h!]
\centering%center}
\subfigure{\includegraphics[width=0.4\textwidth]{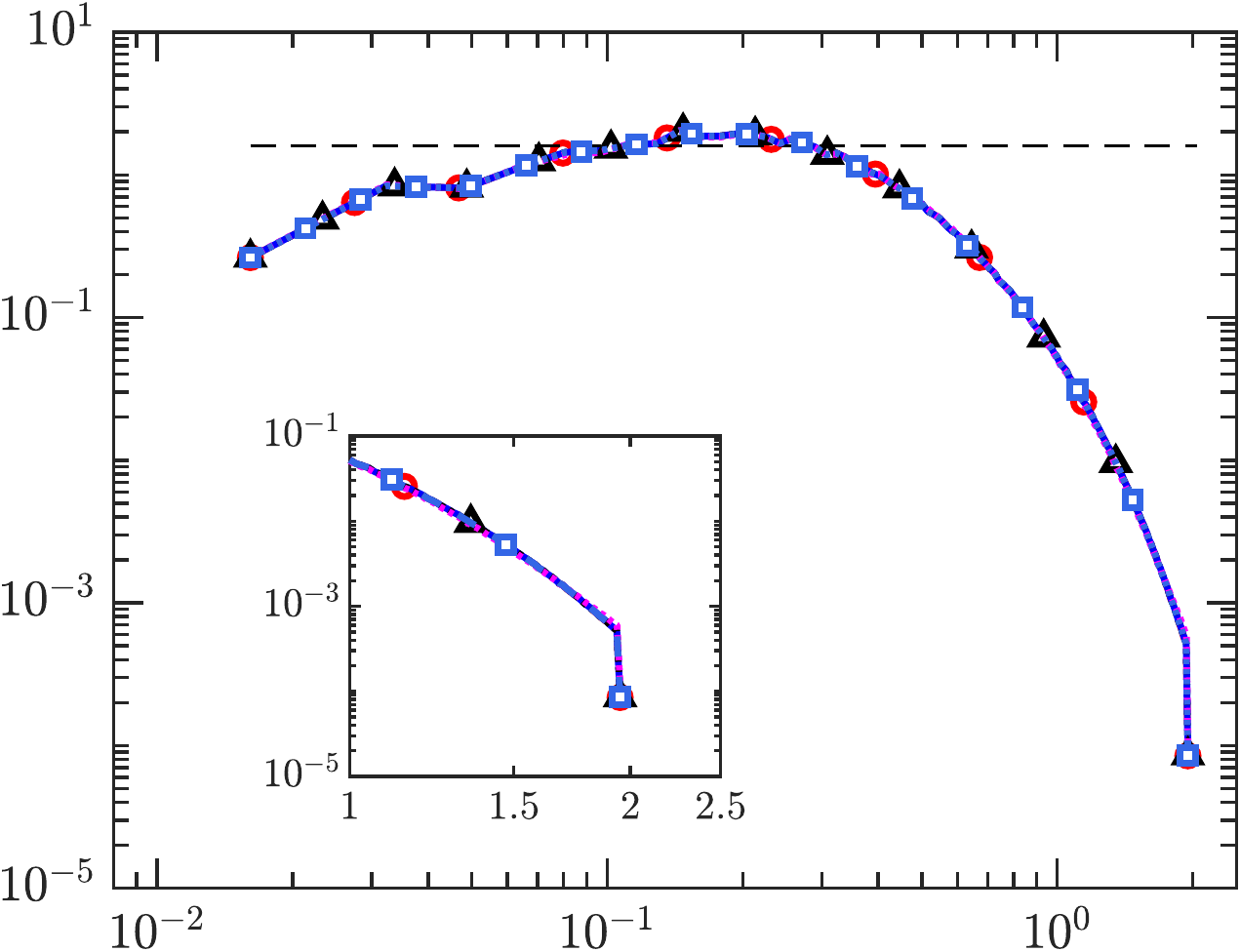}}
\hspace{1cm}
\subfigure{\includegraphics[width=0.41\textwidth]{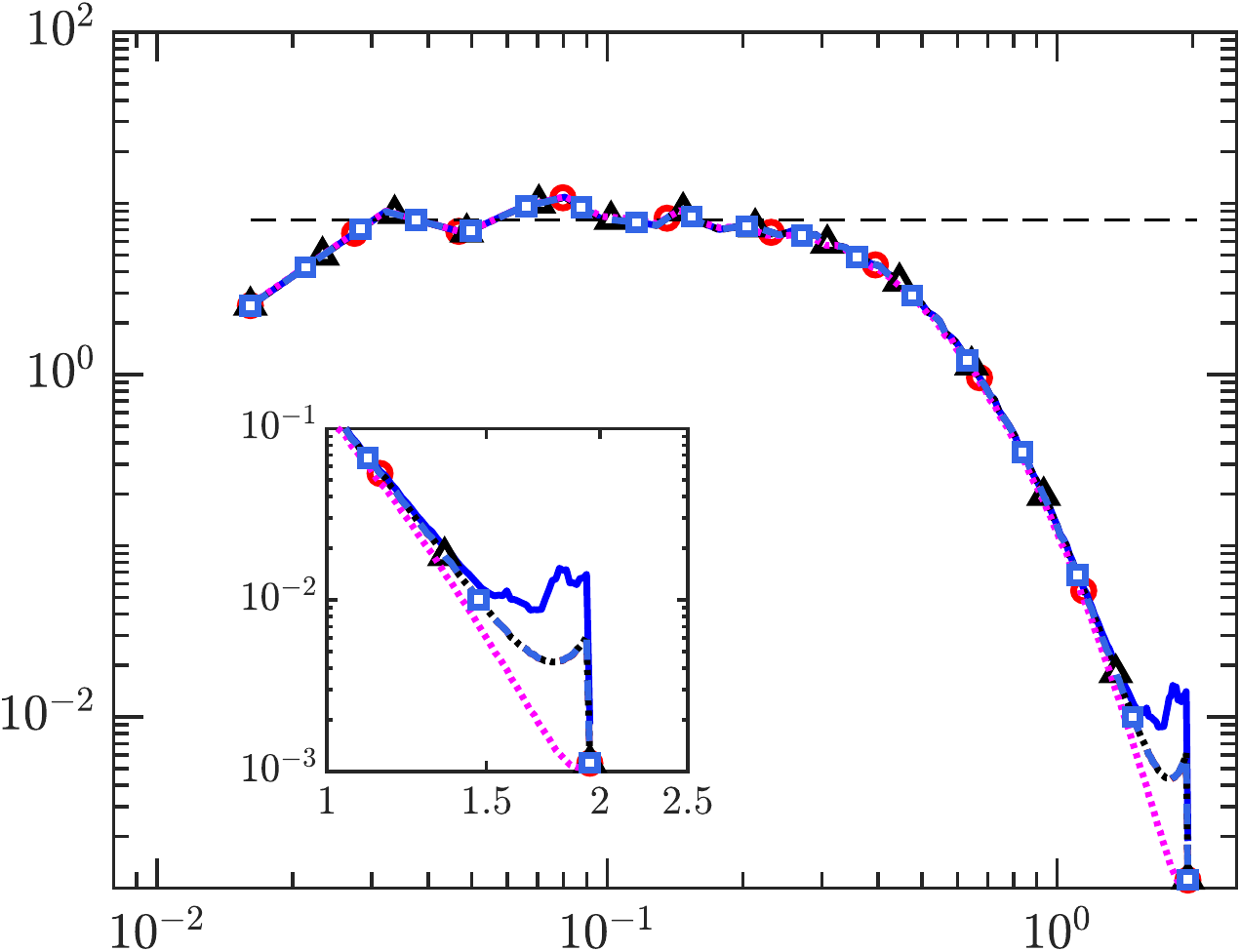}}

\begin{picture}(0,0)
\put(-40,135){$(a)$}
\put(180,135){$(b)$}
\put(-220,50){\rotatebox{90}{$E(\kp)\langle \epsilon\rangle^{-2/3}\kp^{5/3}$}}
\put(0,50){\rotatebox{90}{$E_p(\kp)\langle \epsilon\rangle^{-4/3}\kp^{7/3}$}}
\put(-110,0){$k\eta$}
\put(110,0){$k\eta$}
\end{picture}

\caption{(a) Compensated energy spectrum and
(b) compensated pressure spectrum for $Re_{\lambda}(0)\approx100$
at $t/T_e(0)\approx4$. Faded dashed-blue line with squares is
 the sixth-order asynchronous scheme with random delays and solid blue line is
fourth-order AT scheme with random delays. Rest of the lines are
same as in \rfig{spec}. Insets zoom in on high wavenumbers.
}
\figlabel{ord6}
%\end{center}
\end{figure}
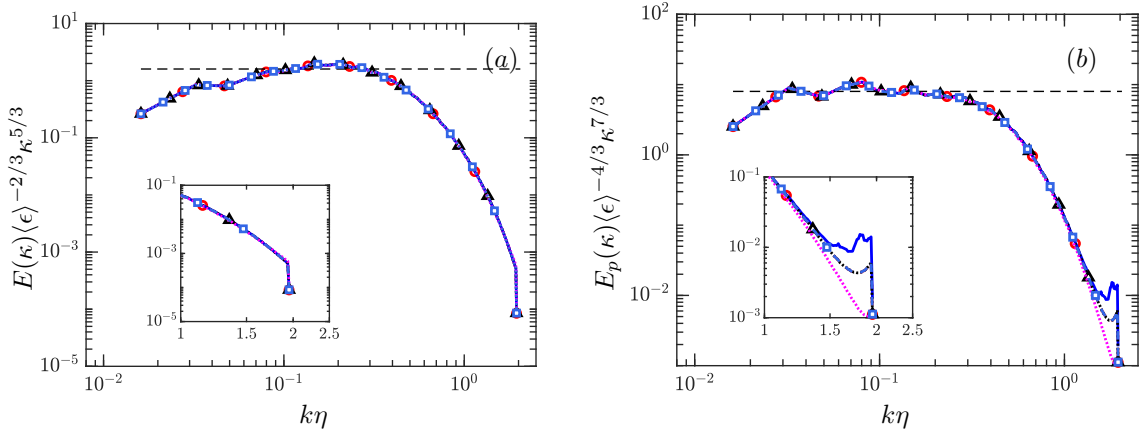

\subsubsection{Statistics of velocity gradients}
%With low order small and large scale statistics in physical
%and spectral space captured with excellent accuracy by asynchronous algorithms,
%we now shift our focus to higher order statistics. %Odd-order moments, for example,
%are sensitive to resolution \cite{DPK2008} and susceptible to numerical
%errors.
An important feature of 3D turbulence is the generation of
vortical motions, often quantified with the so-called
enstrophy $(\Om=\la \omega_i\omega_i \ra$, where
$\boldsymbol{\omega} =\nabla \times \boldsymbol{u}$ is the vorticity vector).
A normalized metric for the production of enstrophy,
which is also representative of the non-linear transfer of energy from large scales to small
scales, is the skewness
of the longitudinal velocity gradient,
%($\partial u_1/\partial x_1$),
$S=\langle (\partial u_1/\partial x_1)^3\rangle /(\langle (\partial u_1/\partial x_1)^2\rangle)^{3/2}$
\cite{Monin1975,Sreeni1997,Davidson2015}.
The negative of the skewness $(-S)$ %asymptotes
is constant at about $\sim 0.5$ as long as the Reynolds number is
not
too low. This has been extensively documented in experiments and numerical simulations
\cite{Kerr1985, Sreeni1997, Gotoh2002, Ish2009}. In \rfig{skew}($a,c$) we show the
time evolution of $-S$ for initial $Re_{\lambda}$ of $100$ and $145$, respectively.
We see that $-S$ is close to $0.5$ and this is consistent
for SFD, CAA and SAA, with some small differences at later times. Despite
odd-order moments being more sensitive to resolution \cite{DPK2008} and susceptible to numerical
errors, we see that the asynchronous algorithms capture skewness
very well and
close to the skewness computed using C10.

%Higher order moments of the velocity gradients have also
%been used to quantify the phenomena of intermittency
Another intrinsic characteristic of turbulent flows is
the phenomena of intermittency
which is a tendency to have localized events of fluctuations that are orders of magnitude
larger than the mean \cite{Kraichnan1967, Sreeni1997,
Davidson2015,DS2013a,PorterEtAl1998,PanEtAl2009}. These events add to the complexity
of the turbulent flows, specifically at the smallest scales.
One way to quantify this phenomena is through the moments of velocity
gradients as most of their contribution stems from the small scales
and it is thus an excellent quantity to check small scale resolution. %measure the frequency of intensity of these extreme events
%is the PDF of velocity gradients,
These moments transition from  %We compute the PDF of the
%longitudinal velocity gradient $(\partial u_1/\partial x_1)$.
Gaussian to anomalous as Reynolds number increases \cite{YakhotDonzis2018,YakhotDonzis2017, Schumacher2014}. %The
%velocity gradient PDF is then characterized by tails that are
% longer or fatter than the Gaussian PDF. % as a
%consequence of these intense fluctuations.
%This departure from the symmetery in Gaussian statistics can be quantified more
%concretely by looking at the normalized third order moment or the skewness
% $(S=\langle (\partial u_1/\partial x_1)^3\rangle /(\langle (\partial u_1/\partial x_1)^2\rangle)^{3/2})$
%of the PDF. The negative of skewness $(-S)$ of the longitudinal velocity gradient
%tends to $\sim 0.5$ as Reynolds number increases and this
%has been extensively observed in experiments and numerical simulations
%\cite{Kerr1985, Sreeni1997, Gotoh2002, Ish2009}. In \rfig{skew}($a,c$) we show the
%time evolution of $-S$ for $Re_{\lambda}(0)\approx 100$ and $145$, respectively.
%We see that $-S$ is close to $0.5$ and this is consistent
%for SFD, CAA and SAA, with some deviations at later times. Despite
%odd-order moments being more sensitive to resolution \cite{DPK2008} and susceptible to numerical
%errors, we see that the asynchronous algorithms capture skewness reasonably well and
%close to the skewness computed by C10.
%Another quantity of interest, widely used to quantify the extreme events in the
In \rfig{skew}($b,d$) we show the normalized fourth-order moment or flatness
($ F=\langle (\partial u_1/\partial x_1)^4\rangle /(\langle (\partial u_1
\allowbreak/\partial x_1)^2\rangle)^{2}$) of the longitudinal velocity gradient. %The flatness
%of the longitudinal velocity gradients is shown in
% \rfig{skew}($b,d$) for both the Reynolds numbers.
The flatness is close to $6$ \cite{Kerr1985, Sreeni1997} at initial times
and tends to decrease because of decrease in Reynolds
number for decaying turbulence. We see an excellent agreement between
synchronous and both the asynchronous simulations with no observable differences
from C10. Even though the computation of the gradient $\partial u_1/\partial x_1$
is directly affected by asynchrony, the higher order moments of the same
exhibit trends similar to SFD and C10.

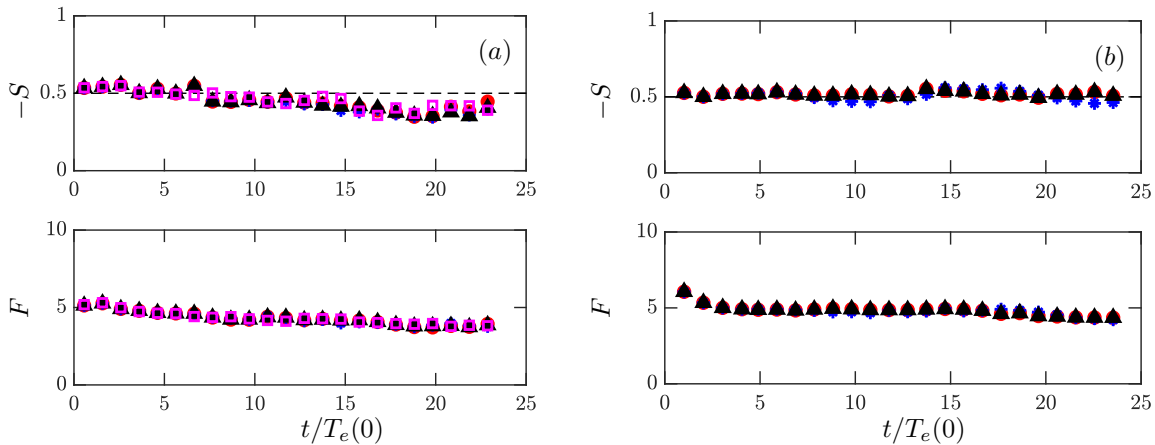
\begin{figure}[h!]
\centering%center}
\subfigure{\includegraphics[width=0.4\textwidth]{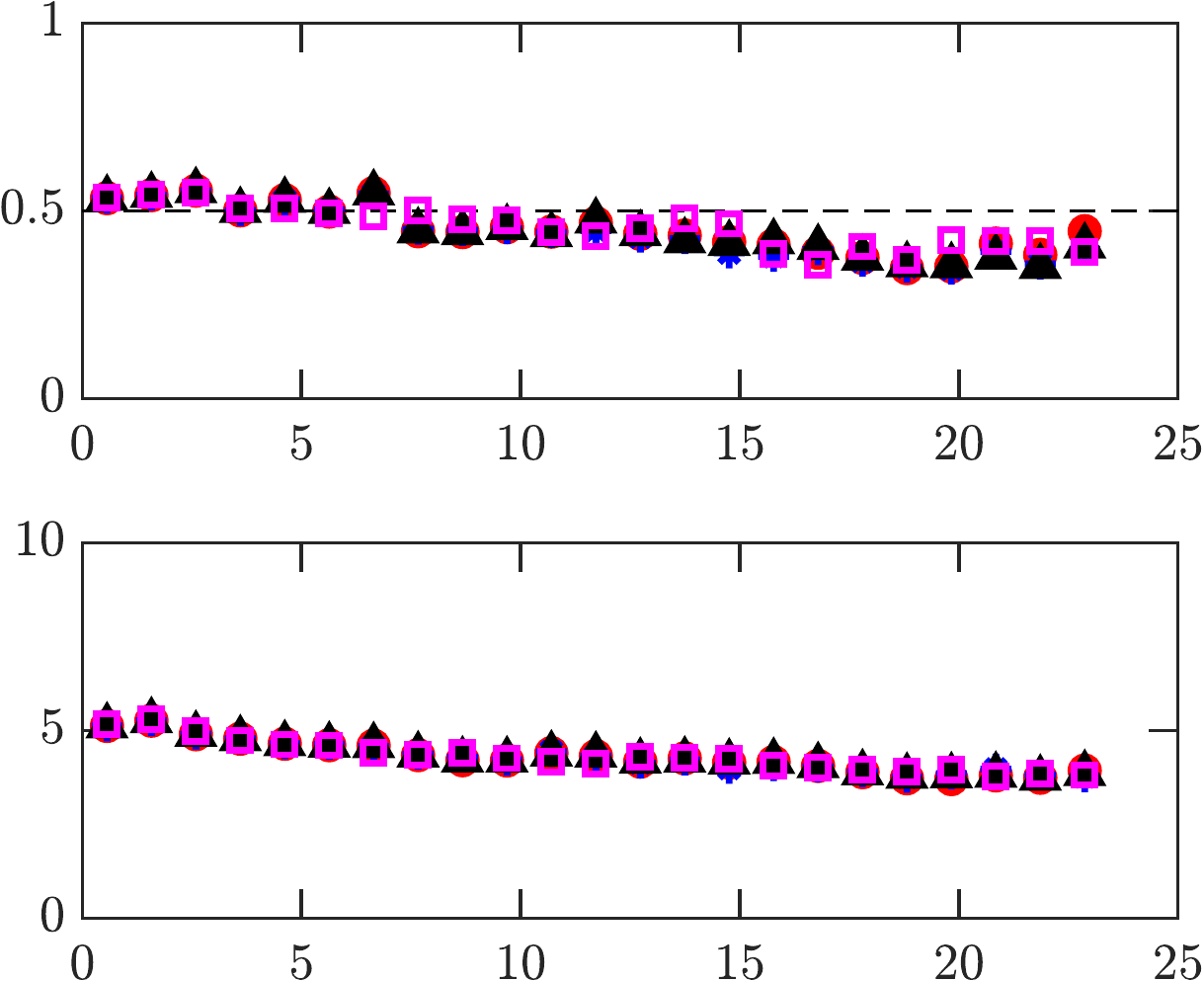}}
\hspace{1cm}
\subfigure{\includegraphics[width=0.42\textwidth]{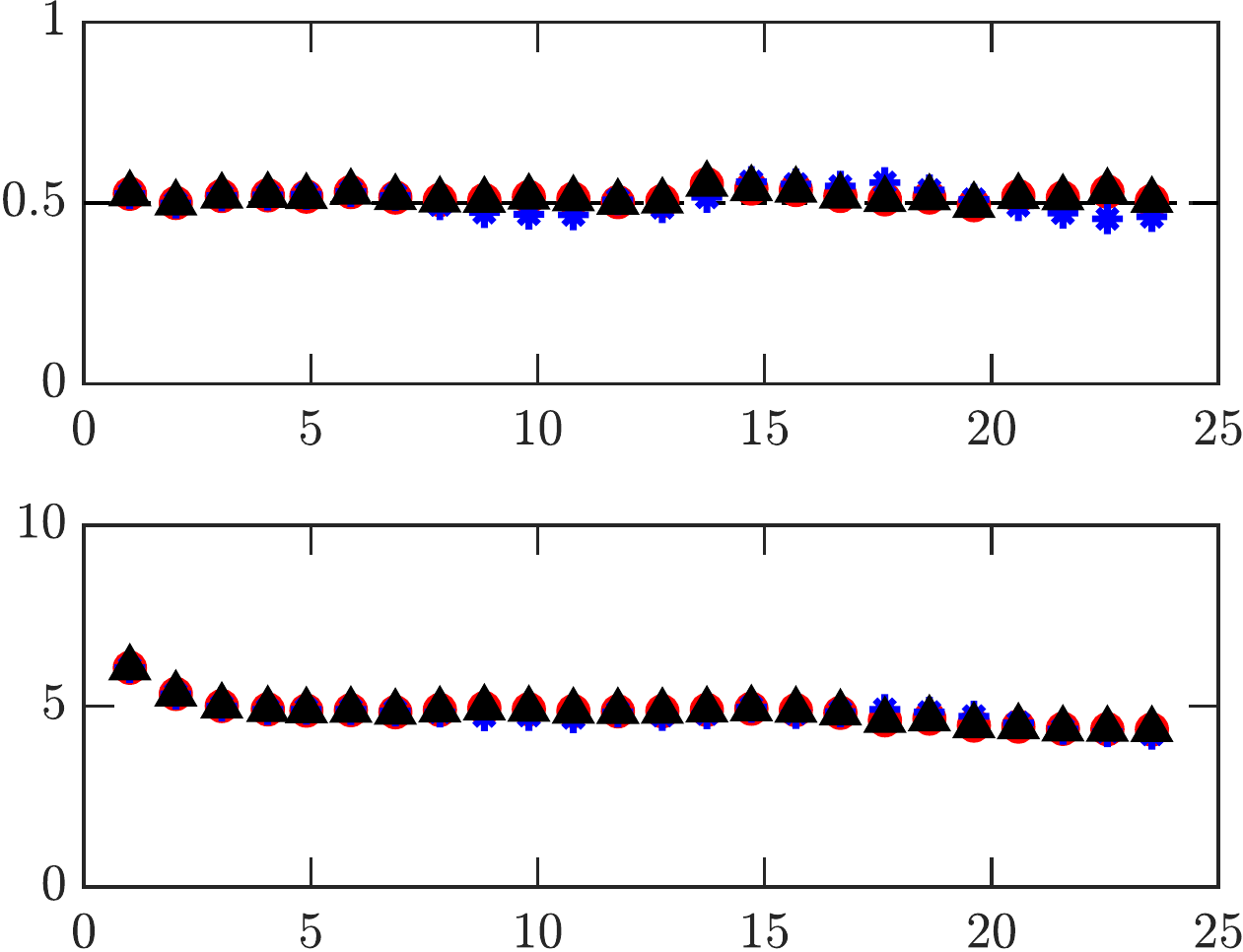}}
\begin{picture}(0,0)
\put(-25,131){$(b)$}
\put(-215,110){\rotatebox{90}{$-S$}}
\put(-215,35){\rotatebox{90}{$F$}}
\put(-105,-10){$t/T_e(0)$}

\put(-258,133){$(a)$}
\put(-435,110){\rotatebox{90}{$-S$}}
\put(-435,35){\rotatebox{90}{$F$}}
\put(-325,-10){$t/T_e(0)$}

\end{picture}

\caption{Negative of skewness (top row) and flatness (bottom row)
of the longitudinal velocity gradient vs.
normalized time for (a) $Re_{\lambda}(0)\approx100$ and
 (b) $Re_{\lambda}(0)\approx145$.
Different symbols are: SFD (red-circle), CAA (black-triangle) and SAA
(blue asterik) with $L=3$. The dashed black line indicates skewness of $0.5$ and
magenta squares in ($a,c$)are C10.
}
\figlabel{skew}
%\end{center}
\end{figure}

\subsubsection{Instantaneous enstrophy field}
The average quantities discussed in the sections
show good agreement between the asynchronous and synchronous simulations.
A stricter test of accuracy would comprise the
instantaneous flow fields which can potentially show some differences
because of different truncation errors for different schemes
in the computation of derivatives
at the boundaries.
As argued above, enstrophy is known to be sensitive to small scale resolution
%and velocity gradients is enstrophy, %($\Om=\omega_j\omega_j$,
%where $\boldsymbol{\omega}$ is the vorticity vector),
and is highly intermittent \cite{DPK2008,PKKRS2015}
and thus provides a stringent test of the numerical performance of schemes.
%For this we will look at the contour
%of a highly intermittent quantity, enstrophy $(\Omega=\omega_{j}\omega_{j})$ where
%$\boldsymbol{\omega}$ is the vorticity vector.
In \rfig{ens} we show the contours of the enstrophy normalized
by its mean ($\Om/\langle \Omega \rangle$) in the
$yz$ plane at $x=\pi$. Qualitatively, all
the large and small structures look identical for SFD, CAA and SAA.
In particular, a concern with asynchronous schemes is the behavior
close to the processor boundaries. If we closely look along these PE boundaries
(faded lines in \rfig{ens}) there are no perceptible
differences between enstrophy contours for SAA, CAA and SFD.
Moreover, even complex structures spanning across multiple PE boundaries,
for example, inside black circle in \rfig{ens}, is consistent for all the three cases.
%While there are practically no differences between the CAA and
%SFD, there are some differences, although small, for SAA. One such
%difference can be seen inside the blue circle in \rfig{ens}.
%Here the intensity of the extreme event is slightly larger for
%SAA as compared SFD.
Besides some very small but not apparent
localized differences in the intensity of enstrophy for SAA,
the asynchronous algorithms accurately resolve the highly intermittent
instantaneous enstrophy field. The instantaneous dissipation field (not shown here)
 exhibits similar behavior and is captured accurately.

\begin{figure}[h!]
\centering%centeri}
%\begin{tikzpicture}
\subfigure{\includegraphics[trim={5cm 1.1cm 8.9cm 2.2cm},clip,width=0.41\linewidth]{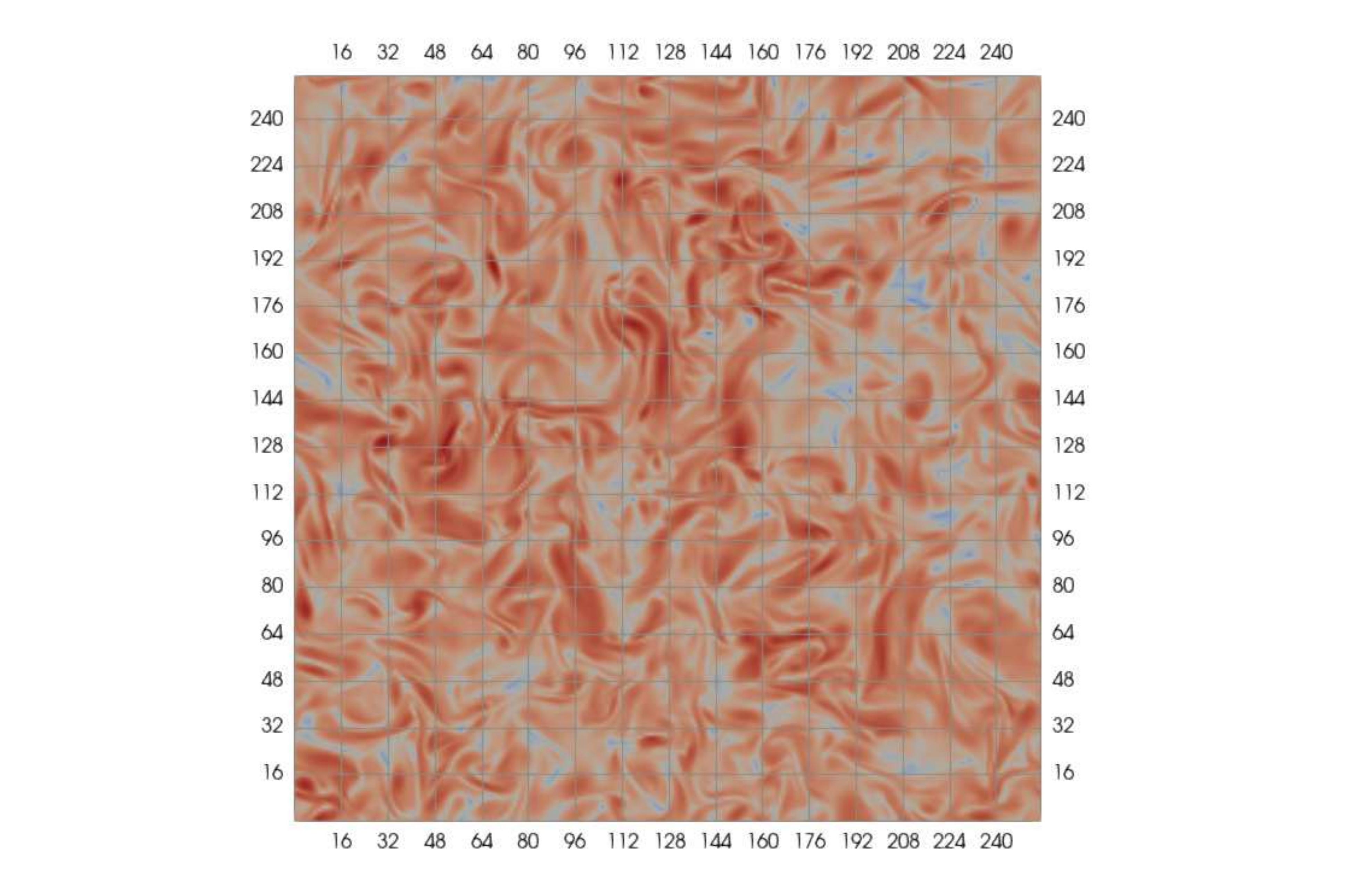}}
\subfigure{\includegraphics[trim={5cm 1.5cm 8.9cm 2.2cm},clip, width=0.423\linewidth]{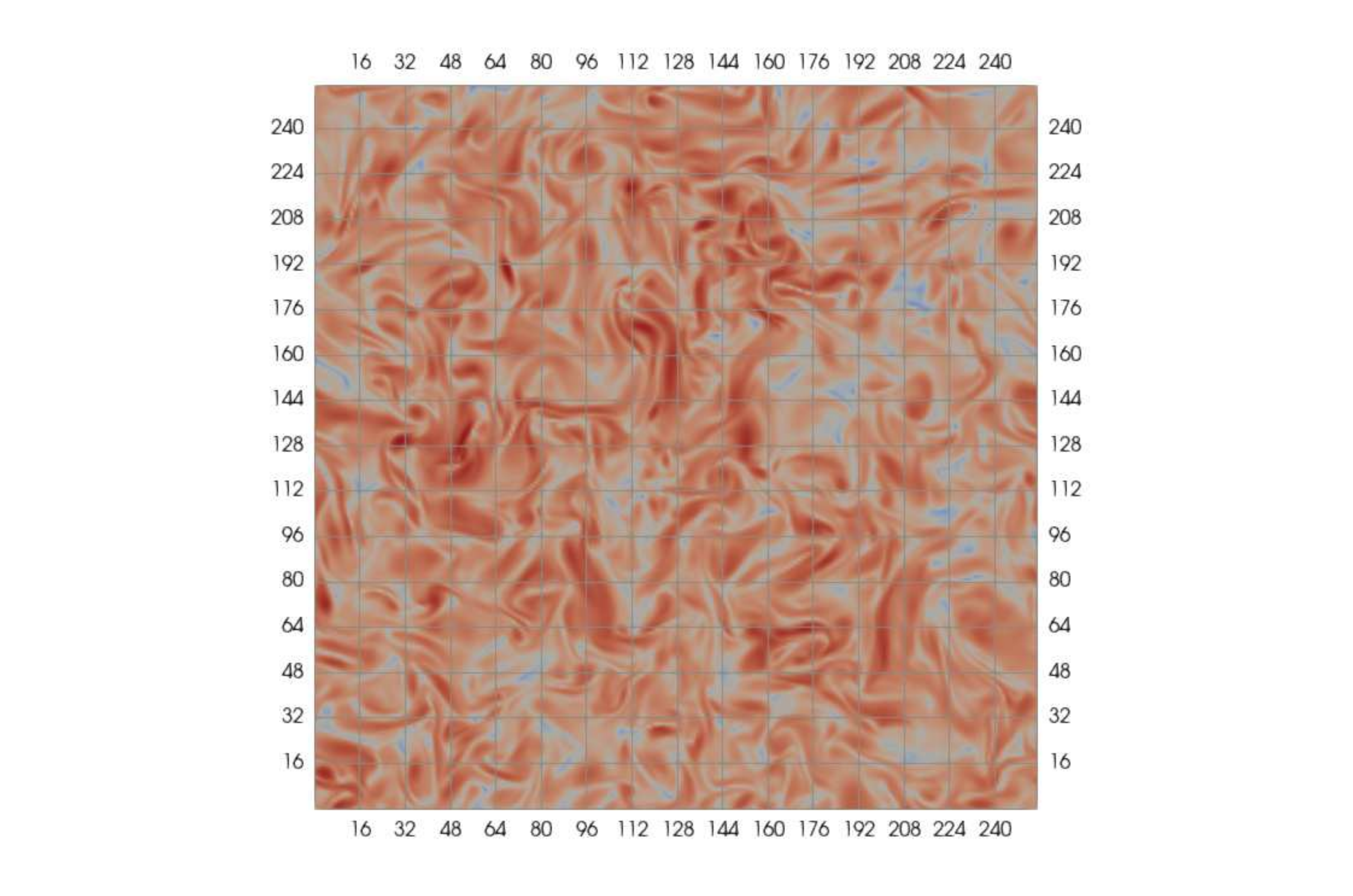}}

\subfigure{\includegraphics[trim={1cm 1.5cm 8.9cm 2.2cm},clip,width=0.5\linewidth]{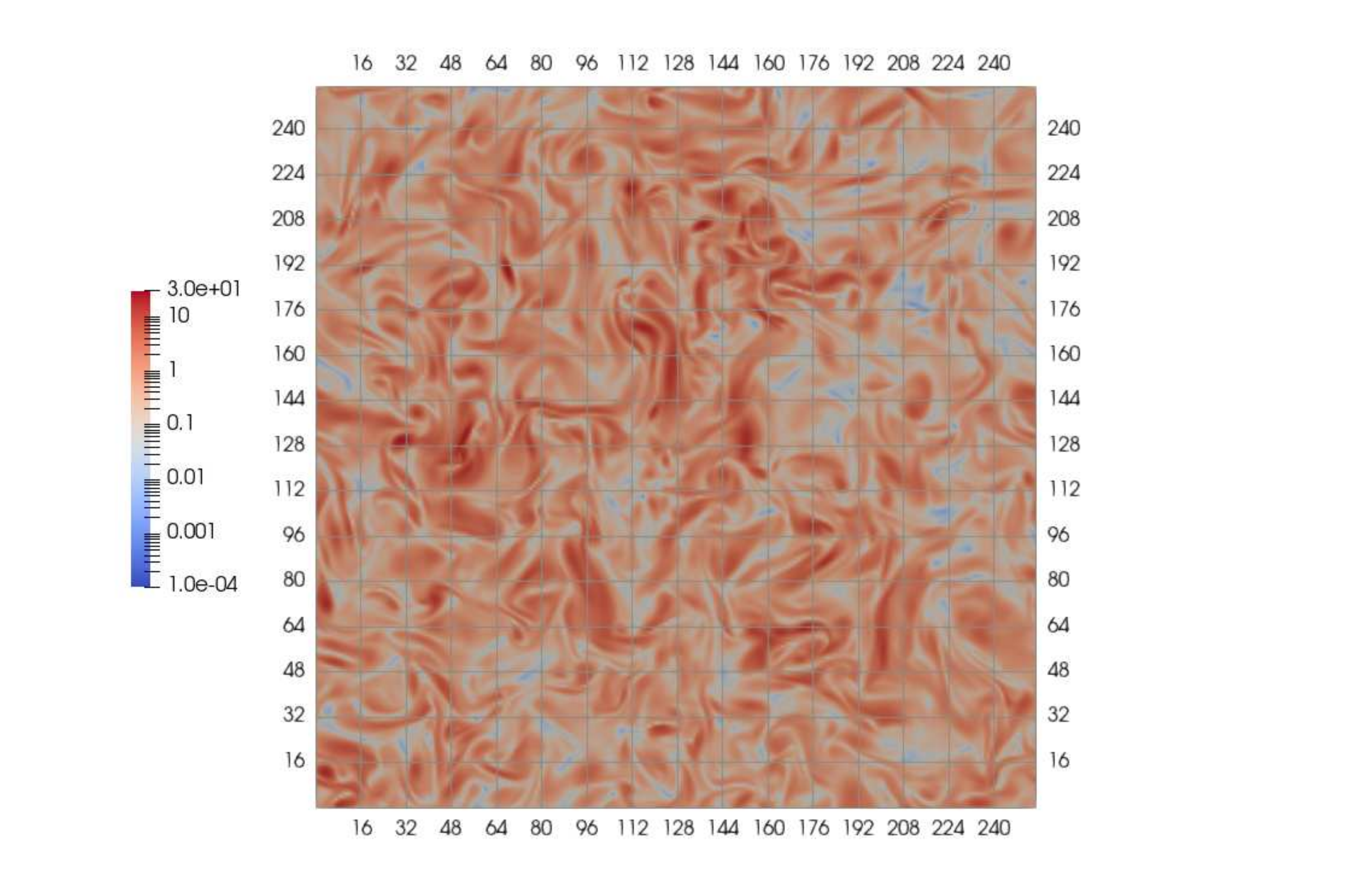}}
\setlength{\unitlength}{1cm}
\begin{picture}(0,0)
%\put(-2.6,3.3){\color{blue}\circle{0.8}}
%\put(-6.65,9.8){\color{blue}\circle{0.8}}
%\put(0.45,9.8){\color{blue}\circle{0.8}}
\put(-2.7,4.8){\circle{1.5}}
\put(-6.9,11.3){\circle{1.5}}
\put(0.2,11.3){\circle{1.5}}
\end{picture}
\caption{Normalized instantaneous enstrophy $(\Omega/\langle
\Omega\rangle)$ field at $t/T_e(0)\approx 4$ for (a) SFD
(b) CAA ($L=3$) and
(c) SAA ($L=3$) in the $yz$ plane at $x=\pi$ for $Re_{\lambda}(0)\approx100$.
The faded lines represent processor boundaries}
\figlabel{ens}
%\end{center}
\end{figure}
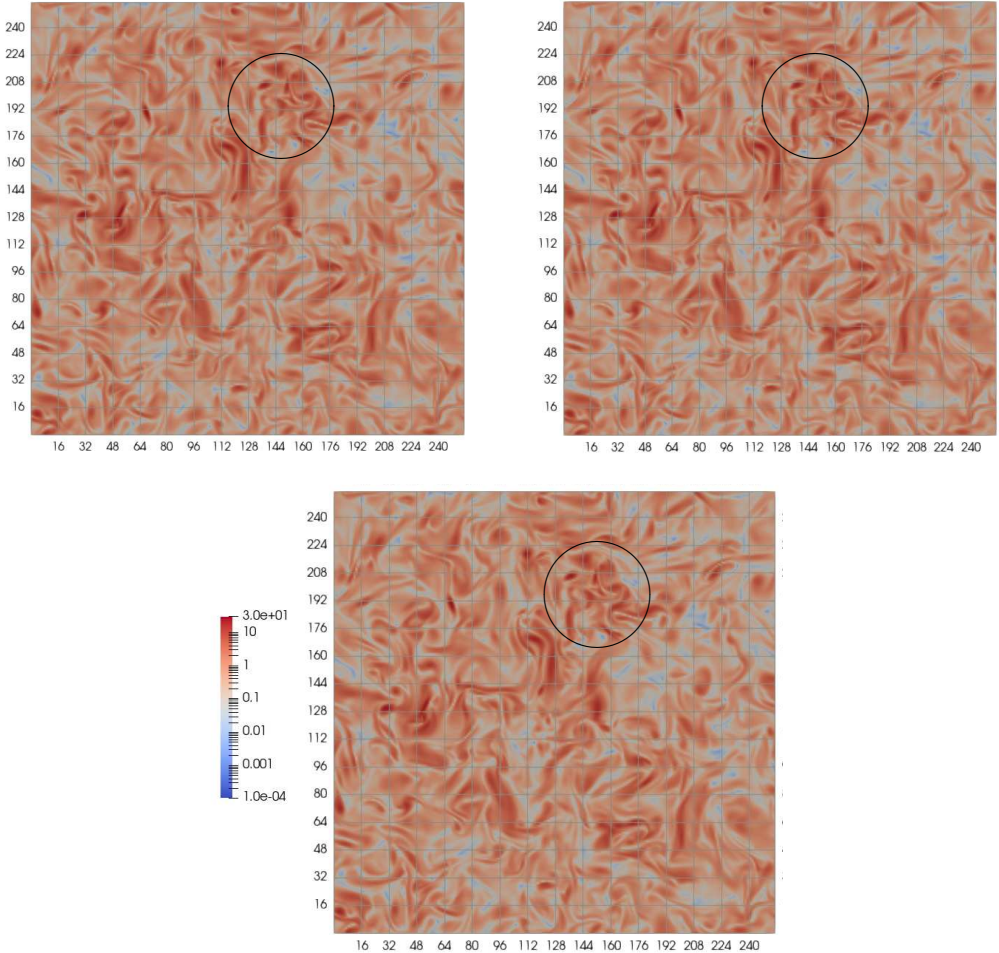

\subsection{Forced Turbulence}
In the preceding section we focused on the DNS of decay
of stationary state initial velocity field
and observed a close agreement between the synchronous and
asynchronous numerical simulations.
In this section we discuss the effect of asynchrony on
forced turbulence. Here, energy is injected at the
large scales, or wavenumbers ($\kappa$) in a spherical shell of
radius $\kappa_f$, where
$\kappa \le \kappa_f,~(\kappa_f=3)$, through the term $f$
in the momentum equation (\eqn{mome}).
The details of the stochastic forcing implemented can be found in
\cite{Pope1988} and has been extensively used in
\cite{PetersenLivescu2010,DS2013,SD2016,DonzisMaqui2016a} for compressible
turbulence. Through the non-linear interactions this injected
energy cascades down to the inertial and small scales, where it
is dissipated into internal energy by the action of viscosity.
One can derive the evolution equation of the mean
turbulent kinetic energy ($K$) by multiplying \eqn{mome} by $u_i$
and taking the mean, which reads as
%performing Reynolds decomposition and taking the mean,
\begin{equation}
\frac{dK}{dt}=\langle p'\theta' \rangle-\langle \epsilon \rangle +
\langle f_iu_i\rangle
\eqnlabel{meantke}
\end{equation}
where $\theta=\partial u_i/\partial x_i$ is the dilatation, $\langle p'\theta' \rangle$
is the mean pressure-dilatation correlation and the mean dissipation
$\langle \epsilon \rangle $. The
external forcing $f$ acts against the dissipative effect of viscosity
to sustain turbulent fluctuations. We can
also write the equation of the mean internal energy ($\langle e \rangle$)
from \eqn{ie} as
\begin{equation}
\frac{d\langle e \rangle}{dt}=-
\langle p'\theta' \rangle-\langle \epsilon \rangle.
\eqnlabel{meanie}
\end{equation}
The pressure-dilatation and viscous dissipation are
responsible for the exchange between kinetic and internal energy.
While the former is a bi-directional exchange depending upon
the value of turbulent Mach number, $M_t$ \cite{DS2013, SD2016},
the latter converts kinetic
energy into internal energy irreversibly. Since no external sink
is added to the energy equation, the internal energy of the system
always increases. %Because of this a true stationary state cannot
%be achieved. %However, after an inital transient state, the mean kinetic energy,
% mean dissipation, mean $M_t$ all approach a stationary state \cite{Kida1990}. %Once
%and equilibrium between $k$ and  $\langle \epsilon \rangle$ is reached,
%the system is said to be in statistically quasi-equilibrium
%state \cite{Kida1990}.
The time evolution of $K$ and
$\langle \epsilon \rangle$, normalized by their initial values,
is plotted in \rfig{tkef}. We can see
that $K$ increases initially, because of the input of energy due to forcing
at large scales. Once the
cascade develops and transfers energy to the smallest dissipative
scales, the mean kinetic energy starts to decrease.
At the same time, dissipation also increases initially, after an initial lag,
 until it reaches an equilibrium. At this point the rate of energy
input is equal to rate of dissipation and a quasi-stationary state is
reached \cite{Kida1990}. In \rfig{tkef}, this state is achieved at
$t/T_e\approx5$ for $Re_{\lambda}\approx35$ (a),
and  $t/T_e\approx6$ for $Re_{\lambda}\approx100$ (b),
 where $T_e$ is the average eddy turnover time. The
average eddy turnover time is computed form
the average taken at ten checkpoints from $t/T_e \ge5$ for $Re_{\lambda}\approx35$
and at fifteen checkpoints from $t/T_e \ge6$ for $Re_{\lambda}\approx100$ .
The net increase in the total energy is, at this point, equal
to the increase in the internal energy.
As in the case of decaying turbulence, we see a good
agreement between the synchronous and asynchronous
simulations in \rfig{tkef} for both high and low $Re_{\lambda}$.

\begin{figure}[h]
\centering%center}
\subfigure{\includegraphics[width=0.4\linewidth]{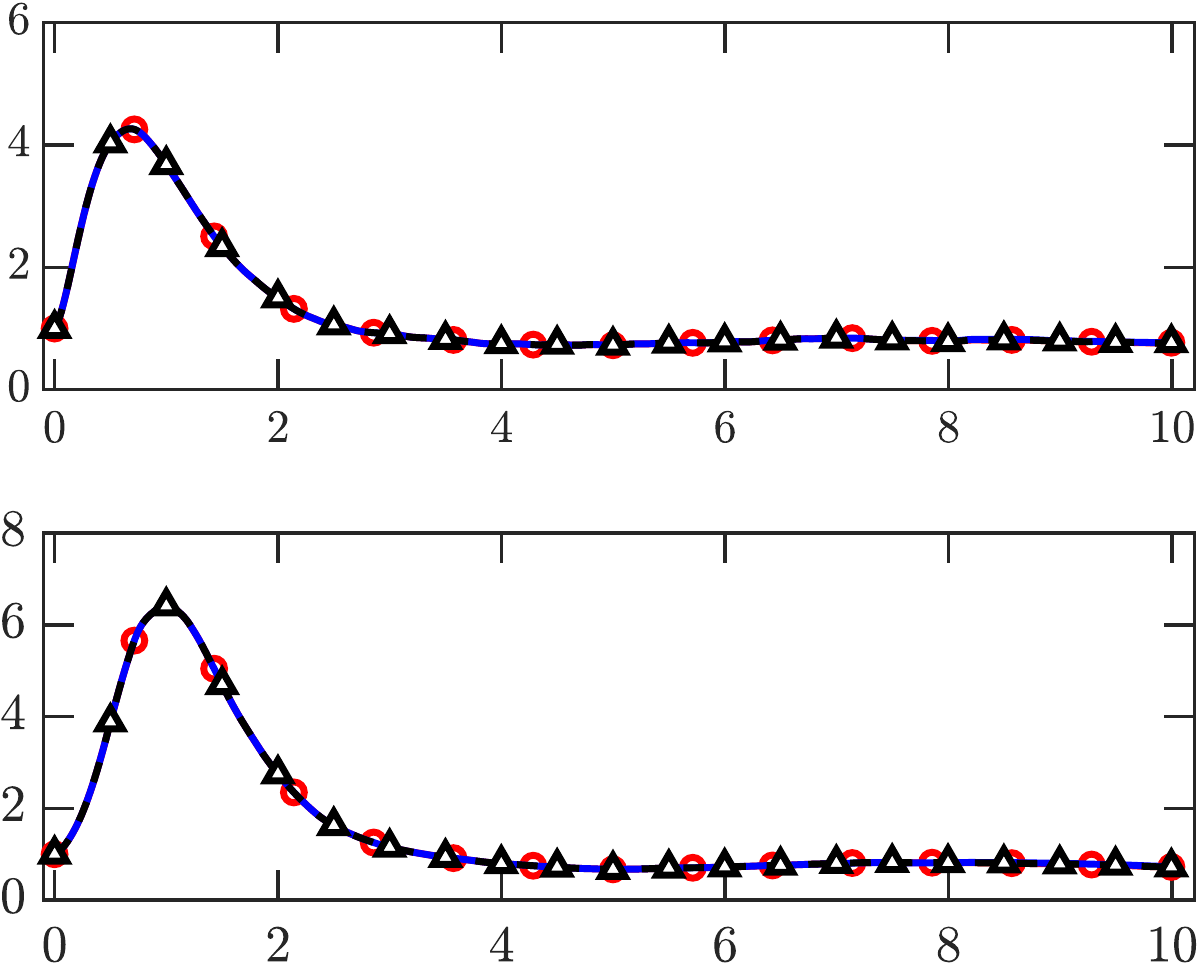}}
\hspace{1cm}
\subfigure{\includegraphics[width=0.4\linewidth]{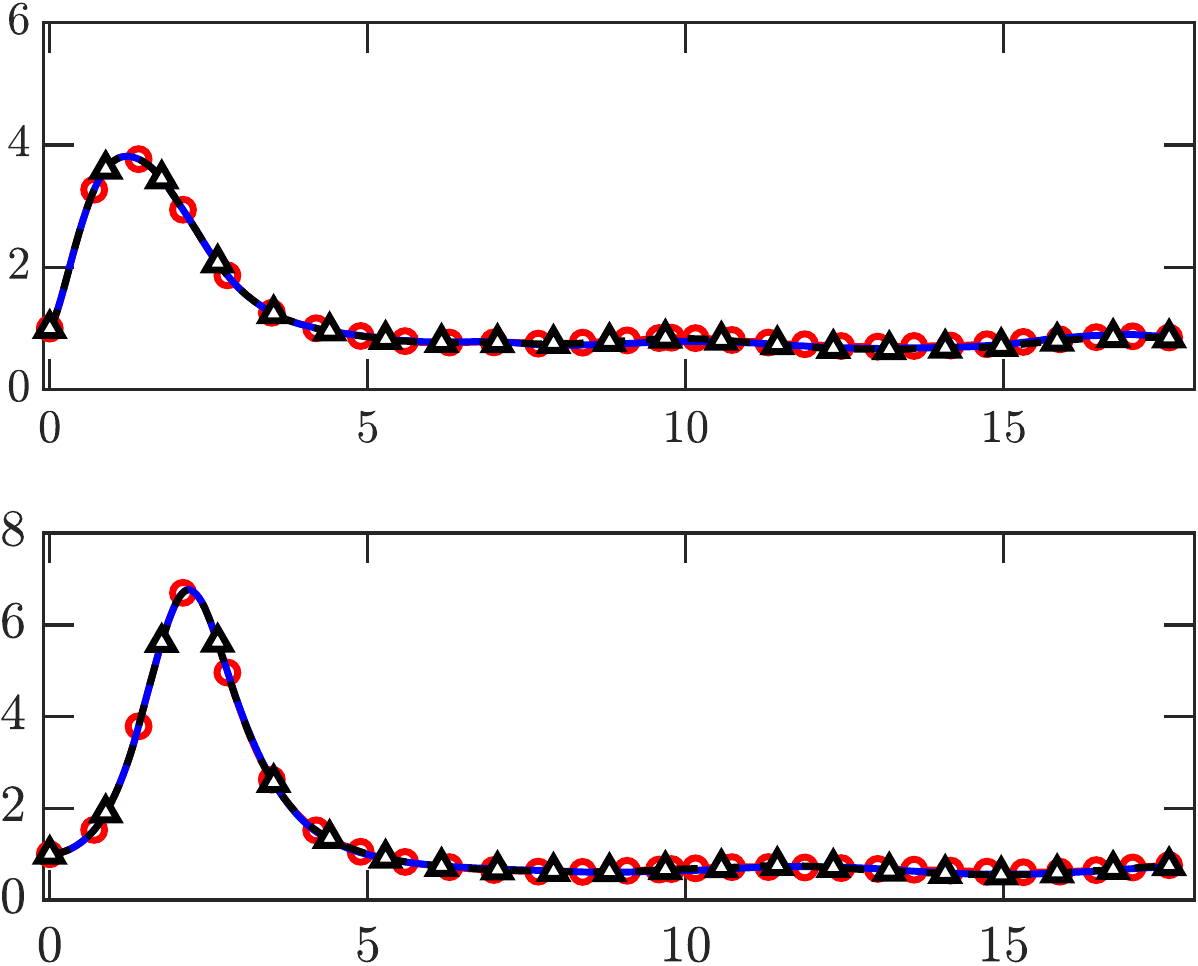}}
\begin{picture}(0,0)
%\put(-50,255){$(a)$}
%\put(155,255){$(b)$}
\put(-430,100){\rotatebox{90}{$K/K_0$}}
\put(-430,25){\rotatebox{90}{$\langle \epsilon \rangle / \langle \epsilon_0\rangle$}}
%\put(-115,145){$t/T_e(0)$}
%\put(85,145){$t/T_e(0)$}

\put(-25,130){$(b)$}
\put(-250,130){$(a)$}
\put(-210,100){\rotatebox{90}{$K/K_0$}}
\put(-210,25){\rotatebox{90}{$\langle \epsilon \rangle / \langle \epsilon_0\rangle$}}
\put(-325,-10){$t/T_e$}
\put(-105,-10){$t/T_e$}
\end{picture}

\caption{Evolution of space averaged
turbulent kinetic energy normalized by the
initial turbulent kinetic energy $K_0$ (top row) and space averaged
dissipation rate normalized by the initial dissipation rate $\eps_0$
(bottom row) for(a) $Re_{\lambda}\approx35$ and
(b) $Re_{\lambda}\approx100$.
Different lines are: SFD (red-circle), CAA (black-triangle) and
SAA (blue) with maximum allowed
delay level of $L=3$. Time is normalized by
the average eddy turnover time ($T_e$).
}
\figlabel{tkef}
%\end{center}
\end{figure}
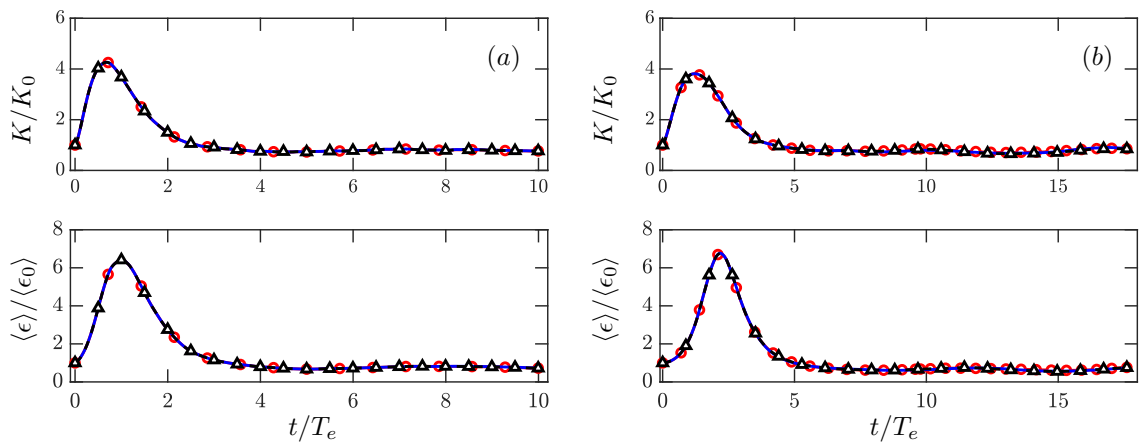

We are also interested in the energy and pressure spectrum, which are
plotted in \rfig{specf}.
These spectra are the average taken for ten and fifteen
checkpoints, respectively for $Re_{\lambda}\approx 35$ and $100$,
 after the quasi-stationary state is reached.
These energy spectra are shown in \rfig{specf}(a)
where we see that both CAA and SAA
simulations are accurately resolved,
with good collapse at all wavenumbers. %and inertial
%range at large $Re_{\lambda}$.
For the pressure
spectrum in \rfig{specf}(b), the CAA and SAA agree equally well with the SFD,
unlike the decaying case where small errors were seen at the large wavenumbers for SAA.
%We do not see an inertial range for
%pressure spectrum  because $Re_{\lambda}$ is not large enough \cite{Gotoh2001}
%and the spectrum does not collapse for high wavenumbers.
These spectra are also consistent with \cite{DS2013} at similar conditions.
% We do see very small deviations at the high wavenumber for both energy and
%presssure spectra,
%which as we discussed in previous section,
%can be eliminated if higher order schemes
%are used.
The higher order moments of
the longitudinal velocity gradients are also plotted in
\rfig{skewf}. We see that $-S$ fluctuates around $0.5$  \cite{Schumacher2014} and
the values are fairly consistent for SAA, CAA and SFD.
Even better agreement is seen for $F$ in \rfig{skewf}($c,d$),
with value close to $6$ for $Re_{\lambda}\approx100$ and smaller
for $Re_{\lambda}\approx35$  \cite{Schumacher2014}.
%
%While there are small disagreements between the
%skewness value obtained for SAA,
%it is still very close to the expected value of $-0.5$ \cite{Schumacher2014}.
%No observable differences are present in the flatness of the
%PDF of the longitudinal velocity gradients and the
%flatness is also consistent with the literature \cite{Schumacher2014}.

\begin{figure}[h]
\centering%center}
\subfigure{\includegraphics[width=0.4\textwidth]{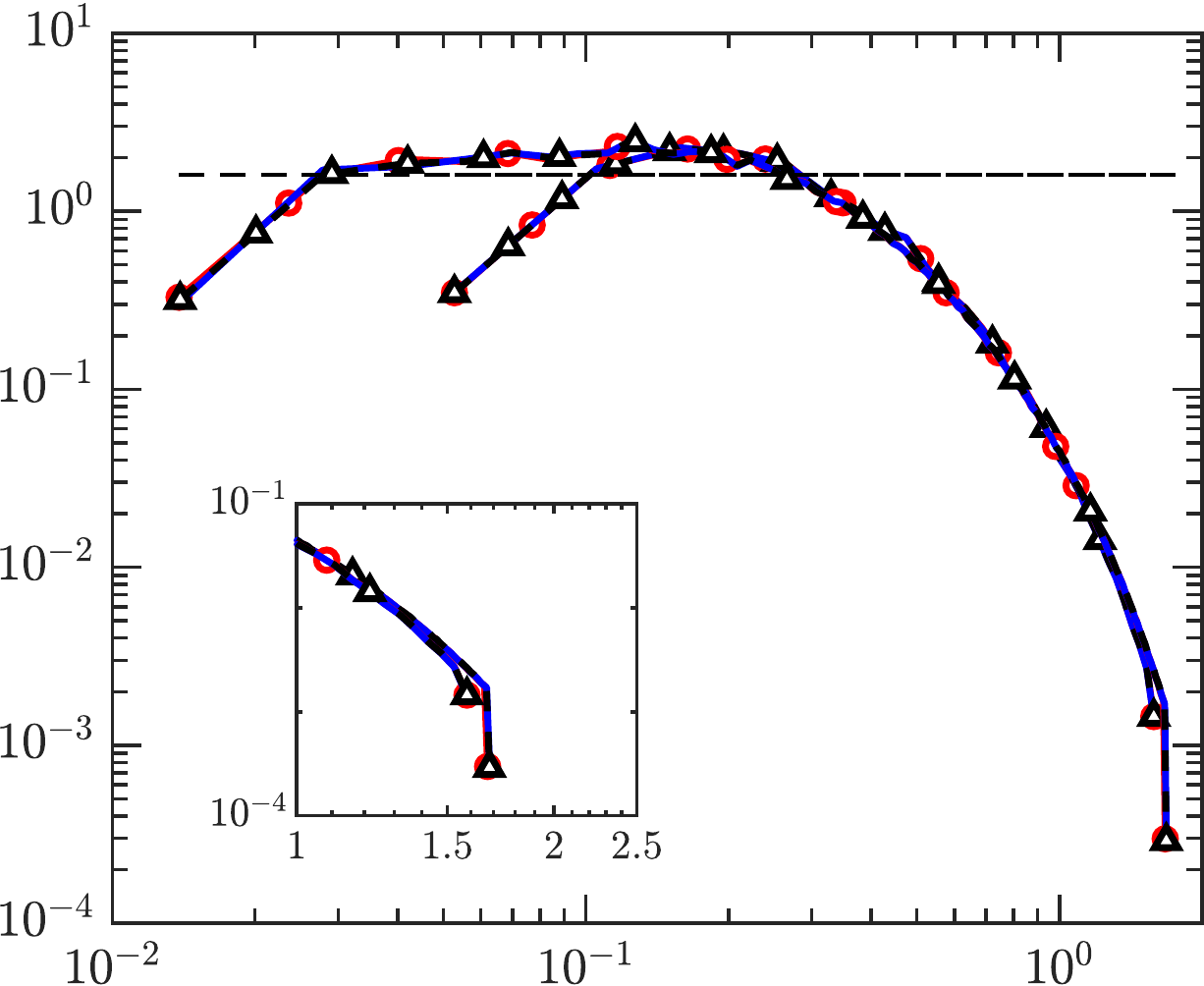}}
\hspace{1cm}
\subfigure{\includegraphics[width=0.4\textwidth]{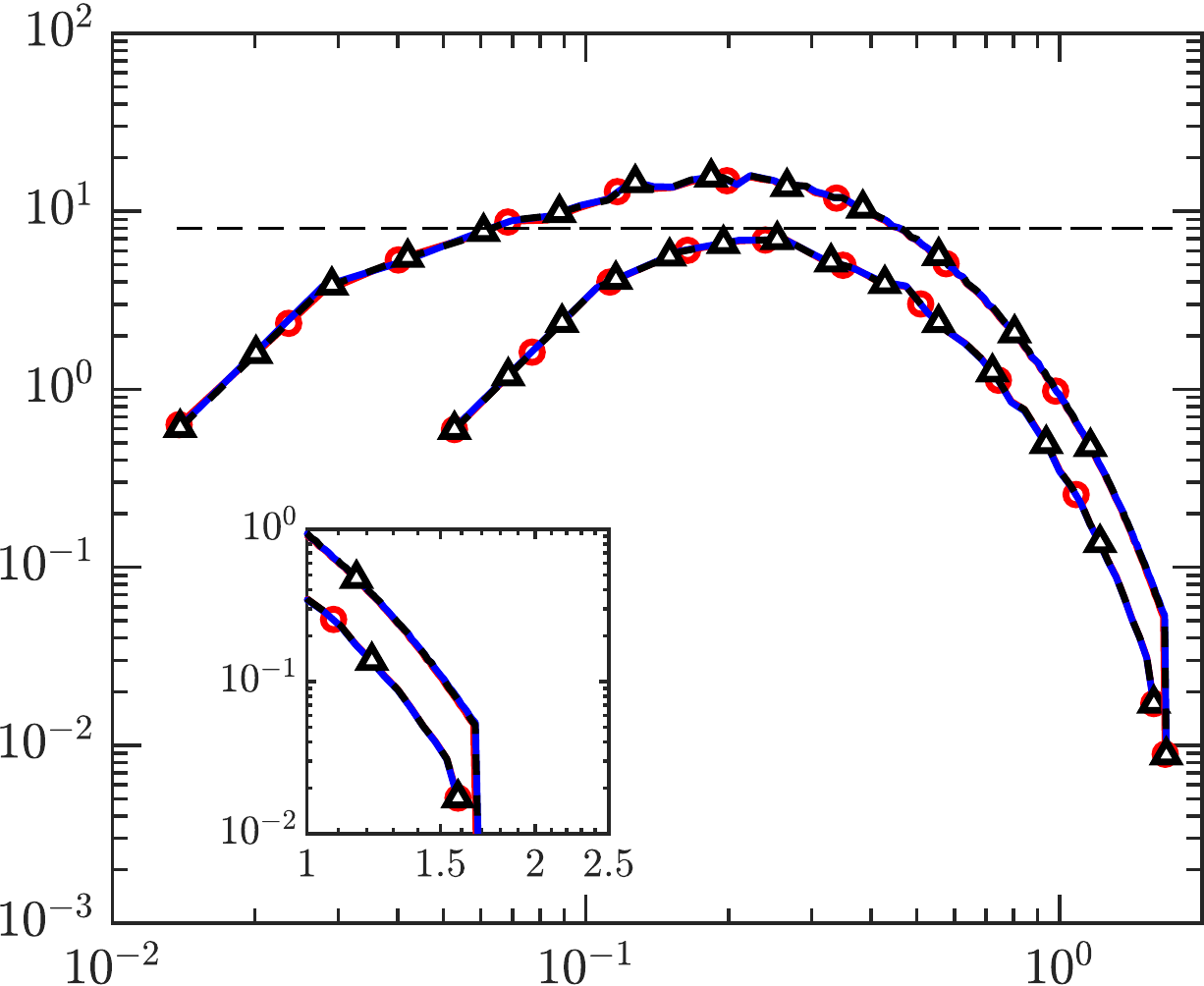}}

\begin{picture}(0,0)
\put(-40,145){$(a)$}
\put(180,145){$(b)$}
\put(-220,55){\rotatebox{90}{$E(\kp)\langle \epsilon\rangle^{-2/3}\kp^{5/3}$}}
\put(0,55){\rotatebox{90}{$E_p(\kp)\langle \epsilon\rangle^{-4/3}\kp^{7/3}$}}
\put(-100,0){$k\eta$}
\put(100,0){$k\eta$}
\put(-110,125){\vector(-1,0){70}}
\put(115,110){\vector(-1,0){75}}

\end{picture}

\caption{(a) Compensated energy spectrum and
(b) compensated pressure spectrum for $Re_{\lambda}\approx35$
and $Re_{\lambda}\approx100$.
Different lines are: SFD (red-circle), ATP (black-triangle) with
$L=3$ and ATR (blue) with $L=2$. Arrow indicated increasing $Re_{\lambda}$.
Dashed line in is Kolmogorov constant $C=1.6$ in (a) and $C_p=8$ in (b).
}
\figlabel{specf}
%\end{center}
\end{figure}
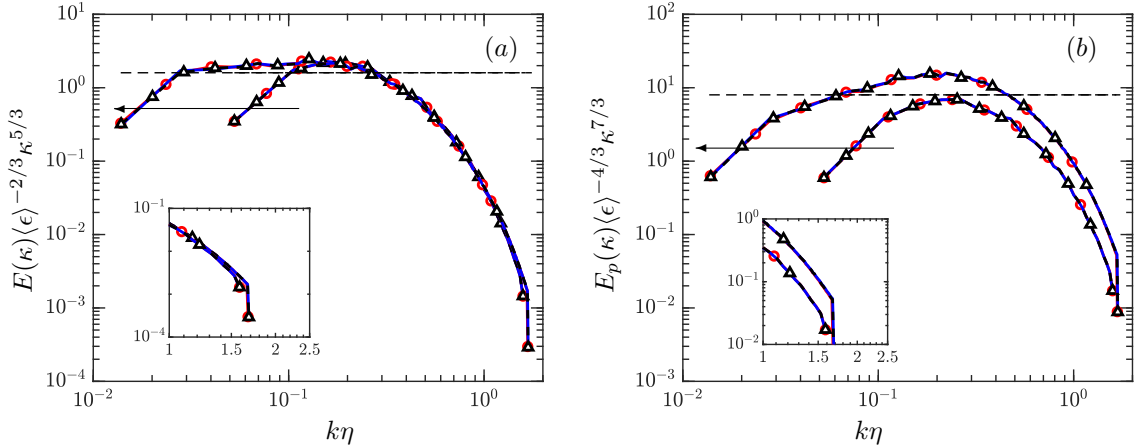

\begin{figure}[h]
\centering%center}
\subfigure{\includegraphics[width=0.4\textwidth]{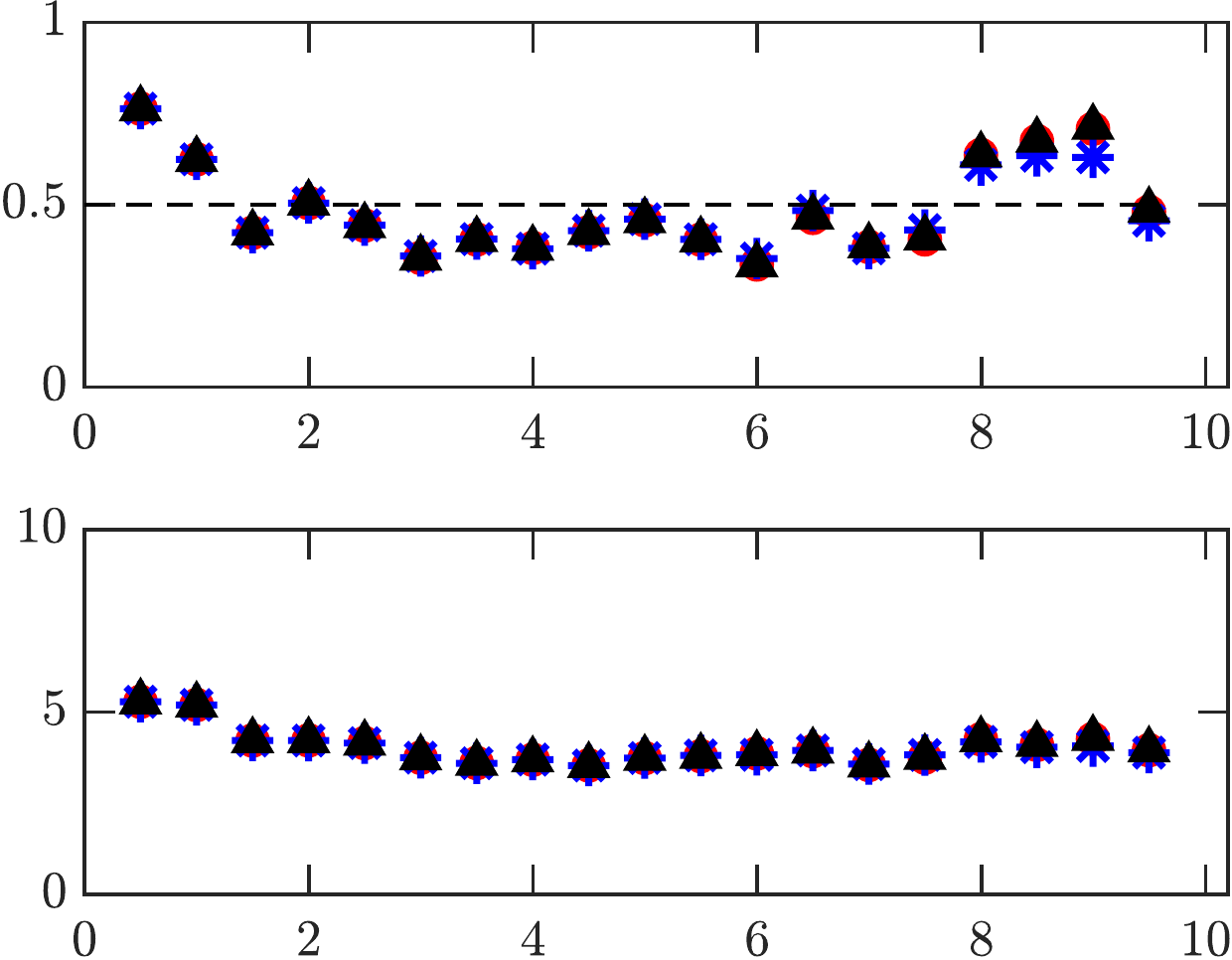}}
\hspace{1cm}
\subfigure{\includegraphics[width=0.4\textwidth]{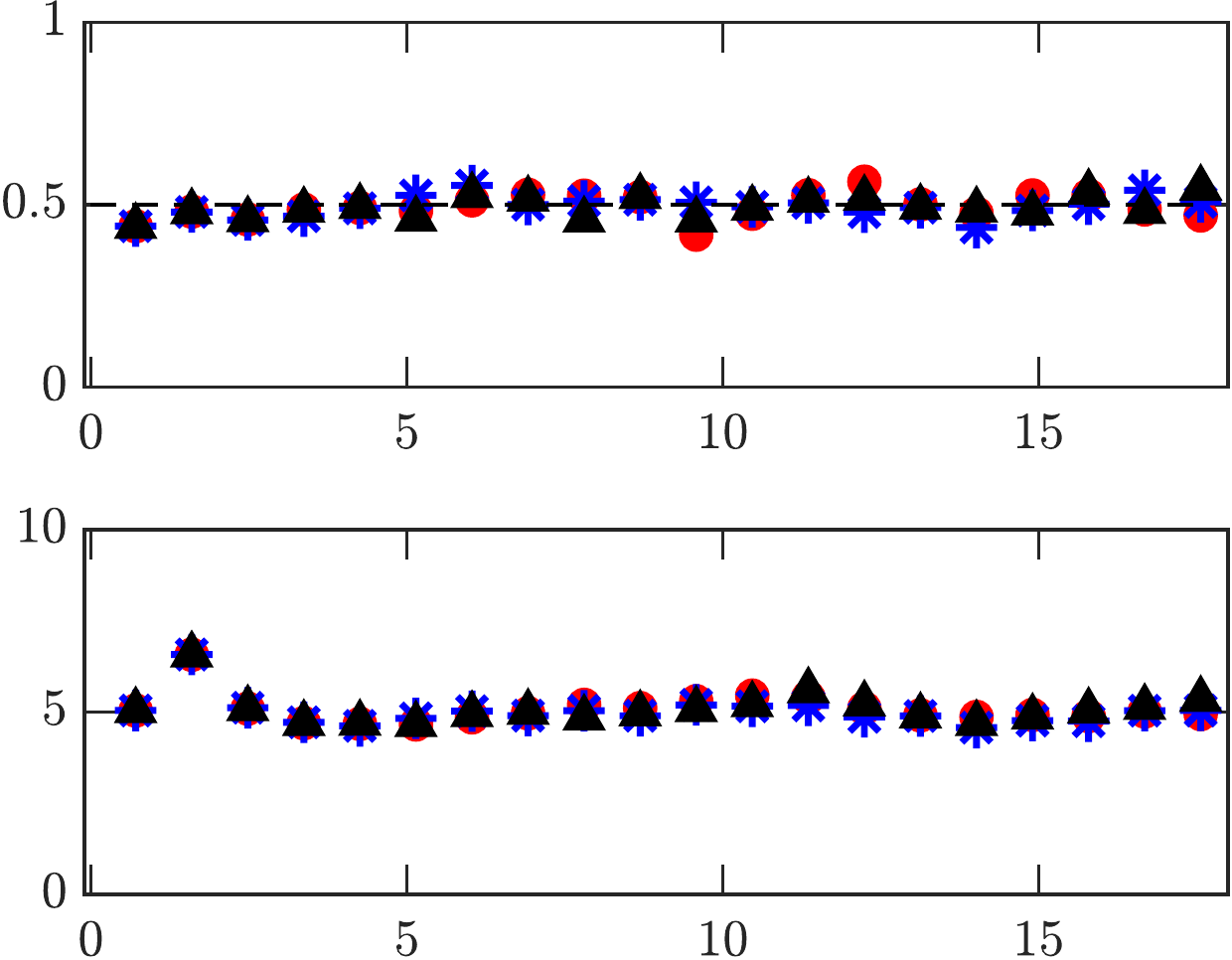}}
\begin{picture}(0,0)
\put(-25,130){$(b)$}
%\put(-25,45){$(d)$}
\put(-205,105){\rotatebox{90}{$-S$}}
\put(-205,35){\rotatebox{90}{$F$}}
\put(-110,-10){$t/T_e(0)$}

\put(-245,130){$(a)$}
%\put(-225,45){$(b)$}
\put(-430,105){\rotatebox{90}{$-S$}}
\put(-430,35){\rotatebox{90}{$F$}}
\put(-320,-10){$t/T_e(0)$}

\end{picture}

\caption{Negative of skewness (top row) and flatness
(bottom row) of the longitudinal velocity gradient vs.
normalized time for (a) $Re_{\lambda}\approx35$ and
 (b) $Re_{\lambda}\approx100$.
Different symbols are: SFD (red-circle), ATP (black-triangle) with
$L=2$ and ATR (blue asterik) with $L=2$. The dashed black line indicates skewness of $0.5$.
}
\figlabel{skewf}
%\end{center}
\end{figure}
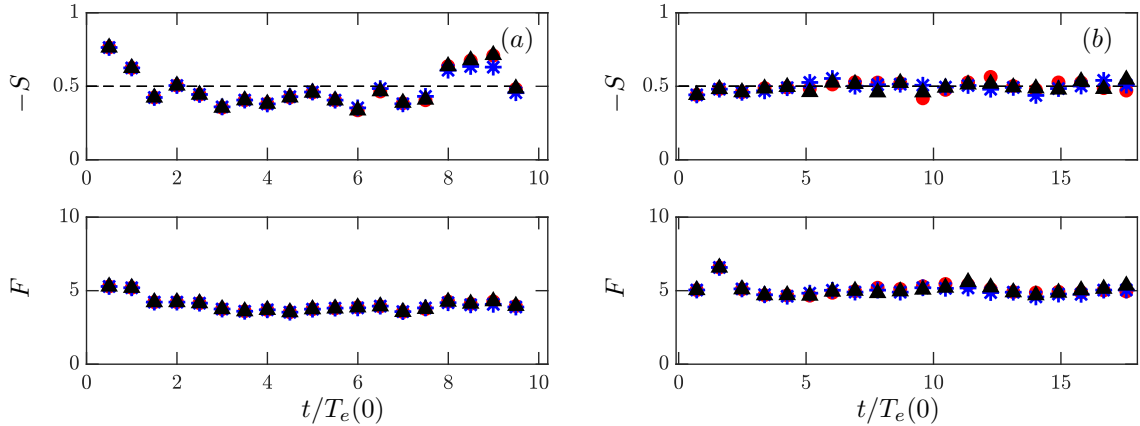

Finally we look at the PDF of enstrophy density ($\Omega$) and dissipation rate $(\epsilon)$
\cite{DPK2008,PKDD2012,PKKRS2015}.
Both dissipation and enstrophy are crucial in the understanding of the
small-scale motions \cite{Sreeni1997} and are highly intermittent.
Because of extreme events in $\epsilon$ and $\omega$, the corresponding
PDFs of the normalized quantities, $\epsilon/\langle \epsilon \rangle$ and
$\langle \omega \rangle$, are characterized by wide tails. % that become
%wider with increasing reynolds number \cite{DPK2008,PKKRS2015}.
 The PDF of $\epsilon/\langle \epsilon \rangle$ and
$\Omega/\langle \Omega \rangle$, averaged over checkpoints as in
case of averaged spectrum, are plotted in \rfig{enst_pdf}($a,b$). We can
clearly see the tails of both the PDFs become wider as Reynolds number is increased from
$38$ to $100$. This suggests that the propensity of events that
are an order of magnitude more intense than the mean,
increases with the Reynolds number \cite{DPK2008,PKKRS2015}.
Furthermore, we also observe that the tails
 for the PDF of $\Omega/\langle \Omega \rangle$ in \rfig{enst_pdf}($b$) are
wider than the tails for PDF of $\epsilon/\langle \epsilon \rangle$ in \rfig{enst_pdf}($a$).
This implies that enstrophy is more intermittent than dissipation and this has been
consistently established in several past studies \cite{Kerr1985,Siggia1981,Sreeni1997,zhou2000,
DPK2008}. These features of the PDF are captured well by
both the asynchronous algorithms with very small differences at the far tails.
Thus, the AT schemes accurately resolve even the finest scales of turbulence including
very highly intermittent events in dissipation and enstrophy.

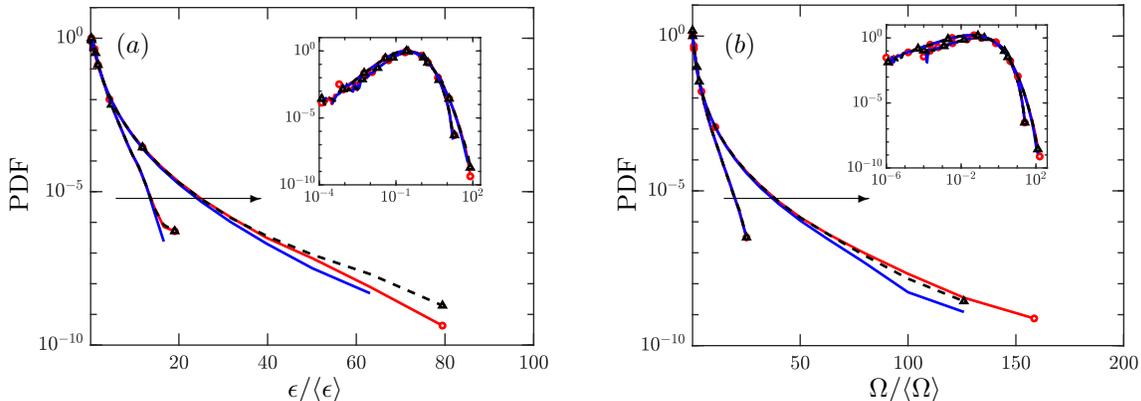
\begin{figure}[h]
\centering%center}
\subfigure{\includegraphics[width=0.41\textwidth]{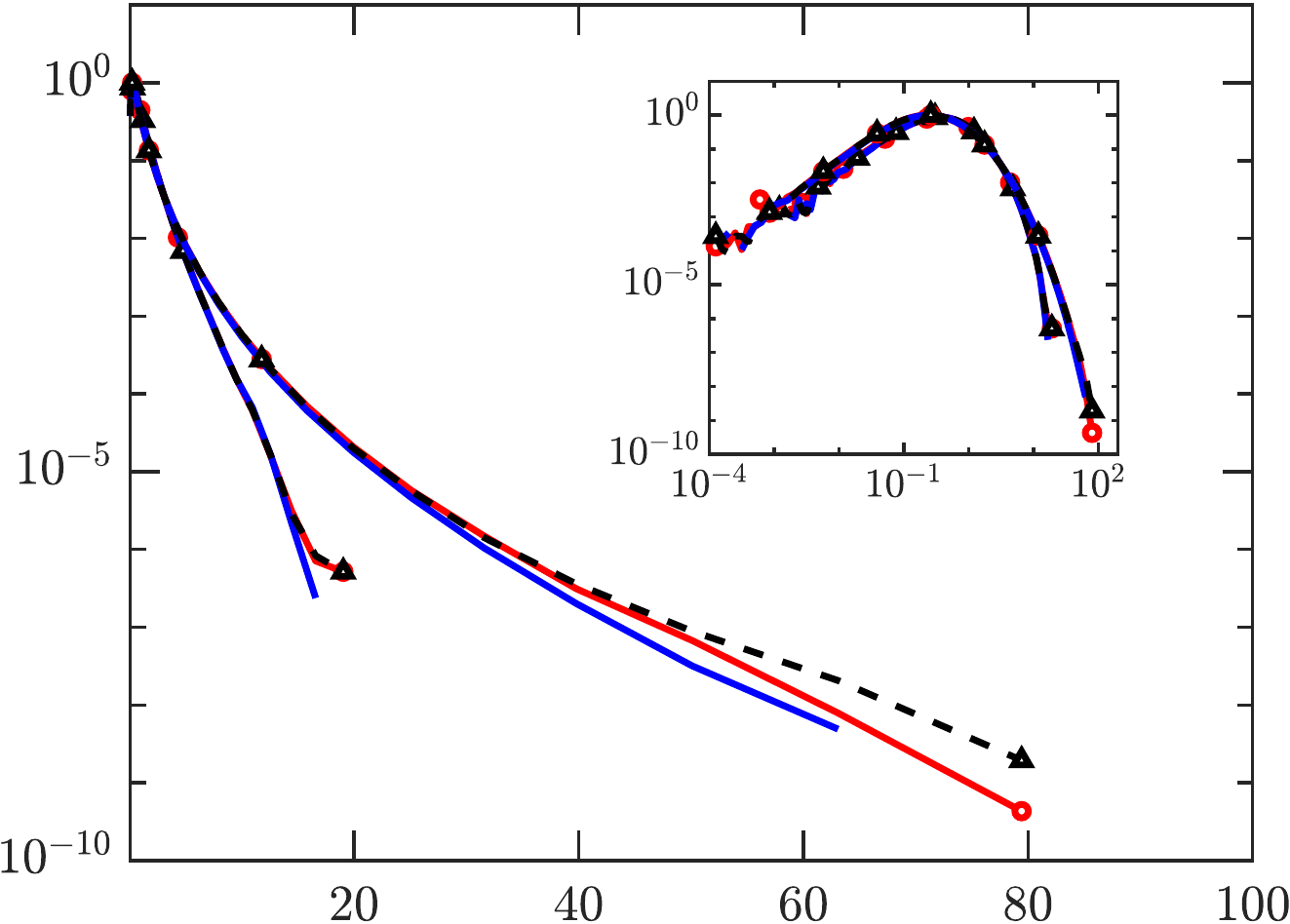}}
\hspace{1cm}
\subfigure{\includegraphics[width=0.4\textwidth]{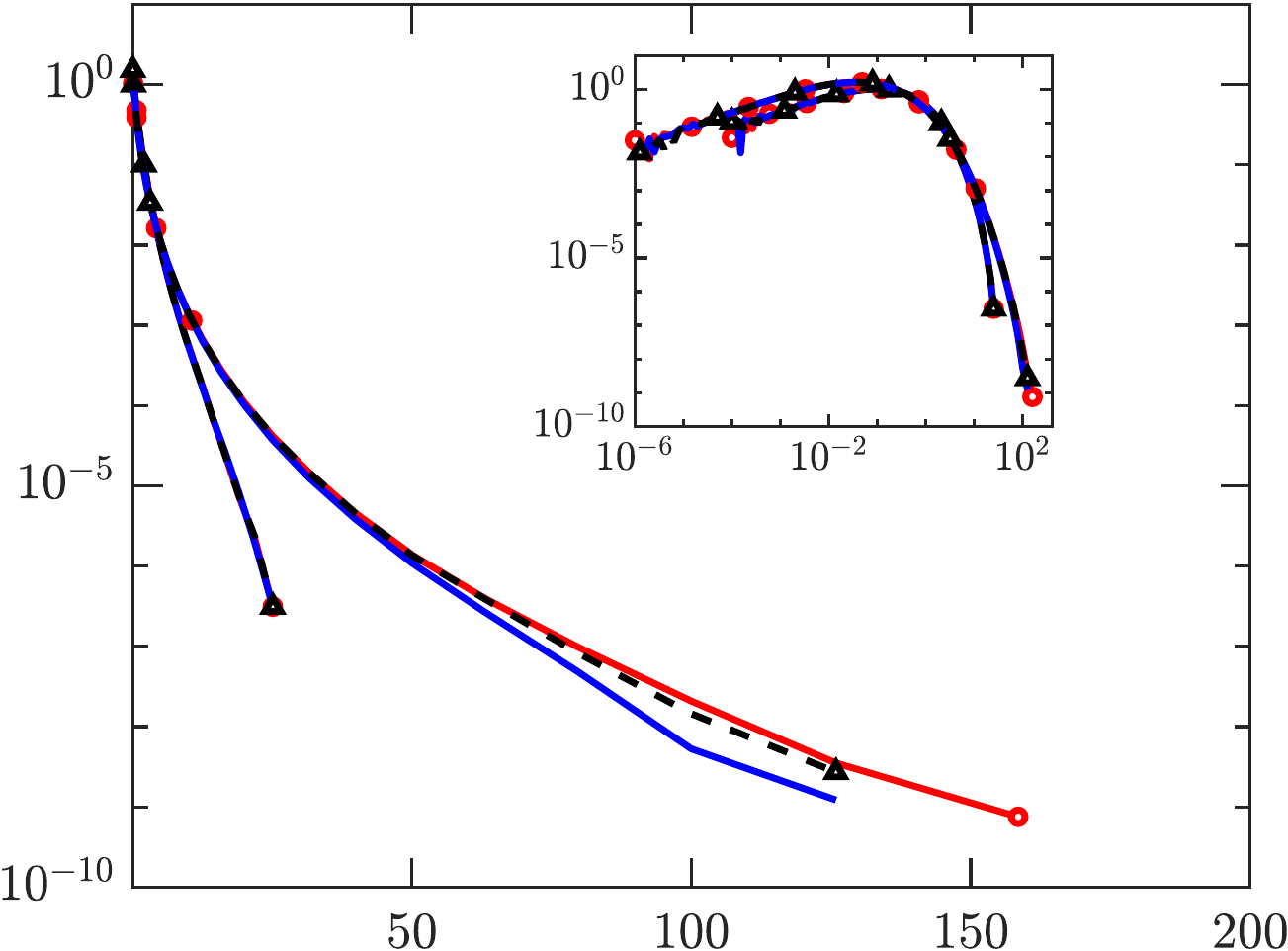}}
\begin{picture}(0,0)
\put(-160,120){$(b)$}
\put(-200,60){\rotatebox{90}{PDF}}
\put(-325,-10){$\epsilon/\langle \epsilon \rangle$}

\put(-390,120){$(a)$}
\put(-430,60){\rotatebox{90}{PDF}}
%\put(-425,20){\rotatebox{90}{$Flatness$}}
\put(-105,-10){$\Omega/\langle \Omega \rangle$}
\put(-390,65){\vector(1,0){55}}
\put(-160,65){\vector(1,0){55}}
\end{picture}

\caption{PDF of (a) normalized dissipation rate ($\eps/\la \eps \ra$) and (b)normalized
enstrophy ($\Om/\la \Om\ra$) in log-linear scale. The insets are the
same PDFs in log-log scale. Different lines are: SFD (solid red with circle),
 CAA (dashed black with triangle) and SAA
(solid blue) with $L=3$. The arrow indicates increasing $Re_{\lambda}$
}
\figlabel{enst_pdf}
%\end{center}
\end{figure}

%\colb{\\++++++++++++++++++++++++++++++++++++++++++++++}

\subsection{Computational performance}
The preceding sections demonstrated the ability of
asynchronous algorithms in resolving important physical
characteristics of turbulent flows including instantaneous field and high order
statistics. Now we show that the asynchronous simulations are
computationally more efficient than their synchronous counterpart.
To study this we look at so-called strong and weak
scaling of the solver. In the former the problem size
remains fixed, while in the
latter the computational work is kept constant.
Ideally, for a fixed problem size, the computation time should decrease linearly
on increasing the processor count. However, with
increasing number of processors, the necessary communications and
synchronizations increase the communication time until it
 eventually dominates the total execution time. This is
essentially the communication bottleneck and is expected
to be a major challenge to scalability \cite{Dongarra2011,JD2012,DA2014,AD2017}.
In \rfig{scaling}(a), we
have plotted the total execution time for synchronous and asynchronous implementations for
our compressible flow solver. These times are an average of five runs of $6000$ steps each
and a maximum allowed delay of $L=4$ for both SAA and CAA.
%Stampede2-Skylake nodes.
For reference we have also plotted ideal scaling
as a dashed black line.
In \rfig{scaling}(a) clear departures
from ideal scaling are seen at $P=512$ for SFD. This, as is evident from \rfig{scaling}(b),
happens because the percentage of communication time (dashed red)
 grows with processor count ($P$) until it becomes comparable to
the computation time. On the other hand, both CAA and SAA
(black and blue lines) are close to the ideal scaling
in \rfig{scaling}(a) for a much larger processor count of $P=8192$.
The improved scaling is attributed to the fact that only a small percentage
($\sim 20\%$) of the overall time is spent on communications. This
percentage (\rfig{scaling}(b)) remains fairly constant on increasing the number of processors
for the asynchronous implementations, whereas grows to larger than $50 \%$
for the synchronous case.

%As a complimentary analysis, we also looked at the scaling for a simple
%constant property 3D Burgers solver, which has very less computations.
%Similar conclusions were drawn from the scaling of Burgers eqution solver as well.
%It must be noted that eventually, the asynchronous case
%also departs from the ideal scaling, which in
%the current case happens at $P=16384$.
%From \rfig{scaling}(a) we can see that ATP departs
%rather abruptly form the ideal scaling at high processor
%count. While ATP reduces the overall number of communication,
%the current implementation forces synchronization at each
%communications. The forced synchronization coupled with the
%large message size (because of multiple time levels) adversally affects
%the scaling. This can be seen from an abrupt increase
%in the percentage of communication time at $P=16384$ for ATP (dashed black
%with triangles) in \rfig{scaling}(b). Also,
%as number of processors is increased, the sum
%of solid and dashed lines in \rfig{scaling}(b) is less
%than one. This is because of additional overheads,
%for example, loops, logical statement etc.
%which are non-parallelizable and account for a small percent of total time,
%that becomes significant at very high processor count, especially when
%the computations involved are very low.

Next we look at the weak scaling, where ideally because of
fixed computational work, the time per step should remain constant
on increasing the processor count. The time per step for
a computational load of $N^3/P=2048$ is plotted in \rfig{weak}.
For the synchronous case, this time per step scaling grows by a factor of
$60\%$ because of increase in communication and synchronization
ovearheads at large core count ($P=262,144$). This
can only be expected to get worse at much higher levels
of parallelism expected in exascale machines.
On the other hand, the asynchronous algorithms show
improved scaling, with a much smaller $21\%$ increase in time per step for
SAA and only $14\%$ increase for CAA on increasing the number
of processors from $P=128$ to $P=262,144$.
This also implies that reduction in the overall volume of
communication (CAA) at extreme scales provides more
improvement in scaling than reducing forced synchronizations (SAA).

Both weak and strong scaling analysis lead
us to the same conclusion that the asynchronous algorithms
remove synchronization and communication overheads, leading to
an effective overlap between communications
and computations and, consequently, an improvement in scaling.

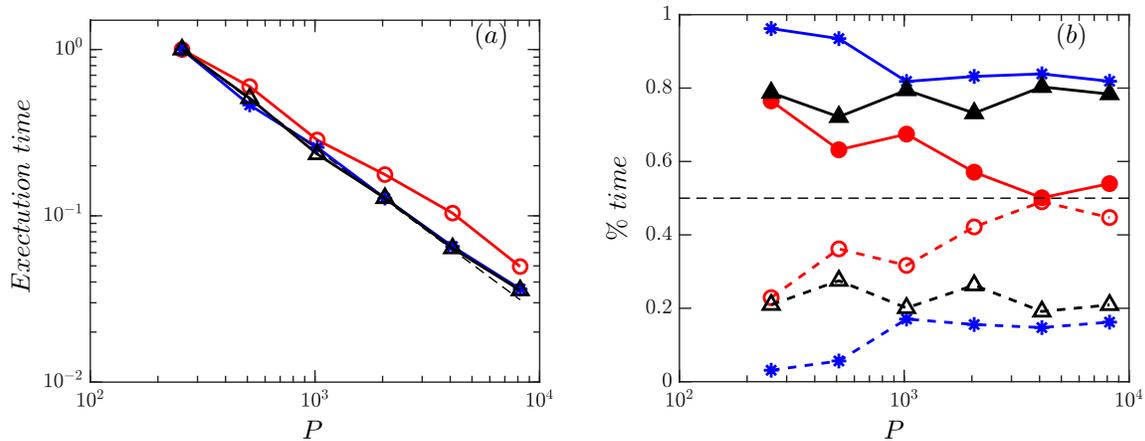
\begin{figure}[h]
\centering%center}
\subfigure{\includegraphics[width=0.41\linewidth]{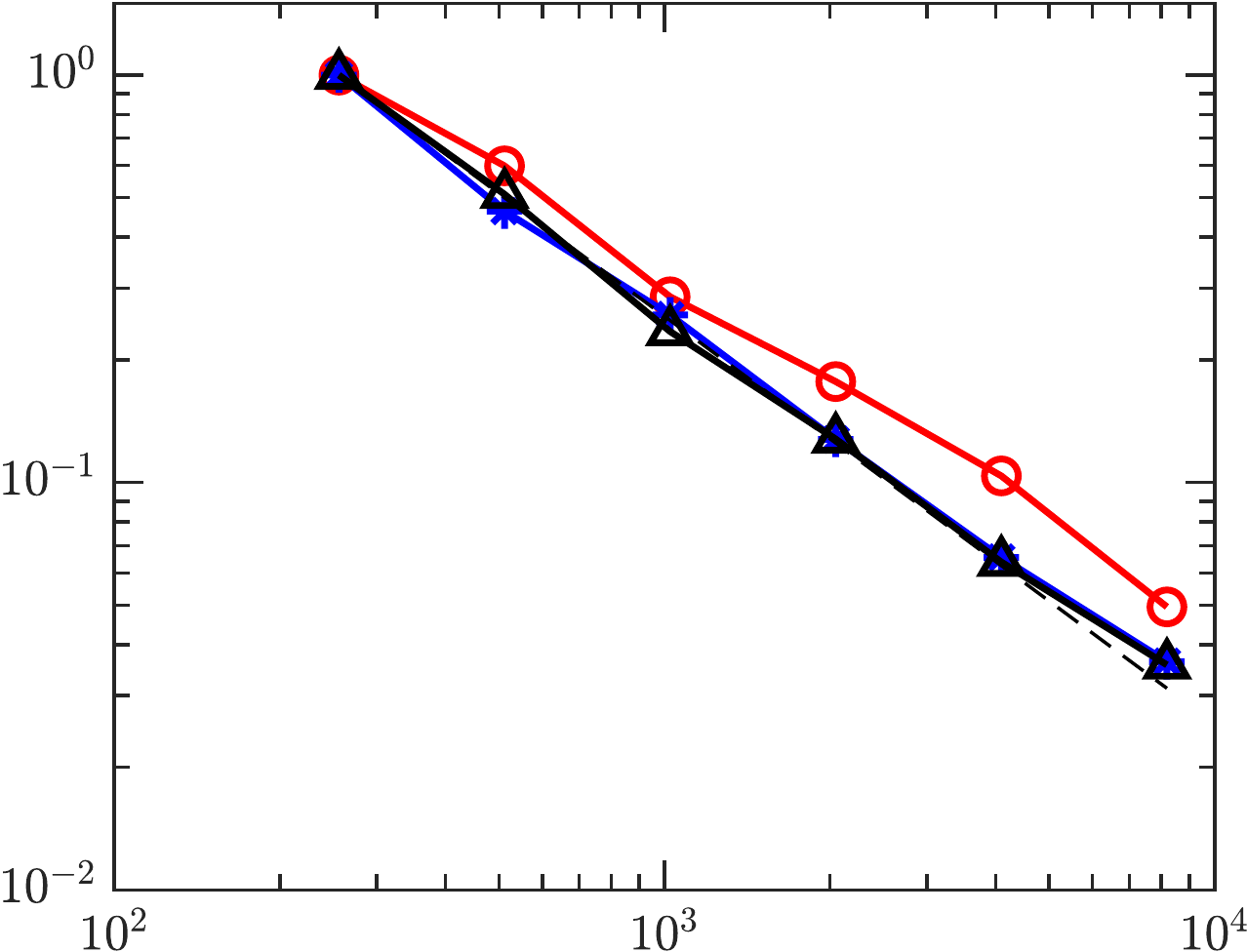}}
\hspace{1cm}
\subfigure{\includegraphics[width=0.4\linewidth]{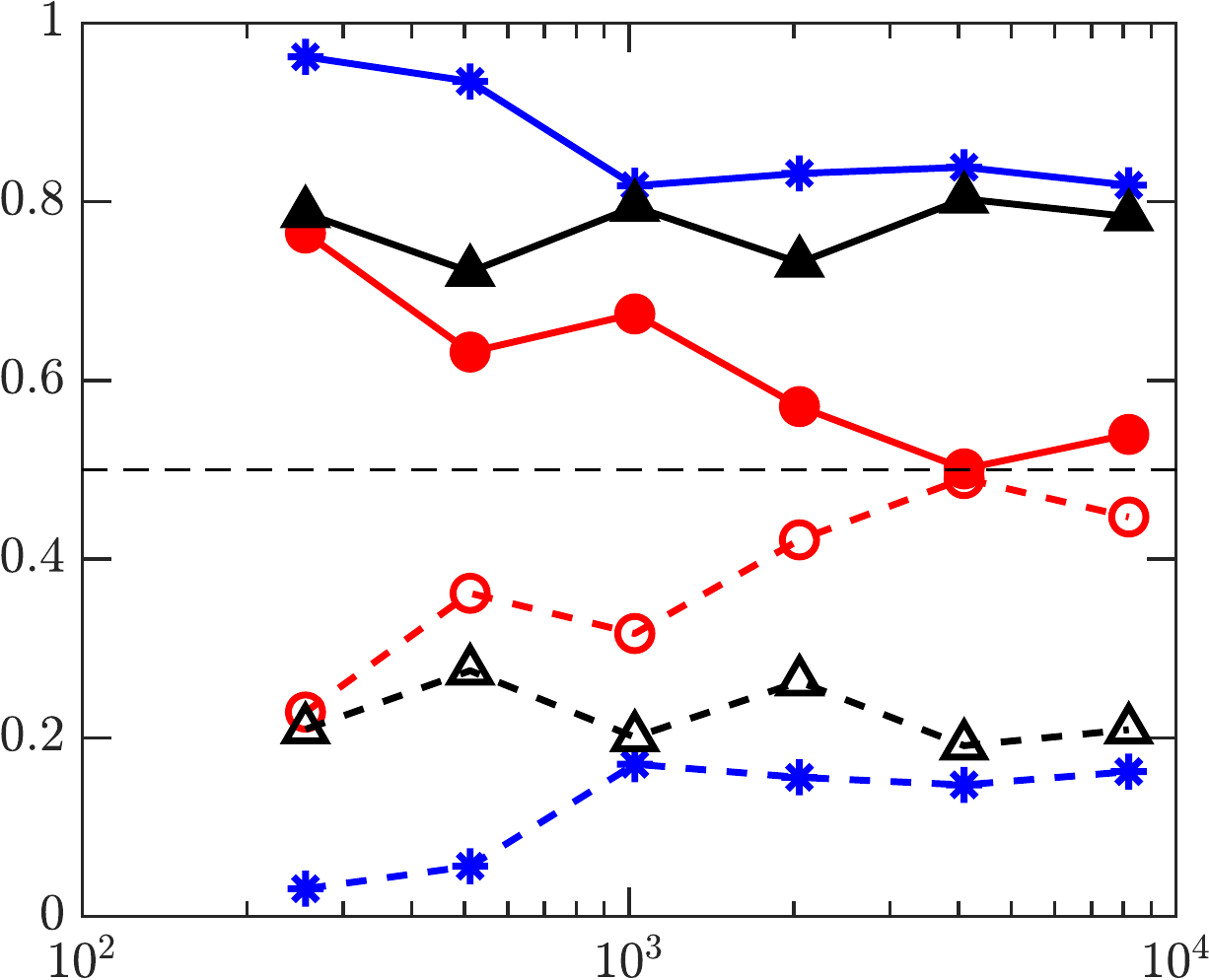}}

%\subfigure{\includegraphics[width=0.35\linewidth]{figures/scaling_128/burg_scl10}}
%\hspace{1cm}
%\subfigure{\includegraphics[width=0.335\linewidth]{figures/scaling_128/burg_ccl10}}
\begin{picture}(0,0)
\put(-45,150){$(a)$}
\put(175,150){$(b)$}
\put(5,75){\rotatebox{90}{$\%~time$}}
\put(-220,55){\rotatebox{90}{$Exectution~ time$}}
\put(-110,0){$P$}
\put(110,0){$P$}

%\put(-225,105){$(c)$}
%\put(-225,240){$(a)$}
%\put(-380,180){\rotatebox{90}{$Total~time$}}
%\put(-380,40){\rotatebox{90}{$Total~time$}}
%\put(-285,-10){$P$}
\end{picture}

\caption{Strong scaling for $N=128$.
(a): Total execution time normalized by the execution
time for $P=256$. (b): Computation time and communication time
as a percentage of the total execution time.
Different lines are: SFD (red), CAA (black) and SAA (blue),
%with $L=3$, %for (a,b) and $L=10$ for (c,d),
dotted black in (a) is ideal scaling
and in (b) is $50\%$ of total time.
Dashed lines with hollow symbols in (b)
is communication time and solid lines with solid symbols is computation
time.
}
\figlabel{scaling}
%\end{center}
\end{figure}

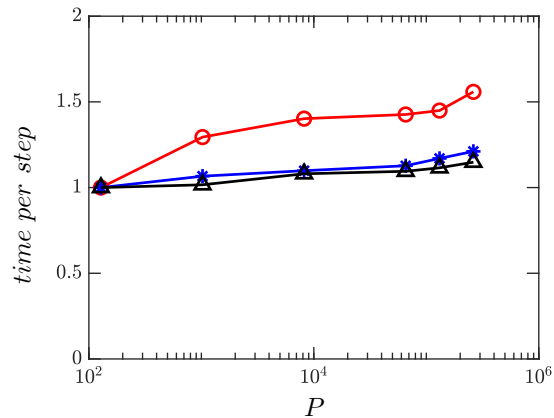
\begin{figure}[h]
\centering%center}
\includegraphics[width=0.4\linewidth]{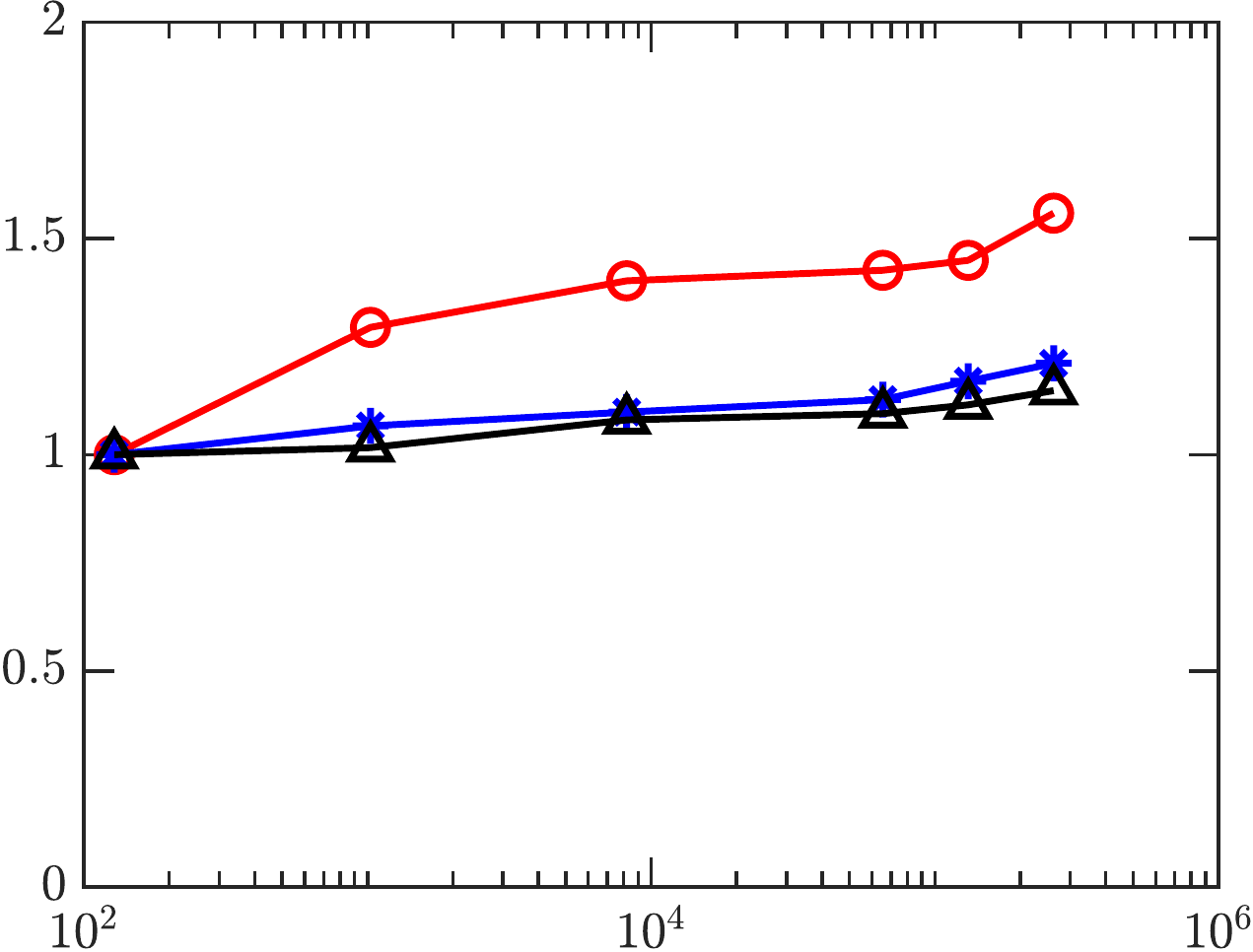}

%\subfigure{\includegraphics[width=0.35\linewidth]{figures/scaling_128/burg_scl10}}
%\hspace{1cm}
%\subfigure{\includegraphics[width=0.335\linewidth]{figures/scaling_128/burg_ccl10}}
\begin{picture}(0,0)
%\put(-45,120){$(a)$}
%\put(155,120){$(b)$}
%\put(5,55){\rotatebox{90}{$\%~time$}}
\put(-110,50){\rotatebox{90}{$time~per~step$}}
\put(0,0){$P$}

%\put(-225,105){$(c)$}
%\put(-225,240){$(a)$}
%\put(-380,180){\rotatebox{90}{$Total~time$}}
%\put(-380,40){\rotatebox{90}{$Total~time$}}
%\put(-285,-10){$P$}
\end{picture}

\caption{Weak scaling: time per step for $N^3/P=2048$ normalized by
time per step for $P=128$. Different lines are: SFD (red), CAA (black) and
SAA (blue).}
\figlabel{weak}
%\end{center}
\end{figure}

\section{Conclusions and future implementation considerations}
Numerical simulations of PDEs, governing complex natural and engineering phenomena,
using standard numerical methods on parallel supercomputers, require PEs
to communicate and synchronize frequently to ensure accuracy.
This synchronization and communication cost and the resulting
PE idling grows with increasing
levels of parallelism and presents a major challenge to scalability to exascale computing.
In order to mitigate this bottleneck, these
constraints were relaxed at a mathematical level to derive the so-called
Asynchrony-Tolerant (AT) of arbitrary order of accuracy in \cite{AD2017}. By allowing
for asynchrony, these AT schemes can be used to allow computations to proceed in a PE without having
to wait for updated values at the boundaries, thus removing synchronizations.

In this work we presented, first of a kind, asynchronous simulations of
compressible turbulence using high-order Asynchrony-Tolerant (AT) schemes to study
the effect of asynchrony on the physics of turbulence at different scales
and on the computational performance of the solver.
%The general methodology of derivation of AT schemes of arbitrary order has been shown in
%the previous work by \cite{AD2017}.
We show analytically that these schemes preserve the conservative property
of standard finite differences up to an
order higher than the order of the scheme. Stability analysis of these schemes shows
that their stability limit is smaller than their synchronous counterpart.
This reduction in the stability limit can be expressed in terms of the
synchronous stability limit and a function of delays that gives
a quantitative measure of the effect of delays.
Numerical data also suggests that these two can be rearranged to
obtain an \textit{effective asynchronous} CFL that is
essentially equal to the known synchronous stability limit and independent of $L$.
%time-step that depends upon the delay and is used to define an asynchronous CFL.
%Results show that the stability can be expressed in terms this new asynchronous CFL.
We introduced two ways to allow for asynchrony, namely, communication avoiding and
synchronization avoiding algorithms (CAA and SAA, respectively).
While the former leads to deterministic delays with a uniform probability
distribution, the latter leads to random delays with a machine specific delay distribution.
% A metric for the measure of points directly affected by asynchrony is
%also defined, which for simulations presented in this work is as high as $58\%$.

The aforementioned asynchronous
algorithms are used for the simulation of decaying and solenoidally
forced turbulence. Important low and high order statistics %such as evolution of turbulent kinetic energy and dissipation, energy and pressure
%spectrum and moments of velocity gradients,
obtained for the asynchronous algorithms
are compared with that for the standard synchronous finite differences (SFD) at the
same resolution and order and also with high-order compact difference schemes (C10).
We found excellent agreement
between SFD and CAA for the time evolution of turbulent kinetic energy and
dissipation for both decaying and forced turbulence, including the transients for the latter.
The distribution of energy at different scales as shown by the
velocity and the pressure spectrum is resolved by CAA with same level of accuracy as SFD and C10,
even at the largest wavenumbers.
Higher-order moments of longitudinal velocity gradient, including skewness and
flatness, also showed excellent agreement between SFD, C10 and CAA.
No observable differences are seen in the complex distribution of
the contours of instantaneous enstrophy field. The PDF of highly intermittent
quantities such as dissipation and enstrophy,
that are also very sensitive to the accuracy of numerical schemes and small
scale resolution, are also captured well by CAA, with some statistical differences at extreme tails.

For SAA as well, the evolution of turbulent kinetic energy and
dissipation for decaying and forced turbulence (including transients) is
in excellent agreement with SFD and C10. While no differences
were seen for the energy spectrum, the pressure spectrum showed
some small differences at high wavenumbers. However,
these differences do not affect the dynamics of the scales of interest
and, as we show, are easily mitigated if higher order AT schemes are used.
Similar to CAA, the instantaneous enstrophy field, the flatness and skewness of
longitudinal velocity gradient and the PDF of dissipation and enstrophy is
shown to be in excellent agreement with the synchronous simulations.

Taken together, the results obtained for both CAA and SAA and their comparison
with synchronous simulations (SFD and C10), clearly show that the
physics of turbulence even at the finest scales is resolved
accurately by the asynchronous algorithms, even though more than $50\%$
of total gridpoints are affected directly by asynchrony.

%with both asynchronous algorithms.
%While some small differences were seen for SAA, for example in the skewness
%of velocity gradient and pressure spectrum at high wavenumbers,
%CAA showed an exact agreement with synchronous. These differences however did not
%affect the dynamics of the scales of interest and as we show
%can be mitigated if higher order schemes are used.

We also presented the effect of asynchrony on computational performance.
In particular, both strong and weak scaling results showed a
 near ideal scaling for the asynchronous algorithms and significant departures from the
same for synchronous case. This improved scaling can be traced back to a
significant reduction in communications (CAA) and synchronizations (SAA), resulting
in an overall lower fraction of communication compared to computation for both
CAA and SAA. We also observed that at very high processor count $(P=262144)$,
the reduction in overall volume of communications (CAA) is more effective in
improving the scaling than relaxing explicit synchronization (SAA).
This improvement in scaling is expected to be more
consequential as we increase the problem size and processor count to levels
anticipated on exascale machines.

In conclusion, asynchronous simulations can accurately resolve
the physics at all scales
and provide better parallel performance as problem size increases. %With their ability to accurately resolve the physics of
%turbulence and their enhanced computational
%performance,
Thus, asynchronous computing presents an effective
alternative to standard computing approaches for simulation of turbulence
and other complex phenomena at unprecedented levels of physical realism on the
next generation exascale machines.

We close by mentioning some important future extensions of the current work. %In terms of the delays, we can also run simulation where
%the delay is periodic but with multiple steps, such that
%the steps are randon.
First, in this work CAA and SAA were presented as two separate algorithms, however,
a combination of the two can also be used. % where both communication and synchronization rate
%are greater than one.
This will potentially lead to further reduction in
overheads associated with the communication and synchronization.
Second, there are generalizations that can be introduced where
the maximum delay level ($L$) is different across different regions in the
domain, depending upon the level of accuracy required. This does require critical analysis
of load balancing to ensure that the processors synchronizing more
often (smaller $L$), have less computational work, so that the synchronization
time in these processors does not affect the total execution time.
Third, in order to further reduce the communication time in CAA,
new AT schemes can be derived which use only
one time level information from the buffer points and multiple delayed levels
 at internal points instead.  %Once such example of a second order
%scheme for second derivative is,
%\be
%\frac{\partial^2 u}{\partial x^2}\bigg\vert_{i}^n\approx
%\frac{\tilde{k} u_{i-1}^{n-\tilde{k}-1}
%-2 \tilde{k} u_{i}^{n-\tilde{k}-1}
%+(\tilde{k}+1) u_{i+1}^{n-\tilde{k}}-2u_i^n+ u_{i+1}^n}
%{{\Delta x}^2 (\tilde{k}+1)},
%\ee
%where instead of using multiple time levels at the buffer points, multiple
%time levels are used at the internal points.
The size of message in CAA for these new AT schemes is the same as that for the
 algorithms which communicate at every time step. The effect of these schemes
on the performance and accuracy is part of our ongoing research. %Furthermore, AT schemes
%can also be derived using spectral error as a metric
%instead of order of accuracy as described in (?).
Lastly, from the performance analysis, we observed that while the asynchronous
algorithms showed an improved scaling compared to the standard
synchronous algorithm, the cache miss rate for the former
was found to be higher. Though this miss rate reduces as the processor
count is increased, optimization in implementation
will help further push the limits of scaling and reduce the overall computation time.
This will be discussed elsewhere.

%We are currently using state-of-the-art
%profiling tools for in-depth performance analysis and optimization of our
%asynchronous solver. In the current
%implementations, synchronization was forced locally for SAA
%only when the delay was larger than $L$. However, due to
%reduced stability on increasing $L$, for large delay at
%consecutive time-steps, instabilities can be triggered locally.
%One way to mitigate this is to monitor local
%$r_c$ or $\tilde{r}_c$ and force synchronization if it crosses
%the stable threshold. While this will add additional overheads,
%it will prevent culmination of the simulation abruptly. This is a part of our
%onging work.

%\section{Conclusions}
%In this paper we presented two asynchronous implementations ..
\section{Acknowledgements}
The authors acknowledge funding from the National Science
Foundation (Grant 1439145), and
XSEDE and
Texas Advanced Computing Center
(TACC) at The University of Texas at Austin for providing
high-performance computing
resources.
%that have contributed to the research results reported
%within this paper.

\appendix
\section{Von-neumann analysis of AT schemes}
\label{sec:Vnat}
Stability analysis in the frequency domain, more commonly
known as the \textit{Von Neumann method}, has been widely
used for linear problems with constant coefficients \cite{hirsch}.
Since this method also requires all the points in the
domain to use the same numerical scheme,
we assume that each processor has only one grid point $(P=N)$
and the AT scheme use same delay $(\kt)$ on both sides for all PEs.
To proceed further, consider a diffusion equation discretized using a second-order
AT scheme in space and forward Euler in time,
\be
u_i^{n+1}=u_i^n+\frac{\alpha\Delta t}{\Delta x^2}
\left((\tilde{k}+1)u_{i-1}^{n-\tilde{k}}
-\tilde{k}u_{i-1}^{n-\tilde{k}-1}
-2u_i^{n}+(\tilde{k}+1)u_{i+1}^{n-\tilde{k}}
-\tilde{k}u_{i+1}^{n-\tilde{k}-1}\right).
\eqnlabel{dis_at1}
\ee
where $\tilde{k}$ is the delay at both left and right boundary.
%This is true when every processor has only one grid point.
Using a Fourier decomposition, $u_i^n=v^ne^{Ii\phi}$
where $I=\sqrt{-1}$ and $\phi=\kp\dx$, %a diffusive CFL $r_d=\alpha \dt/\dx^2$, and allowing
%the delay to be maximum \textit{ie} $\tilde{k}=L$,
we can simplify \eqn{dis_at1} as
\be
v^{n+1}=r_d\left(v^{n-L}(L+1)-v^{n-L-1}L\right)e^{-I\phi}
+(1-2r_d)v^n+
r_d\left(v^{n-L}(L+1)-v^{n-L-1}L\right)e^{I\phi}.
\eqnlabel{vn2}
\ee
where $r_d=\alpha \dt/\dx^2$ is the diffusive CFL and delay
is equal to the maximum allowed delay \textit{ie} $\tilde{k}=L$.
On taking the $Z$-transform of \eqn{vn2} with
\be
v^n=v(z);~~~v^{n-L}=z^{-L}v(z)
\ee
we get a polynomial of order $L+2$ in $z$ which reads as,
\be
z^{L+2}-z^{L+1}(1-2r_d)
-r_d(e^{-I\phi}-e^{I\phi})(L+1)z
-r_d(e^{-I\phi}-e^{I\phi})L=0
\eqnlabel{eqn_z}
\ee
For stability we require the amplitude all the harmonics to
not grow in time. This is possible only if all the roots of
\eqn{eqn_z} are within a unit disc in the complex $Z$-plane and
the root $z=1$ has multiplicity one \cite{hirsch}. The roots of the above
equation can be computed numerically for different $L$ and
these can be used to obtain the maximum $r_d$ such that all roots
lie within a unit disc. This is the largest limit for which
the numerical scheme is stable and as before, we denote it as $\rdml$.
Instead of computing the roots,
the sufficient conditions that guarantee that all the roots lie
within the disc of radius one can also be computed using
the \textit{Schur-Cohn} criteria \cite{Miller1971}. As an
example, we use \textit{Schur-Cohn} criteria to get these conditions for $L=1$.
Since for $L=1$, \eqn{eqn_z} reduces to a cubic polynomial,
we get a total of three conditions,
\be
 4 r_d^2 \cos ^2(\phi )-1<0,
\ee
\be
16r_d^4 \cos ^4(\phi )-4r_d^2 (4r_d (\text{rd}+1)+3) \cos ^2(\phi )+1>0,
\ee
\be
\begin{aligned}
&8 r_d \sin ^2\left(\frac{\phi }{2}\right)
\left(-2r_d^2+2r_d (2r_d \cos (\phi )-r_d \cos (2 \phi )+
\cos (\phi ))+1\right)^2 ~~~\\
&~~~\left(3r_d^3 \cos (3 \phi )+3 \left(3r_d^2-1\right)r_d \cos
(\phi )+(r_d-1) \left(2r_d^2 \cos (2 \phi )+2r_d^2-1\right)\right)>0.
\end{aligned}
\ee
These equations have only two parameters which can be varied
to determine $\rdml$ that satisfies all the three conditions,
for all values of $\phi\in[0,\pi]$. This gives us $\rdm(1)\approx0.25$
which is consistent with the stability limit obtained from the matrix stability analysis
presented in section 3. For larger values of $r_d$, at least one of the above conditions is
violated and thus instabilities can be triggered.
We can see that this value is less
than the traditional stability limit for a synchronous scheme $(\rdm(0)=0.5)$.
It is worthwhile to note that similar constraints can also be written for
$L=0$ or the standard synchronous scheme which give us the
expected limit of $\rdm(0)=0.5$. This idea can be extended to get the constraints
for the stability for higher $L$ as well.
\cite{Miller1971} discusses techniques to extend the \textit{Schur-cohn} criteria
to higher order polynomials efficiently.

The above analysis assumes a worst case scenario with
two-sided delays at all points. A more
practical scenario would be to have delays on either
the left or the right PE boundary. Since the schemes are symmetric,
 the stability limit for an AT
scheme with delays on left boundary, is also applicable to
AT scheme with delays at the right boundary. Repeating the
above procedure, but now considering delay $\tilde{k}=L$
on the left boundary and $\tilde{k}=0$ on the right, gives us
following three conditions for all the roots of the $z$ equation
to lie with a unit disc,
\be
r_d^2-1<0,
\ee
\be
1 - r_d^2 \left(3 + 4 r_d (1 + r_d)\right) +
r_d \left(2 r_d^2 + 4 r_d^3\right) cos(\phi)>0,
\ee
\begin{equation}
\begin{aligned}
8 r_d (-1 + r_d^2)  \sin ^2\left(\frac{\phi }{2}\right)
\big(-1 + r_d + 4 r_d^2 + r_d^3 - 4 r_d^4 + ~~~~~~~~~~\\
   r_d (-1 + 2 r_d) (1 + 2 r_d)^2 cos(\phi) +
   r_d^2 \left(1 - 4 r_d (1 + r_d)\right) cos(2\phi) +
   r_d^3 cos(3\phi) \big) >0.
\end{aligned}
\end{equation}

These new conditions are satisfied for $\rdm(1)\approx0.33$ for all $\phi\in[0,\pi]$, which
is lower than the previously computed bound of $\rdm(1)\approx0.25$.
%The values are also consistent with the $r_d$ stability limit
%obtained using the matrixof the evolution matrix, for both one-sided and two-sided delays.
In \rfig{stability_cfl2} we show the $\rdml$ (solid circles) obtained for varying $L$ for
one-sided delays. Also plotted is $\rdm(0)/f(L)$ (solid line) defined in \eqn{rdml},
where $f(L)\approx0.56L+0.9$, and $\rdma=\rdm(0) \times f(L)$ in hollow circles
with a dashed line corresponding to the synchronous limit $\rdm(0)=0.5$.
\begin{figure}[h]
\begin{center}
\includegraphics[width=0.4\textwidth]{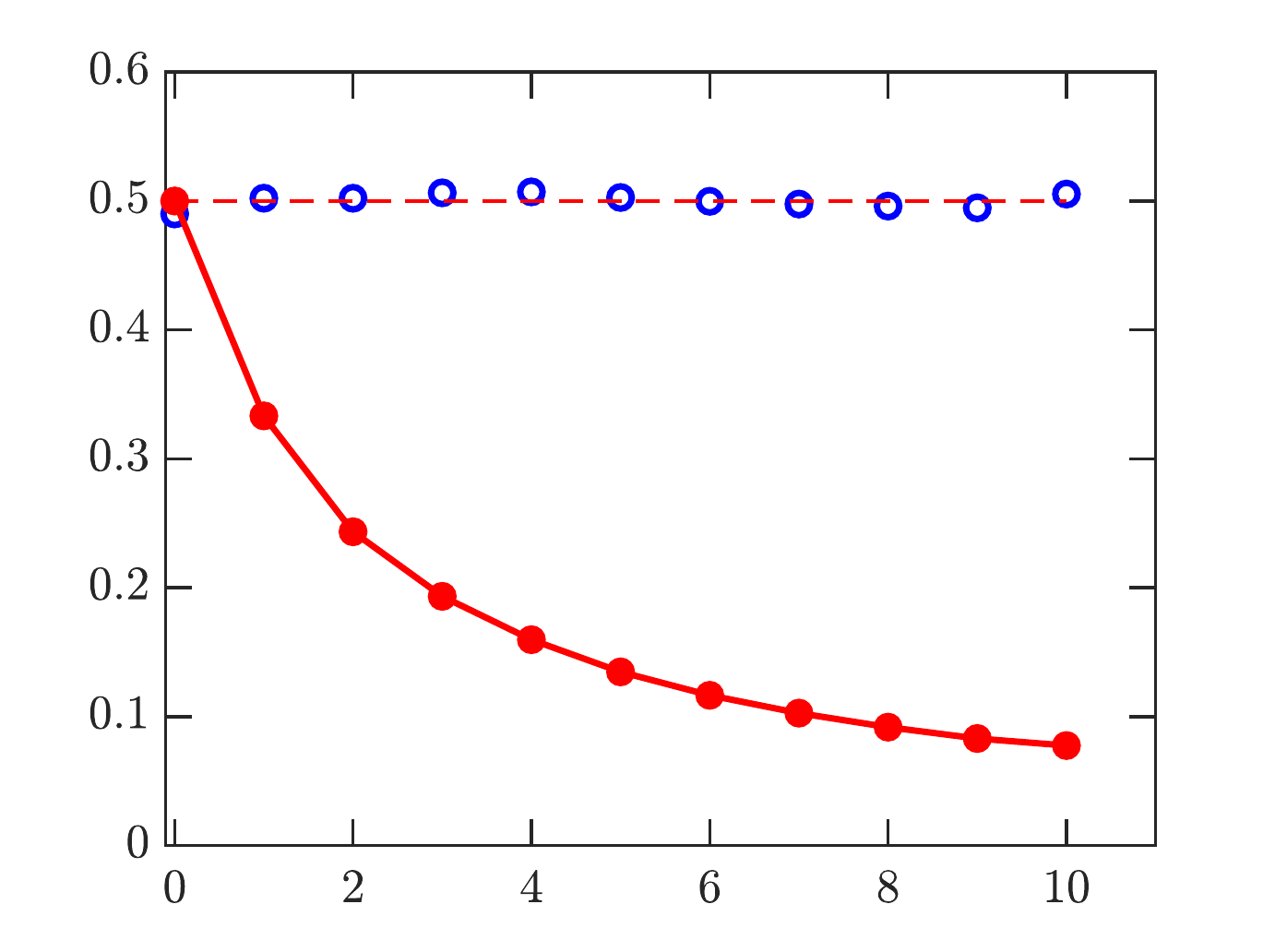}
\begin{picture}(0,0)
\put(-95,-10){$L$}
\put(-210,70){\rotatebox{90}{$\rdml$}}
\end{picture}
\caption{Variation of stability limit $\rdml$ (solid) and $\rdma $ (hollow)
%\approx \rdm(0) \times (0.56L+0.9)$
with $L$ for diffusion equation with delays on only one side.
}
\figlabel{stability_cfl2}
\end{center}
\end{figure}
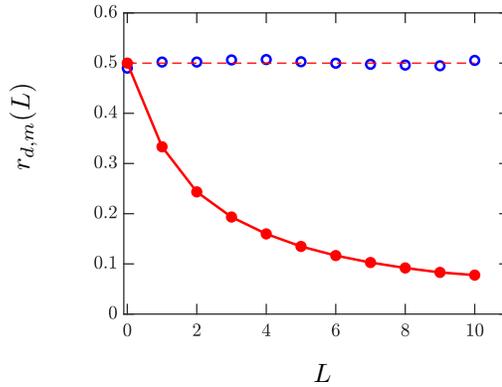

We see that while $\rdml$ decreases as delay increases, the
\textit{effective asynchronous} CFL ($ \rdma$) is close to $0.5$ for all $L$ and supports
the argument presented in section 3.

\section{Asynchrony-tolerant schemes}
\label{sec:ATS}
\begin{table}[h]
\begin{center}
{\tabulinesep=0.5mm
\begin{tabu}{|c|c|c|}
\hline
(Derivative, & Boundary & Scheme\\
Order) & &\\
\hline
\hline
(2,4) & Left & {$\scriptsize\begin{array}{c}
\frac{1}{2}(\k{}^2+3\k{}+2)\left({-\U{i+2}{n} + 16 \U{i+1}{n} - 30 \U{i}{n} + 16\U{i-1}{n-\k{}} - \U{i-2}{n-\k{}}}\right)/{12\dx^2}  \\
 -(\k{}^2+2\k{}) \left({-\U{i+2}{n} + 16 \U{i+1}{n} - 30 \U{i}{n} + 16\U{i-1}{n-\k{}-1} - \U{i-2}{n-\k{}-1}}\right)/{12\dx^2} \\
 +\frac{1}{2}(\k{}^2+\k{}) \left({-\U{i+2}{n} + 16 \U{i+1}{n} - 30 \U{i}{n} + 16\U{i-1}{n-\k{}-2} - \U{i-2}{n-\k{}-2}}\right)/{12\dx^2}
\end{array}$} \\
\hline
(2,4) & Right &{$\scriptsize\begin{array}{c}
%{\tiny $\begin{array} {c}
\frac{1}{2}(\k{}^2+3\k{}+2) \left({-\U{i+2}{n-\k{}} + 16 \U{i+1}{n-\k{}} - 30 \U{i}{n} + 16\U{i-1}{n} - \U{i-2}{n}}\right)/{12\dx^2}  \\
 -(\k{}^2+2\k{}) \left({-\U{i+2}{n-\k{}-1} + 16 \U{i+1}{n-\k{}-1} - 30 \U{i}{n} + 16\U{i-1}{n} - \U{i-2}{n}}\right)/{12\dx^2}  \\
 +\frac{1}{2}(\k{}^2+\k{}) \left({-\U{i+2}{n-\k{}-2} + 16 \U{i+1}{n-\k{}-2} - 30 \U{i}{n} + 16\U{i-1}{n} - \U{i-2}{n}}\right)/{12\dx^2}
\end{array}$}\\
\hline

(1,4) & Left & {$\scriptsize\begin{array}{c}
%{\tiny $\begin{array} {c}
\frac{1}{2}(\k{}^2+3\k{}+2)\left({-\U{i+2}{n} + 8 \U{i+1}{n} - 8\U{i-1}{n-\k{}} + \U{i-2}{n-\k{}}}\right)/{12\dx}  \\
 -(\k{}^2+2\k{})\left({-\U{i+2}{n} + 8 \U{i+1}{n} - 8 \U{i-1}{n-\k{}-1} + \U{i-2}{n-\k{}-1}}\right)/{12\dx} \\
 +\frac{1}{2}(\k{}^2+\k{})\left({-\U{i+2}{n} + 8 \U{i+1}{n} - 8 \U{i-1}{n-\k{}-2} + \U{i-2}{n-\k{}-2}}\right)/{12\dx}
\end{array}$}\\
\hline
(1,4) & Right &
{ $ \scriptsize\begin{array}{c}
%{\tiny $\begin{array} {c}
\frac{1}{2}(\k{}^2+3\k{}+2)\left({-\U{i+2}{n-\k{}} + 8 \U{i+1}{n-\k{}} - 8\U{i-1}{n} + \U{i-2}{n}}\right)/{12\dx}  \\
 -(\k{}^2+2\k{})\left({-\U{i+2}{n-\k{}-1} + 8 \U{i+1}{n-\k{}-1} - 8 \U{i-1}{n} + \U{i-2}{n}}\right)/{12\dx} \\
 +\frac{1}{2}(\k{}^2+\k{})\left({-\U{i+2}{n-\k{}-2} + 8 \U{i+1}{n-\k{}-2} - 8 \U{i-1}{n} + \U{i-2}{n}}\right)/{12\dx}
\end{array}$}\\
\hline
(2,6) & Left &
{ $\scriptsize \begin{array}   {c}
%{\tiny $\begin{array} {c}
\frac{1}{6}(\k{}^3+6\k{}^2+11\k{}+6)
\left({2\U{i+3}{n}-27\U{i+2}{n} + 270 \U{i+1}{n} - 490 \U{i}{n} + 270 \U{i-1}{n-\k{}} - 27\U{i-2}{n-\k{}} + 2\U{i-3}{n-\k{}}}\right)/{180\dx^2}  \\
- \frac{1}{2}(\k{}^3+5\k{}^2+6\k{})
\left({2\U{i+3}{n}-27\U{i+2}{n} + 270 \U{i+1}{n} - 490 \U{i}{n} + 270 \U{i-1}{n-\k{}-1} - 27\U{i-2}{n-\k{}-1} + 2\U{i+3}{n-\k{}-1}}\right)/{180\dx^2}  \\
+ \frac{1}{2}(\k{}^3+4\k{}^2+3\k{})
\left({2\U{i+3}{n}-27\U{i+2}{n} + 270 \U{i+1}{n} - 490 \U{i}{n} + 270 \U{i-1}{n-\k{}-2} - 27\U{i-2}{n-\k{}-2} + 2\U{i+3}{n-\k{}-2}}\right)/{180\dx^2}  \\
- \frac{1}{6}(\k{}^3+3\k{}^2+2\k{})
\left({2\U{i+3}{n}-27\U{i+2}{n} + 270 \U{i+1}{n} - 490 \U{i}{n} + 270 \U{i-1}{n-\k{}-3} - 27\U{i-2}{n-\k{}-3} + 2\U{i+3}{n-\k{}-3}}\right)/{180\dx^2}  \\
\end{array}$}\\
\hline
(2,6) & Right &
{$ \scriptsize\begin{array}{c}
%{\tiny $\begin{array} {c}
\frac{1}{6}(\k{}^3+6\k{}^2+11\k{}+6)
\left({2\U{i+3}{n-\k{}}-27\U{i+2}{n-\k{}} + 270 \U{i+1}{n-\k{}} - 490 \U{i}{n} + 270 \U{i-1}{n} - 27\U{i-2}{n} + 2\U{i-3}{n}}\right)/{180\dx^2}  \\
- \frac{1}{2}(\k{}^3+5\k{}^2+6\k{})
\left({2\U{i+3}{n-\k{}-1}-27\U{i+2}{n-\k{}-1} + 270 \U{i+1}{n-\k{}-1} - 490 \U{i}{n} + 270 \U{i-1}{n} - 27\U{i-2}{n} + 2\U{i-3}{n}}\right)/{180\dx^2}  \\
+ \frac{1}{2}(\k{}^3+4\k{}^2+3\k{})
\left({2\U{i+3}{n-\k{}-2}-27\U{i+2}{n-\k{}-2} + 270 \U{i+1}{n-\k{}-2} - 490 \U{i}{n} + 270 \U{i-1}{n} - 27\U{i-2}{n} + 2\U{i-3}{n}}\right)/{180\dx^2}  \\
- \frac{1}{6}(\k{}^3+3\k{}^2+2\k{})
\left({2\U{i+3}{n-\k{}-3}-27\U{i+2}{n-\k{}-3} + 270 \U{i+1}{n-\k{}-3} - 490 \U{i}{n} + 270 \U{i-1}{n} - 27\U{i-2}{n} + 2\U{i-3}{n}}\right)/{180\dx^2}  \\
\end{array}$}\\
\hline
(1,6) & Left &
{ $\scriptsize \begin{array}   {c}
%{\tiny $\begin{array} {c}
\frac{1}{6}(\k{}^3+6\k{}^2+11\k{}+6)
\left({\U{i+3}{n}-9\U{i+2}{n} + 45 \U{i+1}{n}  -45 \U{i-1}{n-\k{}} +9\U{i-2}{n-\k{}} -\U{i-3}{n-\k{}}}\right)/{60\dx}  \\
- \frac{1}{2}(\k{}^3+5\k{}^2+6\k{})
\left({\U{i+3}{n}-9\U{i+2}{n} + 45 \U{i+1}{n} - 45 \U{i-1}{n-\k{}-1} +9\U{i-2}{n-\k{}-1} -\U{i+3}{n-\k{}-1}}\right)/{60\dx}  \\
+ \frac{1}{2}(\k{}^3+4\k{}^2+3\k{})
\left({\U{i+3}{n}-9\U{i+2}{n} + 45 \U{i+1}{n} - 45 \U{i-1}{n-\k{}-2} +9\U{i-2}{n-\k{}-2} -\U{i+3}{n-\k{}-2}}\right)/{60\dx}  \\
- \frac{1}{6}(\k{}^3+3\k{}^2+2\k{})
\left({\U{i+3}{n}-9\U{i+2}{n} + 45 \U{i+1}{n} -45 \U{i-1}{n-\k{}-3} +9\U{i-2}{n-\k{}-3} - \U{i+3}{n-\k{}-3}}\right)/{60\dx}  \\
\end{array}$}\\
\hline
(1,6) & Right &
{$ \scriptsize\begin{array}{c}
%{\tiny $\begin{array} {c}
\frac{1}{6}(\k{}^3+6\k{}^2+11\k{}+6)
\left({\U{i+3}{n-\k{}}-9\U{i+2}{n-\k{}} + 45 \U{i+1}{n-\k{}}  -45 \U{i-1}{n} +9\U{i-2}{n} -\U{i-3}{n}}\right)/{60\dx}  \\
- \frac{1}{2}(\k{}^3+5\k{}^2+6\k{})
\left({\U{i+3}{n-\k{}-1}-9\U{i+2}{n-\k{}-1} + 45 \U{i+1}{n-\k{}-1} -45 \U{i-1}{n} +9\U{i-2}{n} -\U{i-3}{n}}\right)/{600\dx}  \\
+ \frac{1}{2}(\k{}^3+4\k{}^2+3\k{})
\left({\U{i+3}{n-\k{}-2}-9\U{i+2}{n-\k{}-2} + 45 \U{i+1}{n-\k{}-2} -45 \U{i-1}{n} +9\U{i-2}{n} -\U{i-3}{n}}\right)/{60\dx}  \\
- \frac{1}{6}(\k{}^3+3\k{}^2+2\k{})
\left({\U{i+3}{n-\k{}-3}-9\U{i+2}{n-\k{}-3} + 45 \U{i+1}{n-\k{}-3} -45 \U{i-1}{n} +9\U{i-2}{n} -\U{i-3}{n}}\right)/{60\dx}  \\
\end{array}$}\\
\hline
\end{tabu}}
\caption{Asynchrony-tolerant (AT) schemes for left and right boundary
used in numerical simulations (in \rsec{results}) for first and second derivative.}
\label{tab:atschemes}
\end{center}
\end{table}

% References
%
% Following citation commands can be used in the body text:
% Usage of \cite is as follows:
%   \cite{key}          ==>>  [#]
%   \cite[chap. 2]{key} ==>>  [#, chap. 2]
%   \citet{key}         ==>>  Author [#]

% References with bibTeX database:
\newpage
\bibliographystyle{model1-num-names}
\bibliography{main}

% Authors are advised to submit their bibtex database files. They are
% requested to list a bibtex style file in the manuscript if they do
% not want to use model1-num-names.bst.

% References without bibTeX database:


% \bibitem must have the following form:
%   \bibitem{key}...
%


% \end{thebibliography}

\end{document}